\documentclass[reprint,aps,pra,nofootinbib,superscriptaddress]{revtex4-2}
\usepackage{graphicx}
\usepackage {amsmath, amsthm, amssymb, mathtools,
 braket, tikz, xcolor}
\usetikzlibrary{positioning,arrows.meta}
\graphicspath{{fig/}}

\newcommand{\Tr}{\mathrm{Tr}}
\newcommand{\SU}{\mathrm{SU}}

\newcommand{\mH}{\mathcal{H}}

\newcommand{\mL}{\mathcal{L}}
\newcommand{\mU}{\mathcal{U}}

\newcommand{\mK}{\mathcal{K}}

\newcommand{\necase}{\ne}
\newcommand{\ME}{\mathrm{ME}}
\newcommand{\Y}{\mathcal{Y}}
\newcommand{\Ucal}{\mathcal{U}}
\newcommand{\Vcal}{\mathcal{V}}
\newcommand{\Hcal}{\mathcal{H}}
\newcommand{\C}{\mathbb{C}}
\newcommand{\Id}{I}
\newcommand{\Hom}{\operatorname{Hom}}
\newcommand{\ydiagram}[1]{%
  \tikz[baseline=-0.55ex,x=1.25em,y=1.25em,line width=0.35pt]{#1}}
\newcommand{\ydold}[2]{%
  \filldraw[draw=black,fill=black!12] (#1,-#2) rectangle ++(1,-1);}
\newcommand{\ydblank}[2]{%
  \draw[black] (#1,-#2) rectangle ++(1,-1);}
\newcommand{\ydnew}[3]{%
  \filldraw[draw=black,fill=blue!10] (#1,-#2) rectangle ++(1,-1);
  \path (#1,-#2) ++(0.5,-0.5) node {\scriptsize #3};}

\newtheorem{theorem}{Theorem}
\newtheorem{proposition}{Proposition}
\newtheorem{lemma}{Lemma}

\makeatletter
\let\uc@PackageWarning\PackageWarning
\def\PackageWarning#1#2{%
  \def\uc@warningpkg{#1}%
  \def\uc@namerefpkg{nameref}%
  \ifx\uc@warningpkg\uc@namerefpkg
  \else
    \uc@PackageWarning{#1}{#2}%
  \fi
}
\makeatother
\usepackage[
  colorlinks=true,
  linkcolor=blue,
  citecolor=blue,
  urlcolor=blue
]{hyperref}
\makeatletter
\let\PackageWarning\uc@PackageWarning
\makeatother
\makeatletter
\def\@bibdataout@aps{%
  \immediate\write\@bibdataout{%
    @CONTROL{%
      apsrev42Control%
      \longbibliography@sw{%
        ,author="48",editor="1",pages="0",title="0",year="1"%
      }{%
        ,author="48",editor="1",pages="0",title="",year="1"%
      }%
    }%
  }%
  \if@filesw
    \immediate\write\@auxout{\string\citation{apsrev42Control}}%
  \fi
}%
\makeatother

\DeclareRobustCommand{\reviewchange}[1]{\textcolor{black}{#1}}

\begin{document}
\preprint{APS/123-QED}

\title{Comparison of unknown unitary channels with multiple queries}

\author{Yutaka Hashimoto}
\affiliation{Department of Physics, Graduate School of Science, The University of Tokyo, 7-3-1 Hongo, Bunkyo City, Tokyo 113-0033, Japan}
\author{Akihito Soeda}
\affiliation{Principles of Informatics Research Division, National Institute of Informatics, 2-1-2 Hitotsubashi, Chiyoda-ku, Tokyo 101-8430, Japan}
\affiliation{\reviewchange{Informatics Program, Graduate Institute for Advanced Studies, SOKENDAI}, 2-1-2 Hitotsubashi, Chiyoda-ku, Tokyo 101-8430, Japan}
\affiliation{Department of Physics, Graduate School of Science, The University of Tokyo, 7-3-1 Hongo, Bunkyo City, Tokyo 113-0033, Japan}
\author{Mio Murao}
\affiliation{Department of Physics, Graduate School of Science, The University of Tokyo, 7-3-1 Hongo, Bunkyo City, Tokyo 113-0033, Japan}
\affiliation{Trans-Scale Quantum Science Institute, The University of Tokyo, 7-3-1 Hongo, Bunkyo City, Tokyo 113-0033, Japan}

\date{\today}

\begin{abstract}
  Comparison of quantum objects is the task of determining relational properties, such
  as whether two unknown objects are the same or different, without identifying the
  objects themselves.  It is a basic information-processing primitive for quantum
  states, channels, and measurements.  Multiple copies of unknown states or multiple
  queries to unknown channels are natural resources for improving comparison, and the
  optimal strategy for pure-state comparison with multiple copies is known.  For
  unitary-channel comparison with multiple queries, however, the optimal strategy has been
  unclear because different queries can be arranged in parallel, sequentially,
  adaptively, or in more general causal structures.  We study comparison of two unknown
  $d$-dimensional unitary channels using $N_1$ and $N_2$ queries to the two channels,
  respectively, under the promise that they are either identical Haar-random unitaries or
  independent Haar-random unitaries.  We optimize over valid general testers,
  a class containing
  ordinary quantum testers, classically controlled causal orders, and
  indefinite-causal-order processes.  We characterize
  the optimal minimum-error and one-sided unambiguous comparison probabilities for
  arbitrary finite query numbers.  The upper bound in fact holds for a larger positive
  relaxation, while the optimum is attained by a valid parallel tester.  The optimum is
  determined by a finite
  representation-theoretic parameter built from Young diagrams and
  Littlewood--Richardson coefficients.
  For fixed $N_1$, the performance is saturated for every
  $N_2\geq(d-1)N_1$.  By contrast, the corresponding pure-state comparison parameter
  has no finite-$N_2$ saturation and reaches its known-reference limit only
  asymptotically, yielding continued success-probability improvement in the nontrivial
  decision regime.  This highlights a sharp difference between comparison tasks for
  states and channels, analogous to distinctions known in quantum discrimination.
\end{abstract}

% REVTeX warns when keywords are provided without the showkeys option.
% Keep the technical version visually unchanged by omitting hidden keywords here.

\maketitle

\section{Introduction}

Efficiently learning properties of unknown quantum objects is a fundamental task in
quantum mechanics and quantum information. Commonly investigated target objects are
quantum states and quantum channels, but they are not restricted to these. There are
different settings and strategies for learning depending on properties to learn, prior
information about the object, and given resources.

Quantum state discrimination
\cite{holevoStatisticalDecisionTheory1973,
  helstromQuantumDetectionEstimation1969,
  yuenOptimumTestingMultiple1975}
is one of the settings to learn the identity of a quantum state when a set of candidate
states and a distribution of the candidates are given. When the figure of merit for
optimization is given by the success probability of learning the correct candidate, it
is called minimum-error quantum state discrimination
\cite{holevoStatisticalDecisionTheory1973}. The number of available copies of target
states is a resource that can improve the success probability or the fidelity of quantum
state discrimination. The pioneering works on quantum state discrimination contributed
to establishing the field of quantum information, and similarly, quantum channel
discrimination has been investigated
\cite{acinStatisticalDistinguishabilityUnitary2001,
  sacchiOptimalDiscriminationQuantum2005,
  childsQuantumInformationPrecision2000,
  pirandolaFundamentalLimitsQuantum2019}.

Quantum discrimination tasks are learning tasks for a \textit{single} quantum object in
question. When there are \textit{two} unknown quantum objects, and we want to learn the
relationship between the objects, we \textit{compare} the two target objects. It is
always possible to first identify the description of each unknown object by using
quantum state or channel \reviewchange{tomography} and then compare the descriptions of the
objects. However, such an approach requires correctly identifying both objects and
therefore incurs the combined errors of the two identification procedures. It also
provides unnecessary information about the identity of each object.
In contrast, only the relationship between the objects is necessary for comparison. A
method to directly compare two objects without identifying their descriptions is
preferable for more efficient learning of the difference between the target objects,
especially when the number of available copies or queries to each target object is limited.
A canonical example of such a direct comparison method is the swap test, used for
quantum fingerprinting and quantum digital signatures
\cite{buhrmanQuantumFingerprinting2001,gottesmanQuantumDigitalSignatures2001},
which estimates the overlap of two unknown quantum states without identifying the
states themselves.

A simple but fundamental task to compare two objects is to determine whether the two
objects are the same or not. This decision task of \textit{comparison}%
\footnote{In Ref.~\cite{gourComparisonQuantumChannels2019}, the term ``comparison of
quantum channels'' is used in a different context from that of the present work.}
of two pure states \reviewchange{is} introduced and analyzed in
Ref.~\cite{barnettComparisonTwoUnknown2003}.
Extensions to mixed states and multiple-copy settings \reviewchange{are} studied in
Refs.~\cite{jexComparingStatesMany2004,
  cheflesUnambiguousComparisonStates2004,
  kleinmannGeneralizationQuantumstateComparison2005,
  sedlakUnambiguousComparisonEnsembles2008,
  pangComparisonMixedQuantum2011,
  hayashiQuantumstateComparisonDiscrimination2018,
  fanizzaSwapTestOptimal2020,
  badescuQuantumStateCertification2019}.
Comparison of quantum measurements \reviewchange{is} studied in
Ref.~\cite{zimanUnambiguousComparisonQuantum2009}. Related to comparison, equivalence
determination
\cite{shimboEquivalenceDeterminationUnitary2018,soedaOptimalQuantumDiscrimination2021}
is the decision problem for an unknown unitary channel $U_?$ which is equal to either of
two candidates $U_1$ and $U_2$.

For quantum channels, comparison of two unknown unitary channels on a qubit $(d=2)$
system \reviewchange{is} considered in Ref.~\cite{anderssonComparisonUnitaryTransforms2003}, where
one-query optimal strategies \reviewchange{are} obtained in the \textit{unambiguous} sense. Here,
\textit{unambiguous} refers to a setting in which erroneous conclusive outcomes are
forbidden while inconclusive outcomes are allowed
\cite{cheflesUnambiguousDiscriminationLinearlyIndependent1998,
  fengUnambiguousDiscriminationMixed2004,
  wangUnambiguousDiscriminationQuantum2006}.
The optimal unambiguous comparison strategy for a general $d$-dimensional system \reviewchange{is}
derived in Ref.~\cite{sedlakUnambiguousComparisonUnitary2009}. Decision problems in which an
unknown unitary black box is compared with reference or program unitaries \reviewchange{are} studied
in both minimum-error and unambiguous settings in
Ref.~\cite{hilleryDecisionProblemsQuantum2010}. Single-query comparison of
general random quantum channels and measurements \reviewchange{is} analyzed in
Ref.~\cite{markiewiczSingleShotComparisonRandom2025}.
Although the no-cloning theorem for quantum channels
\cite{chiribellaOptimalCloningUnitary2008} forbids copying an unknown unitary channel
with a single query, multiple queries to an unknown unitary channel are natural resources
that can be obtained by running the same experimental setup repeatedly.

However, allowing multiple queries introduces additional complexity, since queries to
each black box can be combined in parallel, sequentially, adaptively, or through more
general higher-order quantum processes.
Prior work establishes the relevant landscape. For quantum channel discrimination,
adaptive strategies can outperform nonadaptive ones
\cite{harrowAdaptiveNonadaptiveStrategies2010}, while memory can matter in general but is
unnecessary for optimally discriminating and estimating certain sets of unitary channels
\cite{chiribellaMemoryEffectsQuantum2008}.
More recently, sequential and indefinite-causal-order strategies have been shown
to outperform parallel ones for some unitary-channel discrimination tasks
\cite{bavarescoUnitaryChannelDiscrimination2022}. These results motivate treating the allowed
causal structure explicitly rather than assuming parallel queries from the outset.

The \textit{quantum tester} formalism
\cite{chiribellaTransformingQuantumOperations2008,
  chiribellaQuantumCircuitsArchitecture2008,
  chiribellaTheoreticalFrameworkQuantum2009,
  zimanProcessPOVMMathematical2008}
provides a general description of strategies involving causally ordered queries to
channels, with closely related formalisms for multi-round quantum interactions
developed independently in theoretical computer science
\cite{gutoskiTowardGeneralTheory2007}. This framework accommodates both parallel- and
sequential-query strategies for channel comparison, as well as strategies beyond fixed
causal order
\cite{oreshkovQuantumCorrelationsNo2012,
  chiribellaQuantumComputationsDefinite2013,
  baumelerSpaceLogicallyConsistent2016,
  wechsQuantumCircuitsClassical2021a}. In this paper, the admissible operational class
is the set of valid general testers
\cite{bavarescoStrictHierarchy2021,
  bavarescoUnitaryChannelDiscrimination2022,
  andreStrategyOptimizationBayesian2026},
which encompasses fixed-order quantum
testers (quantum combs), classically controlled causal orders, and
indefinite-causal-order processes.

We investigate how multiple queries to each quantum channel can improve
the success probability and what role the causal order of the queries plays in efficient
property learning of quantum channels. We analyze optimal strategies of quantum channel
comparison of general $d$-dimensional unitary channels $U_1$ and $U_2$ when multiple but
finite $N_1$ and $N_2\geq N_1$ queries to channels $U_1$ and $U_2$, respectively, are
provided. We set the probability that $U_1=U_2$ to be $p$ and the probability that $U_1$ and
$U_2$ are independently distributed to be $1-p$, and we use the uniform
distribution on $\SU(d)$ for the candidate channels.

\begin{figure*}[t]
  \centering
  \includegraphics[width=0.94\linewidth]{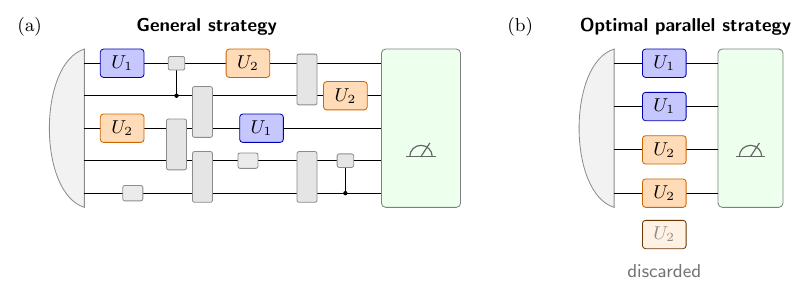}
  \caption{Conceptual illustration of the optimal parallel strategy for $d=2$,
    $N_1=2$, and $N_2=3$. Although a general strategy may insert two queries to
    $U_1$ and three queries to $U_2$ into an arbitrary quantum process, the optimum
    is achieved by a parallel tester with two information-carrying queries to each
    unitary, and the extra query to $U_2$ can be discarded.}
  \label{fig:concept-parallel-saturation}
\end{figure*}

Our main contribution is twofold.
First, we introduce a finite representation-theoretic parameter
$\Theta_d(N_1,N_2)$ and show that it completely determines the optimal
minimum-error and one-sided unambiguous comparison probabilities for all
finite query numbers.
We further show that these optimal probabilities can be achieved by a valid parallel
tester.
Second, we prove a saturation phenomenon. For fixed $N_1$, increasing
$N_2$ beyond $(d-1)N_1$ gives no further improvement
in the optimal comparison probabilities.
Figure~\ref{fig:concept-parallel-saturation} provides a schematic overview of the
comparison setting.

This paper is organized as follows. In Section~\ref{sec:unitary-comparison} we review
the general tester formulation and present our setting of unitary comparison with
multiple queries to the two unknown channels. In Section~\ref{sec:optimal-comparison} we
obtain the optimal minimum-error strategy for all finite query numbers. We extend the
analysis to the one-sided unambiguous setting in
Section~\ref{sec:optimal-unambiguous-comparison}. In Section~\ref{sec:saturation} we
prove the saturation formula and discuss the relation to the comparison problem when one
of the unitary channels is known. We present the summary in
Section~\ref{sec:Conclusion}.

\section{Unitary comparison}
\label{sec:unitary-comparison}

\subsection{Notations}
All systems are finite-dimensional with $d\geq2$. A unitary channel is denoted by
$\mU$, and a corresponding unitary operator by $U$ (similarly $\mU_i$ and $U_i$). Since
$U$ and $e^{i\phi}U$ define the same channel, equality of unitary channels is
understood up to a global phase of the operator. Unitary operators are sampled from the
Haar measure on $\SU(d)$.

We use the unnormalized Choi convention
\cite{choiCompletelyPositiveLinear1975,jamiolkowskiLinearTransformationsWhich1972}.
For a unitary channel from $\Hcal_{\mathrm{in}}\cong\C^d$ to
$\Hcal_{\mathrm{out}}\cong\C^d$, its Choi operator
$C_U\in\mL(\Hcal_{\mathrm{in}}\otimes\Hcal_{\mathrm{out}})$ is defined by
\[
  C_U=(\Id\otimes U)|\Phi\rangle\langle\Phi|(\Id\otimes U^\dagger),
  \qquad
  |\Phi\rangle=\sum_{r=1}^{d}|r\rangle|r\rangle,
\]
so that $\Tr_{\mathrm{out}}C_U=\Id_{\mathrm{in}}$.

For $r\geq0$, let $\Y_{r,d}$ denote the set of Young diagrams with $r$ boxes and at most
$d$ rows.  For $\lambda\in\Y_{r,d}$, let $\Ucal_\lambda$ be the corresponding polynomial
$\mathrm{U}(d)$ irreducible representation, and let
\[
  D_\lambda:=\dim\Ucal_\lambda.
\]
We label the Schur--Weyl sectors by Young diagrams and compute their tensor-product
multiplicities using the standard polynomial $\mathrm{U}(d)$ Littlewood--Richardson
coefficients $c_{\lambda\mu}^{\nu}$,
\begin{equation}
  \Ucal_\lambda\otimes\Ucal_\mu
  \cong
  \bigoplus_\nu \Ucal_\nu\otimes\C^{c_{\lambda\mu}^{\nu}}.
  \label{eq:LR-definition}
\end{equation}
This is only a convenient bookkeeping convention: the Haar-averaged Choi operators are
insensitive to the global phase of the unitary, so the same sector labels and
multiplicities apply to the corresponding $\SU(d)$ representations relevant to the
channel problem.\footnote{Within a fixed tensor degree, two restrictions to $\SU(d)$
are inequivalent unless the Young diagrams are identical, so the Schur blocks appearing
in a fixed tensor power remain unambiguously labelled by Young diagrams.}

\subsection{Problem setting}

\textit{Unitary comparison} is a task of determining whether two unknown unitary
channels $\mU_1$ and $\mU_2$ are the same or different under a promise on
$\mU_1$ and $\mU_2$, by querying $\mU_1$ and $\mU_2$ multiple times.
We consider strategies with $N_1$ queries to the first channel and $N_2$ queries to the
second channel.
Without loss of generality, we assume $N_2\geq N_1$ in this paper, as we can
always choose the unitary channel with a smaller number of queries to be $\mU_1$
when $N_2\neq N_1$.
We consider the promise that one of the following two cases occurs with probability
$p$ and $1-p$, respectively.
\begin{description}
  \item[Case 1 $U_1=U_2$] In the perfectly correlated case, denoted by $H_=$, $U_1$ is
        chosen randomly over $\SU(d)$. $U_2$ is the same as $U_1$.
  \item[Case 2 $U_1\neq U_2$] In the independently distributed case, denoted by
        $H_{\necase}$, $U_1$ and $U_2$ are chosen randomly over $\SU(d)$, independently.
\end{description}
Although Case 2 contains the case of $U_1=U_2$, we call Case 2 the $U_1\neq U_2$ case
since $U_1=U_2$ only happens with probability zero in this setting. Below we use $H_=$
and $H_{\necase}$ as shorthand for Case 1 and Case 2, respectively.

The objective of unitary channel comparison is to obtain the \textit{optimal} strategy
of determining whether $H_=$ or $H_{\necase}$ holds using $N_1$ queries to $U_1$ and
$N_2$ queries to $U_2$ under a given figure of merit. For minimum-error comparison, the
figure of merit is the average success probability
\begin{align}
  P_{\ME}
   & :=p\Pr(\text{decide }H_=\mid H_=)\nonumber                 \\
   & \quad+(1-p)\Pr(\text{decide }H_{\necase}\mid H_{\necase}).
  \label{eq:average-success-probability}
\end{align}
Later in Section~\ref{sec:optimal-unambiguous-comparison}, we introduce another figure
of merit for one-sided unambiguous unitary comparison.

For fixed actual query counts $a,b\geq1$, define the two averaged Choi operators
\begin{align}
  M_{=}^{a,b}
   & :=
  \int_{\SU(d)} dU\, C_U^{\otimes a}\otimes C_U^{\otimes b},
  \label{eq:M-same} \\
  M_{\necase}^{a,b}
   & :=
  \int_{\SU(d)} dU_1
  \int_{\SU(d)} dU_2\,
  C_{U_1}^{\otimes a}\otimes C_{U_2}^{\otimes b}.
  \label{eq:M-diff}
\end{align}
The first operator corresponds to the identical-unitary hypothesis, and the second to
the independent-unitary hypothesis.

\subsection{General testers}

Quantum tester formalism describes general measurement processes of quantum channels
implementable by quantum circuits in which the causal order of the channel queries
is predefined before execution of the quantum circuits
\cite{chiribellaTransformingQuantumOperations2008,
  chiribellaQuantumCircuitsArchitecture2008,
  zimanProcessPOVMMathematical2008}.
The process-matrix formalism extends this description to more general cases where the
causal order of channel queries can be adaptively determined or even indefinite
\cite{oreshkovQuantumCorrelationsNo2012,
  chiribellaQuantumComputationsDefinite2013,
  araujoWitnessingCausalNonseparability2015,
  oreshkovCausalCausallySeparable2016,
  wechsQuantumCircuitsClassical2021a}.
Following Refs.~\cite{bavarescoStrictHierarchy2021,
  bavarescoUnitaryChannelDiscrimination2022,
  andreStrategyOptimizationBayesian2026},
we call the resulting most general class general testers.

For fixed actual query counts $a,b$, put $n:=a+b$ and label the open slots by
$j=1,\ldots,n$, with input and output systems $I_j$ and $O_j$.  A general tester
for comparison is a collection of positive semidefinite operators $\{\Pi_x\}_x$ on the
Choi space of these slots.  We order the Choi factors as
$(I_1O_1)\otimes\cdots\otimes(I_nO_n)$. Whenever convenient below, the same tensor
product is canonically regrouped into all input factors
$\mK:=I_1\otimes\cdots\otimes I_n$ and all output factors
$\mH:=O_1\otimes\cdots\otimes O_n$.  If arbitrary CPTP maps
$\mathcal E_1,\ldots,\mathcal E_n$ with Choi operators
$C_{\mathcal E_1},\ldots,C_{\mathcal E_n}$, in the same convention as $C_U$, are
inserted, the outcome probability is
\begin{equation}
  \Pr(x\mid\mathcal E_1,\ldots,\mathcal E_n)
  =
  \Tr\!\left[
    \Pi_x\bigotimes_{j=1}^{n}C_{\mathcal E_j}
    \right].
  \label{eq:process-born-rule}
\end{equation}
For the comparison problem, the inserted maps are
$U_1,\ldots,U_1,U_2,\ldots,U_2$.  The sum $T:=\sum_x\Pi_x$ is the deterministic operator
associated with the tester.  For a valid general tester, this operator is
normalized on every choice of local CPTP maps,
\begin{equation}
  \Tr\!\left[
    T\bigotimes_{j=1}^{n}C_{\mathcal E_j}
    \right]
  =
  1
  \qquad
  \forall \mathcal E_1,\ldots,\mathcal E_n\ \mathrm{CPTP}.
  \label{eq:valid-process-normalization}
\end{equation}
Equivalently, $T$ satisfies the following standard linear constraints for general testers
\cite{araujoWitnessingCausalNonseparability2015,
  oreshkovCausalCausallySeparable2016}.
Readers interested primarily in the comparison bound may skip the explicit constraints
in Eqs.~\eqref{eq:process-trace-normalization}--\eqref{eq:process-linear-constraints}.
The upper-bound proof uses only positivity and the repeated-unitary normalization in
Eq.~\eqref{eq:deterministic-normalization-pointwise}, while the achievability proof uses
the parallel-tester normalization in Eqs.~\eqref{con1-para} and~\eqref{con2-para}.
For any subsystem
$X$, define the trace-and-replace map
\begin{equation}
  \mathcal R_X(A)
  :=
  \frac{\Id^X}{d_X}\otimes\Tr_X A,
  \label{eq:trace-and-replace}
\end{equation}
where $d_X=\dim X$ and the identity is reinserted in the original tensor position.
Since these maps commute on distinct subsystems, a positive operator $T$ is a valid
deterministic process matrix if and only if
\begin{equation}
  \Tr T
  =
  \prod_{j=1}^{n}d_{O_j},
  \label{eq:process-trace-normalization}
\end{equation}
and, for every nonempty $S\subseteq\{1,\ldots,n\}$,
\begin{align}
  \mathcal L_S
   & :=
  \prod_{j\in S}(\operatorname{id}-\mathcal R_{O_j}),
  \label{eq:process-LS} \\
  \mathcal Q_S
   & :=
  \prod_{k\notin S}\mathcal R_{I_k}\mathcal R_{O_k},
  \label{eq:process-QS} \\
  (\mathcal L_S\mathcal Q_S)(T)
   & =
  0.
  \label{eq:process-linear-constraints}
\end{align}
Equations~\eqref{eq:process-trace-normalization} and
\eqref{eq:process-linear-constraints}, together with positivity, define the set of valid
general testers.  These constraints are obtained by expanding arbitrary local CPTP
Choi operators into normalized identity components plus output-traceless components.
Normalization for all CPTP insertions is equivalent to removing every term in which a
nonempty set of output systems is left traceless while all outside parties are
trace-and-replaced, together with the trace normalization
Eq.~\eqref{eq:process-trace-normalization}.  They in particular imply normalization for
every repeated-unitary input appearing in this comparison problem.
\begin{equation}
  \Tr\!\left[
    T\left(C_{U_1}^{\otimes a}\otimes C_{U_2}^{\otimes b}\right)
    \right]
  =
  1
  \qquad
  \forall U_1,U_2\in\SU(d).
  \label{eq:deterministic-normalization-pointwise}
\end{equation}
A fixed-order quantum comb, a classically controlled causal-order strategy, a parallel
tester, and an indefinite-causal-order tester are all valid subclasses of
this general-tester set.  Throughout this paper, ``valid'' means precisely
positivity together with the general-tester normalization conditions above. It does not
assert a concrete laboratory implementation for every indefinite-causal-order process
matrix.

For the upper-bound proofs, however, we use a larger relaxation that is not necessarily a
valid general tester. This relaxation allows any positive operators
$\{\Pi_x\}_x$ whose sum $T$ satisfies only
Eq.~\eqref{eq:deterministic-normalization-pointwise}. The constraints
Eqs.~\eqref{eq:process-trace-normalization}--\eqref{eq:process-linear-constraints} are
not imposed.  We refer to this larger class as the unitary-normalized positive
relaxation.  Since it contains all valid general testers, an upper bound proved
for the relaxation is automatically an upper bound for all operational classes listed
above. Averaging Eq.~\eqref{eq:deterministic-normalization-pointwise} over the
two hypotheses gives
\begin{equation}
  \Tr(T M_{=}^{a,b})=\Tr(T M_{\necase}^{a,b})=1.
  \label{eq:deterministic-normalization}
\end{equation}
Positivity and Eq.~\eqref{eq:deterministic-normalization} are the only tester properties
used in the upper-bound arguments below.

A parallel tester is the valid tester whose deterministic operator has the form
$T=S^{\mK}\otimes\Id^{\mH}$, where $S^{\mK}$ is a unit-trace positive semidefinite
operator on the joint input Hilbert space $\mK$ and $\Id^{\mH}$ is the identity on the
joint output Hilbert space $\mH$.  Its normalization conditions are
\begin{align}
  \sum_x\Pi_x & =S^{\mK}\otimes\Id^{\mH},\label{con1-para} \\
  \Tr S^{\mK} & =1.\label{con2-para}
\end{align}
All channel queries are then made in parallel.

% REVTeX defers full-width floats by one page, so declare this roadmap before the
% section it guides; otherwise readers encounter the proof before the figure.
\begin{figure*}[t]
  \centering
  \begin{tikzpicture}[
      x=1cm,
      y=1cm,
      >=Latex,
      line cap=round,
      line join=round,
      gate/.style={
        draw,
        rounded corners=1pt,
        minimum width=0.48cm,
        minimum height=0.32cm,
        inner sep=0.6pt,
        font=\scriptsize\sffamily
      },
      uone/.style={gate, fill=blue!22, draw=blue!70!black},
      utwo/.style={gate, fill=orange!28, draw=orange!80!black},
      usame/.style={gate, fill=black!12, draw=black!55},
      inputstate/.style={draw=black!45, fill=black!5},
      arr/.style={->, line width=0.65pt, draw=black!50}
    ]
    % (a) Haar-averaged hypotheses
    \node[anchor=base west, font=\small] at (0.00,3.28) {(a)};
    \node[anchor=base west, font=\scriptsize] at (0.48,3.28)
      {Haar-averaged hypotheses};
    \draw[line width=0.55pt] (0.78,2.48) -- (2.52,2.48);
    \node[anchor=east, font=\scriptsize] at (0.70,2.48) {$H_=$};
    \node[usame] at (1.35,2.48) {$U$};
    \node[usame] at (2.08,2.48) {$U$};
    \node[anchor=west, font=\scriptsize] at (2.58,2.48) {$M_=$};
    \draw[line width=0.55pt] (0.78,1.58) -- (2.52,1.58);
    \node[anchor=east, font=\scriptsize] at (0.70,1.58) {$H_{\necase}$};
    \node[uone] at (1.35,1.58) {$U$};
    \node[utwo] at (2.08,1.58) {$V$};
    \node[anchor=west, font=\scriptsize] at (2.58,1.58) {$M_{\necase}$};

    \draw[arr] (3.55,2.03) -- (4.38,2.03);

    % (b) Sector comparison
    \node[anchor=base west, font=\small] at (4.45,3.28) {(b)};
    \node[anchor=base west, font=\scriptsize] at (4.93,3.28)
      {Sector comparison};
    \node[font=\scriptsize] at (5.36,2.96) {$H_=$};
    \foreach \i in {0,1,2} {
      \foreach \j in {0,1,2} {
        \filldraw[fill=blue!22, draw=blue!65!black]
          ({4.88+0.32*\j}, {2.52-0.32*\i}) rectangle ++(0.28,0.28);
      }
    }
    \node[font=\scriptsize, text=black!50] at (5.36,1.38) {coherent};
    \node[font=\scriptsize] at (7.22,2.96) {$H_{\necase}$};
    \foreach \i in {0,1,2} {
      \foreach \j in {0,1,2} {
        \ifnum\i=\j
        \else
          \filldraw[fill=black!4, draw=black!28]
            ({6.74+0.32*\j}, {2.52-0.32*\i}) rectangle ++(0.28,0.28);
        \fi
      }
    }
    \filldraw[fill=blue!16, draw=blue!70!black]
      (6.74,2.52) rectangle ++(0.28,0.28);
    \filldraw[fill=orange!22, draw=orange!80!black]
      (7.06,2.20) rectangle ++(0.28,0.28);
    \filldraw[fill=green!18, draw=green!55!black]
      (7.38,1.88) rectangle ++(0.28,0.28);
    \node[font=\scriptsize, text=black!50] at (7.22,1.38) {path-diagonal};

    \draw[arr] (8.12,2.03) -- (8.98,2.03);

    % (c) Relative bound
    \node[anchor=base west, font=\small] at (9.05,3.28) {(c)};
    \node[anchor=base west, font=\scriptsize] at (9.53,3.28)
      {Relative bound};
    \node[
        draw=black!40,
        fill=black!4,
        rounded corners=1.4pt,
        inner sep=3.2pt,
        font=\scriptsize
      ] at (9.88,2.28) {$M_=$};
    \node[font=\scriptsize] at (10.82,2.28) {$\le\Theta_d$};
    \node[
        draw=black!40,
        fill=black!4,
        rounded corners=1.4pt,
        inner sep=3.2pt,
        font=\scriptsize
      ] at (11.82,2.28) {$M_{\necase}$};
    \node[font=\scriptsize, text=black!50] at (10.88,1.78)
      {Lemma~\ref{lem:relative-operator-inequality}};
    \node[font=\scriptsize] at (10.88,1.28)
      {$P_{\ME}^{\star}=\max\{1-p,\,1-(1-p)/\Theta_d\}$};

    \draw[arr] (12.78,2.03) -- (13.55,2.03);

    % (d) Parallel achievability
    \node[anchor=base west, font=\small] at (13.62,3.28) {(d)};
    \node[anchor=base west, font=\scriptsize] at (14.10,3.28)
      {Parallel achievability};
    \filldraw[inputstate]
      (14.18,1.18)
      .. controls (13.42,1.42) and (13.42,2.58) ..
      (14.18,2.82)
      -- (14.18,1.18) -- cycle;
    \foreach \y in {2.64,2.32,1.68,1.36} {
      \draw[line width=0.5pt] (14.18,\y) -- (15.78,\y);
    }
    \node[uone] at (14.82,2.64) {$U_1$};
    \node[uone] at (14.82,2.32) {$U_1$};
    \node[utwo] at (14.82,1.68) {$U_2$};
    \node[utwo] at (14.82,1.36) {$U_2$};
    \filldraw[draw=black!50, fill=green!8, rounded corners=1.2pt]
      (15.78,1.18) rectangle (16.78,2.82);
    \begin{scope}[shift={(16.28,2.08)}]
      \draw[black!60, line width=0.55pt] (-0.18,-0.07) -- (0.18,-0.07);
      \draw[black!60, line width=0.55pt]
        (-0.14,-0.07) arc[start angle=180, end angle=0, radius=0.14];
      \draw[black!60, line width=0.55pt] (0.00,-0.07) -- (0.11,0.12);
    \end{scope}
  \end{tikzpicture}
  \caption{Logical structure of the minimum-error argument.
    (a)~The two hypotheses are the Haar-averaged Choi operators $M_=^{a,b}$
    (identical unitary) and $M_{\necase}^{a,b}$ (independent unitaries).
    (b)~Schur--Weyl duality reduces the comparison to each outer sector $\nu$:
    the identical-unitary average is a coherent sum of coupling paths, while
    the independent-unitary average is path-diagonal
    (see Fig.~\ref{fig:coupling-paths}).
    (c)~A weighted Cauchy--Schwarz estimate yields the relative operator bound
    of Lemma~\ref{lem:relative-operator-inequality}, which determines
    $P_{\ME}^{\star}$.
    (d)~The bound is attained by a parallel tester that selects a maximizing
    sector.}
  \label{fig:section-iii-logic-flow}
\end{figure*}

\section{Optimal comparison in the general tester formalism}
\label{sec:optimal-comparison}

The two hypotheses are represented by the Haar-averaged Choi operators
$M_{=}^{a,b}$ and $M_{\necase}^{a,b}$.  Our upper bound is governed by the
smallest constant for which the identical-unitary average is dominated by
the independent-unitary average.  After resolving the Haar-twirled operators
into Schur--Weyl sectors, this constant can be evaluated independently in
each outer irreducible sector.  We denote the largest sector-wise value by
$\Theta_d(a,b)$.  For $a,b\geq1$,
\begin{equation}
  \Theta_d(a,b)
  :=
  \max_{\nu\in\Y_{a+b,d}}
  \frac{
    \displaystyle
    \sum_{\lambda\in\Y_{a,d}}
    \sum_{\mu\in\Y_{b,d}}
    c_{\lambda\mu}^{\nu}D_\lambda D_\mu
  }{D_\nu}.
  \label{eq:Theta-definition}
\end{equation}
Here the Littlewood--Richardson coefficient $c_{\lambda\mu}^{\nu}$ counts the
admissible ways in which the two
Schur sectors $\lambda$ and $\mu$ can combine into the outer sector $\nu$,
while the representation dimensions determine their relative statistical
weights.  For fixed $d,a,b$, this is a finite and algorithmically computable
rational number.  Appendix~\ref{appendix:theta-computation} gives a concrete
algorithm and small worked examples.

Figure~\ref{fig:section-iii-logic-flow} summarizes the logical structure of this
section. Representation theory is used only to compare the two Haar-averaged
hypotheses sector by sector. The output is a single relative operator bound,
which then determines both the optimal success probability and an explicit
parallel tester.

The following lemma converts the sector-wise weights summarized by
$\Theta_d(a,b)$ into a global operator inequality. It is the bridge
between the representation-theoretic decomposition and the comparison
problem.

Here, $\lambda$ and $\mu$ label the Schur sectors of the two query groups, $\nu$
labels a Schur sector of the combined queries, and $\alpha$ labels a
Littlewood--Richardson coupling path from $(\lambda,\mu)$ to $\nu$. Under $H_=$, all
paths leading to the same $\nu$ are added coherently, whereas under $H_{\necase}$ they
appear only through path-diagonal terms.

\begin{lemma}
  \label{lem:relative-operator-inequality}
  For every $a,b\geq1$,
  \begin{equation}
    M_{=}^{a,b}\leq \Theta_d(a,b)M_{\necase}^{a,b}.
    \label{eq:relative-operator-inequality}
  \end{equation}
\end{lemma}

  \begin{figure*}[t]
    \centering
    \begin{tikzpicture}[
        x=1cm,
        y=1cm,
        >=Latex,
        line cap=round,
        line join=round,
        pathbox/.style={
          draw,
          rounded corners=1.4pt,
          align=center,
          inner sep=3.0pt,
          minimum width=1.72cm,
          minimum height=0.56cm,
          font=\scriptsize
        },
        arr/.style={->, line width=0.65pt, draw=black!50}
      ]
      % (a) coupling paths
      \node[anchor=base west, font=\small] at (0.00,3.28) {(a)};
      \node[anchor=base west, font=\scriptsize] at (0.48,3.28)
        {Coupling paths into $\nu$};
      \node[pathbox, fill=blue!16, draw=blue!70!black] (p1)
        at (1.15,2.48) {$\lambda_1\otimes\mu_1$};
      \node[pathbox, fill=orange!20, draw=orange!80!black] (p2)
        at (1.15,1.72) {$\lambda_2\otimes\mu_2$};
      \node[pathbox, fill=green!16, draw=green!55!black] (p3)
        at (1.15,0.96) {$\lambda_3\otimes\mu_3$};
      \node[font=\scriptsize, text=black!55] at (0.18,2.48) {$1$};
      \node[font=\scriptsize, text=black!55] at (0.18,1.72) {$2$};
      \node[font=\scriptsize, text=black!55] at (0.18,0.96) {$3$};
      \node[
          draw=black!45,
          fill=black!6,
          rounded corners=2pt,
          inner sep=5pt,
          font=\small
        ] (nu) at (4.45,1.72) {$\nu$};
      \draw[arr] (p1.east) -- (nu.north west);
      \draw[arr] (p2.east) -- (nu.west);
      \draw[arr] (p3.east) -- (nu.south west);

      % (b) coherent
      \node[anchor=base west, font=\small] at (5.70,3.28) {(b)};
      \node[anchor=base west, font=\scriptsize] at (6.18,3.28)
        {Identical: coherent sum};
      \foreach \i in {0,1,2} {
        \foreach \j in {0,1,2} {
          \filldraw[fill=blue!22, draw=blue!65!black]
            ({6.55+0.42*\j}, {2.48-0.42*\i}) rectangle ++(0.36,0.36);
        }
      }
      \node[font=\scriptsize] at (7.18,1.00)
        {$|\psi\rangle\langle\psi|$};
      \node[font=\scriptsize, text=black!55] at (7.18,0.62)
        {$|\psi\rangle=\sum_i|i\rangle$};

      % (c) path-diagonal
      \node[anchor=base west, font=\small] at (10.55,3.28) {(c)};
      \node[anchor=base west, font=\scriptsize] at (11.03,3.28)
        {Independent: path-diagonal};
      \foreach \i in {0,1,2} {
        \foreach \j in {0,1,2} {
          \ifnum\i=\j
          \else
            \filldraw[fill=black!4, draw=black!28]
              ({11.55+0.42*\j}, {2.48-0.42*\i}) rectangle ++(0.36,0.36);
          \fi
        }
      }
      \filldraw[fill=blue!16, draw=blue!70!black]
        (11.55,2.48) rectangle ++(0.36,0.36);
      \filldraw[fill=orange!22, draw=orange!80!black]
        (11.97,2.06) rectangle ++(0.36,0.36);
      \filldraw[fill=green!18, draw=green!55!black]
        (12.39,1.64) rectangle ++(0.36,0.36);
      \node[font=\scriptsize] at (12.18,1.00)
        {$\sum_i w_i^{-1}|i\rangle\langle i|$};
      \node[font=\scriptsize, text=black!55] at (12.18,0.62)
        {$w_i=D_{\lambda_i}D_{\mu_i}$};
    \end{tikzpicture}
    \caption{Schematic of the comparison inside a fixed outer Schur sector $\nu$.
      (a)~Each admissible Littlewood--Richardson coupling
      $(\lambda,\mu)\to\nu$ is a path feeding that sector; the multiplicity
      $c_{\lambda\mu}^{\nu}$ counts parallel copies of the same pair.
      (b)~Under the identical-unitary hypothesis the paths are added coherently,
      producing a rank-one block with off-diagonal coherences.
      (c)~Under the independent-unitary hypothesis the same paths appear only as
      a diagonal mixture with inverse dimension weights $w_i^{-1}$, where
      $w_i=D_{\lambda_i}D_{\mu_i}$. Lemma~\ref{lem:relative-operator-inequality}
      bounds the coherent block by this diagonal weight.}
    \label{fig:coupling-paths}
  \end{figure*}

\begin{proof}
  We prove the inequality in four steps. Step~1 writes both Haar averages in
  Schur--Weyl form. Step~2 refines the product of the two query groups into common
  outer sectors labelled by $\nu$. Step~3 identifies, within each fixed $\nu$
  sector, the coherent subspace on which the identical-unitary hypothesis is
  supported. Step~4 compares this coherent rank-one contribution with the
  corresponding diagonal part of the independent-unitary average by a weighted
  Cauchy--Schwarz inequality.

  \emph{Step 1 (Schur--Weyl form).}
  We write the input side of the Choi space as $K$ and the output side as $H$.
  On the input side, the representation is the conjugate one.
  Schur--Weyl duality
  \cite{schurUeberKlasseMatrizen1901,weylClassicalGroupsTheir1939,
    fultonRepresentationTheory2004}
  gives, for each $r$,
  \begin{align}
    \Hcal_r^K
     & \cong
    \bigoplus_{\lambda\in\Y_{r,d}}
    \overline{\Ucal_\lambda}^{\,K}\otimes\Vcal_\lambda^K,
    \label{eq:SW-input} \\
    \Hcal_r^H
     & \cong
    \bigoplus_{\lambda\in\Y_{r,d}}
    \Ucal_\lambda^H\otimes\Vcal_\lambda^H.
    \label{eq:SW-output}
  \end{align}
  With this convention,
  \begin{equation}
    |\Phi\rangle^{\otimes r}
    =
    \sum_{\lambda\in\Y_{r,d}}
    |\Phi_{\overline{\Ucal_\lambda}^{\,K},\Ucal_\lambda^H}\rangle
    \otimes
    |\Phi_{\Vcal_\lambda^K,\Vcal_\lambda^H}\rangle.
    \label{eq:Phi-SW}
  \end{equation}
  For readability, we write $\Ucal_\lambda^K$ for the conjugate input copy
  $\overline{\Ucal_\lambda}^{\,K}$.

  By Schur's lemma, the one-register Haar average is
  \begin{align}
    \int dU\,C_U^{\otimes r}
     & =
    \sum_{\lambda\in\Y_{r,d}}
    \frac{\Id_{\Ucal_\lambda^K}\otimes\Id_{\Ucal_\lambda^H}}{D_\lambda}
    \otimes
    |\Phi_{\Vcal_\lambda^K,\Vcal_\lambda^H}\rangle
    \langle\Phi_{\Vcal_\lambda^K,\Vcal_\lambda^H}|.
    \label{eq:single-twirl}
  \end{align}
  The independent-unitary average therefore factors as
  \[
    M_{\necase}^{a,b}
    =
    \Bigl(\int dU\,C_U^{\otimes a}\Bigr)
    \otimes
    \Bigl(\int dU\,C_U^{\otimes b}\Bigr),
  \]
  and expanding both factors gives
  \begin{align}
    M_{\necase}^{a,b}
     & =
    \sum_{\lambda\in\Y_{a,d}}
    \sum_{\mu\in\Y_{b,d}}
    \frac{1}{D_\lambda D_\mu}
    \nonumber \\
     & \quad\times
    \Bigl(
    \Id_{\Ucal_\lambda^K}\otimes\Id_{\Ucal_\lambda^H}
    \otimes
    |\Phi_{\Vcal_\lambda^K,\Vcal_\lambda^H}\rangle
    \langle\Phi_{\Vcal_\lambda^K,\Vcal_\lambda^H}|
    \Bigr)
    \nonumber \\
     & \quad\otimes
    \Bigl(
    \Id_{\Ucal_\mu^K}\otimes\Id_{\Ucal_\mu^H}
    \otimes
    |\Phi_{\Vcal_\mu^K,\Vcal_\mu^H}\rangle
    \langle\Phi_{\Vcal_\mu^K,\Vcal_\mu^H}|
    \Bigr).
    \label{eq:independent-explicit-block-form}
  \end{align}
  Haar averaging therefore organizes the remaining information into irreducible
  sectors and multiplicity spaces.

  \emph{Step 2 (outer sectors).}
  To compare the two hypotheses sector by sector, we refine the product of the two
  Schur blocks into irreducible sectors of the combined $a+b$ queries. The
  Littlewood--Richardson coefficient $c_{\lambda\mu}^{\nu}$ records the multiplicity
  of each allowed coupling $(\lambda,\mu)\to\nu$,
  \begin{equation}
    \Ucal_\lambda^H\otimes\Ucal_\mu^H
    \cong
    \bigoplus_{\nu\in\Y_{a+b,d}}
    \Ucal_\nu^H\otimes\C^{c_{\lambda\mu}^{\nu}},
    \label{eq:LR-output-blocks}
  \end{equation}
  and similarly on the conjugate input side.
  The multiplicity space of an outer sector $\nu$ then decomposes as
  \[
    \Vcal_\nu
    \cong
    \bigoplus_{\lambda\in\Y_{a,d}}
    \bigoplus_{\mu\in\Y_{b,d}}
    \C^{c_{\lambda\mu}^{\nu}}\otimes\Vcal_\lambda\otimes\Vcal_\mu.
  \]
  The identical-unitary average has the same Schur--Weyl form as
  Eq.~\eqref{eq:single-twirl}, now for $a+b$ queries:
  \begin{equation}
    M_{=}^{a,b}
    =
    \sum_{\nu\in\Y_{a+b,d}}
    \frac{\Id_{\Ucal_\nu^K}\otimes\Id_{\Ucal_\nu^H}}{D_\nu}
    \otimes
    |\Phi_{\Vcal_\nu^K,\Vcal_\nu^H}\rangle
    \langle\Phi_{\Vcal_\nu^K,\Vcal_\nu^H}|.
    \label{eq:same-block-form}
  \end{equation}
  The independent average is block diagonal with respect to the pair of
  outer labels $(\nu^K,\rho^H)$.
  Indeed, after applying Eq.~\eqref{eq:LR-output-blocks} and its conjugate-input
  counterpart to each summand of Eq.~\eqref{eq:independent-explicit-block-form},
  the representation factors split as
  \[
    \begin{aligned}
      \Ucal_\lambda^K\otimes\Ucal_\mu^K
       & \cong
      \bigoplus_\nu \Ucal_\nu^K\otimes\C^{c_{\lambda\mu}^{\nu}}, \\
      \Ucal_\lambda^H\otimes\Ucal_\mu^H
       & \cong
      \bigoplus_\rho \Ucal_\rho^H\otimes\C^{c_{\lambda\mu}^{\rho}}.
    \end{aligned}
  \]
  Projecting Eq.~\eqref{eq:independent-explicit-block-form} onto the
  $(\nu^K,\rho^H)$ sector therefore leaves the identity on
  $\Ucal_\nu^K\otimes\Ucal_\rho^H$ and the corresponding identity operators on the
  multiplicity spaces $\C^{c_{\lambda\mu}^{\nu}}$ and $\C^{c_{\lambda\mu}^{\rho}}$.
  Since $M_=^{a,b}$ has support only on diagonal sectors $\nu^K=\rho^H$, sectors with
  $\nu^K\ne\rho^H$ are irrelevant for the comparison.
  We therefore fix $\nu$ and compare the two operators on the $(\nu^K,\nu^H)$ sector.

  \emph{Step 3 (matched subspace).}
  Under the identical-unitary hypothesis, the same outer label appears on the input
  and output sides, and the corresponding coupling paths are added coherently. By
  contrast, the independent-unitary average is diagonal in the coupling-path labels
  and also contains additional unmatched positive contributions. Therefore, for an
  upper bound, it is sufficient to retain only the matched diagonal part of the
  independent average.
  Figure~\ref{fig:coupling-paths} depicts this distinction: several
  Littlewood--Richardson paths feed the same outer sector $\nu$, they are
  superposed coherently under the identical-unitary hypothesis, and they appear
  only as a path-diagonal mixture under the independent-unitary hypothesis.
  For each pair $(\lambda,\mu)$ with $c_{\lambda\mu}^{\nu}>0$, choose an orthonormal
  basis $\{|\alpha\rangle:\alpha=1,\ldots,c_{\lambda\mu}^{\nu}\}$ of
  $\C^{c_{\lambda\mu}^{\nu}}$, and write
  \[
    |\Phi_{\lambda\mu\alpha}\rangle
    :=
    |\alpha\rangle_K|\alpha\rangle_H
    \otimes
    |\Phi_{\Vcal_\lambda^K\otimes\Vcal_\mu^K,\,\Vcal_\lambda^H\otimes\Vcal_\mu^H}\rangle.
  \]
  Then
  \[
    |\Phi_{\Vcal_\nu^K,\Vcal_\nu^H}\rangle
    =
    \sum_{\lambda,\mu,\alpha}|\Phi_{\lambda\mu\alpha}\rangle,
  \]
  where the sum runs over $\lambda\in\Y_{a,d}$, $\mu\in\Y_{b,d}$, and
  $\alpha=1,\ldots,c_{\lambda\mu}^{\nu}$, omitting vanishing Littlewood--Richardson
  coefficients. The $\nu$ block of $M_=^{a,b}$ is
  \begin{equation}
    M_{=}^{(\nu)}
    =
    \frac{\Id_{\Ucal_\nu^K}\otimes\Id_{\Ucal_\nu^H}}{D_\nu}
    \otimes
    |\Phi_{\Vcal_\nu^K,\Vcal_\nu^H}\rangle
    \langle\Phi_{\Vcal_\nu^K,\Vcal_\nu^H}|.
    \label{eq:same-nu-block}
  \end{equation}
  The corresponding block of the independent average is
  \begin{align}
    M_{\necase}^{(\nu)}
     & =
    \Id_{\Ucal_\nu^K}\otimes\Id_{\Ucal_\nu^H}
    \nonumber \\
     & \quad\otimes
    \sum_{\lambda,\mu}
    \frac{1}{D_\lambda D_\mu}
    \bigl(\Id_{\C^{c_{\lambda\mu}^{\nu}}}^K\otimes\Id_{\C^{c_{\lambda\mu}^{\nu}}}^H\bigr)
    \nonumber \\
     & \qquad\otimes
    |\Phi_{\Vcal_\lambda^K\otimes\Vcal_\mu^K,\,\Vcal_\lambda^H\otimes\Vcal_\mu^H}\rangle
    \langle
    \Phi_{\Vcal_\lambda^K\otimes\Vcal_\mu^K,\,\Vcal_\lambda^H\otimes\Vcal_\mu^H}|.
    \label{eq:diff-nu-block}
  \end{align}
  Split the multiplicity identities into matched and unmatched parts,
  \begin{align}
    \Id_{\C^{c_{\lambda\mu}^{\nu}}}^K\otimes\Id_{\C^{c_{\lambda\mu}^{\nu}}}^H
     & =
    \sum_\alpha
    |\alpha\rangle\langle\alpha|^K
    \otimes
    |\alpha\rangle\langle\alpha|^H
    \nonumber \\
     & \quad+
    \sum_{\alpha\ne\beta}
    |\alpha\rangle\langle\alpha|^K
    \otimes
    |\beta\rangle\langle\beta|^H .
    \label{eq:multiplicity-matched-unmatched}
  \end{align}
  Every unmatched summand is positive semidefinite, so
  \begin{equation}
    M_{\necase}^{(\nu)}
    \geq
    \Id_{\Ucal_\nu^K}\otimes\Id_{\Ucal_\nu^H}
    \otimes
    \sum_{\lambda,\mu,\alpha}
    \frac{1}{D_\lambda D_\mu}
    |\Phi_{\lambda\mu\alpha}\rangle\langle\Phi_{\lambda\mu\alpha}|.
    \label{eq:diff-lower-Z}
  \end{equation}

  \emph{Step 4 (weighted comparison).}
  The remaining comparison is between a coherent sum over coupling paths and an
  incoherent weighted sum over the same paths. The required factor is therefore the
  total weight of all admissible paths in the fixed $\nu$ sector.
  The vectors $|\Phi_{\lambda\mu\alpha}\rangle$ are mutually orthogonal, and both
  $|\Phi_{\Vcal_\nu^K,\Vcal_\nu^H}\rangle\langle\Phi_{\Vcal_\nu^K,\Vcal_\nu^H}|$
  and the matched part of $M_{\necase}^{(\nu)}$ are supported on their span.
  Let $|\xi\rangle=\sum_{\lambda,\mu,\alpha}|\xi_{\lambda\mu\alpha}\rangle$ be a
  vector in that span. Then
  $\langle\Phi_{\Vcal_\nu^K,\Vcal_\nu^H}|\xi\rangle
    =\sum_{\lambda,\mu,\alpha}\langle\Phi_{\lambda\mu\alpha}|\xi_{\lambda\mu\alpha}\rangle$,
  and the weighted Cauchy--Schwarz inequality gives
  \begin{align}
    |\langle\Phi_{\Vcal_\nu^K,\Vcal_\nu^H}|\xi\rangle|^2
     & =
    \Bigl|\sum_{\lambda,\mu,\alpha}
    \langle\Phi_{\lambda\mu\alpha}|\xi_{\lambda\mu\alpha}\rangle\Bigr|^2
    \nonumber \\
     & \leq
    \Bigl(\sum_{\lambda,\mu,\alpha}D_\lambda D_\mu\Bigr)
    \Bigl(
    \sum_{\lambda,\mu,\alpha}
    \frac{|\langle\Phi_{\lambda\mu\alpha}|\xi_{\lambda\mu\alpha}\rangle|^2}
    {D_\lambda D_\mu}
    \Bigr)
    \nonumber \\
     & =
    \Bigl(\sum_{\lambda,\mu,\alpha}D_\lambda D_\mu\Bigr)
    \nonumber \\
     & \qquad\times
    \langle\xi|
    \Bigl(
    \sum_{\lambda,\mu,\alpha}
    \frac{1}{D_\lambda D_\mu}
    |\Phi_{\lambda\mu\alpha}\rangle\langle\Phi_{\lambda\mu\alpha}|
    \Bigr)
    |\xi\rangle.
    \label{eq:weighted-CS}
  \end{align}
  Therefore
  \begin{multline}
    |\Phi_{\Vcal_\nu^K,\Vcal_\nu^H}\rangle
    \langle\Phi_{\Vcal_\nu^K,\Vcal_\nu^H}|
    \leq
    \Bigl(\sum_{\lambda,\mu,\alpha}D_\lambda D_\mu\Bigr)
    \\
    \times
    \sum_{\lambda,\mu,\alpha}
    \frac{1}{D_\lambda D_\mu}
    |\Phi_{\lambda\mu\alpha}\rangle\langle\Phi_{\lambda\mu\alpha}|.
    \label{eq:rank-one-bound}
  \end{multline}
  Combining Eqs.~\eqref{eq:same-nu-block}, \eqref{eq:diff-lower-Z}, and
  \eqref{eq:rank-one-bound}, we obtain
  \begin{equation}
    M_{=}^{(\nu)}
    \leq
    \frac{\sum_{\lambda,\mu,\alpha}D_\lambda D_\mu}{D_\nu}\,
    M_{\necase}^{(\nu)}.
    \label{eq:block-relative-bound}
  \end{equation}
  The numerator is
  \[
    \sum_{\lambda,\mu,\alpha}D_\lambda D_\mu
    =
    \sum_{\lambda\in\Y_{a,d}}
    \sum_{\mu\in\Y_{b,d}}
    c_{\lambda\mu}^{\nu}D_\lambda D_\mu,
  \]
  so the sector-wise ratio is at most $\Theta_d(a,b)$.
  Thus $\Theta_d(a,b)$ is precisely the largest coherent-to-incoherent weight ratio
  among all outer Schur sectors.
  Taking the maximum over $\nu$ proves Eq.~\eqref{eq:relative-operator-inequality}.
\end{proof}

Once Lemma~\ref{lem:relative-operator-inequality} is available, the optimization
over arbitrary general testers reduces to a constraint on two conditional
acceptance probabilities. The remaining optimization is elementary; the nontrivial
part is to show that the resulting bound is attained by a valid parallel tester.

\begin{theorem}
  \label{thm:minimum-error}
  With $N_1$ and $N_2$ queries to two unknown $d$-dimensional unitary channels, the
  optimal minimum-error success probability over valid general testers is
  \begin{equation}
    P_{\ME}^{\star}
    =
    \begin{cases}
      1-p,
       & 0\leq p\leq \frac{1}{1+\Theta_d(N_1,N_2)}, \\[1.1ex]
      1-\frac{1-p}{\Theta_d(N_1,N_2)},
       & \frac{1}{1+\Theta_d(N_1,N_2)}<p\leq1.
    \end{cases}
    \label{eq:ME-main}
  \end{equation}
\end{theorem}

Operationally, the optimal strategy has only two regimes. For a sufficiently
small prior probability of equality, no measurement can outperform always
answering ``different.'' Above the threshold, an optimal parallel tester accepts
identical unitaries with certainty and has false-acceptance probability
$1/\Theta_d(N_1,N_2)$ for independently sampled unitaries.

\begin{proof}
  \emph{Upper bound.}
  We prove the upper bound for the relaxed tester class. This is sufficient for all
  valid general testers.
  Choose a relaxed tester $\{\Pi_=,\Pi_{\necase}\}$, where $\Pi_=$ is the effect for
  answering ``same,'' $\Pi_{\necase}$ is the effect for answering ``different,'' and
  $T=\Pi_=+\Pi_{\necase}$ is deterministic.
  The two conditional acceptance probabilities are
  \begin{align*}
    \Tr(\Pi_=M_=^{N_1,N_2})
     & =
    \Pr(=\mid H_=),
    \\
    \Tr(\Pi_=M_{\necase}^{N_1,N_2})
     & =
    \Pr(=\mid H_{\necase}).
  \end{align*}
  The minimum-error success probability being optimized is
  \begin{align}
    P_{\ME}
     & =
    p\Pr(=\mid H_=)+(1-p)\Pr(\necase\mid H_{\necase})
    \nonumber \\
     & =
    p\Tr(\Pi_=M_=^{N_1,N_2})
    +(1-p)\bigl(1-\Tr(\Pi_=M_{\necase}^{N_1,N_2})\bigr).
    \label{eq:PME-xy}
  \end{align}
  The relaxed optimum is the supremum of this quantity over all positive effects
  $0\leq\Pi_=,\Pi_{\necase}\leq T$ satisfying $\Pi_=+\Pi_{\necase}=T$ and
  Eq.~\eqref{eq:deterministic-normalization-pointwise}.

  Lemma~\ref{lem:relative-operator-inequality} gives
  $M_=^{N_1,N_2}\leq\Theta_d(N_1,N_2)M_{\necase}^{N_1,N_2}$.
  Since $\Pi_=\geq0$, every relaxed tester satisfies
  \begin{equation}
    \Tr(\Pi_=M_=^{N_1,N_2})
    \leq
    \Theta_d(N_1,N_2)\Tr(\Pi_=M_{\necase}^{N_1,N_2}).
    \label{eq:fixed-query-affine-bound}
  \end{equation}
  Because $T-\Pi_=\,=\Pi_{\necase}\geq0$, we have $0\leq\Pi_=\leq T$, and therefore
  \[
    0\leq\Tr(\Pi_=M_=^{N_1,N_2})\leq\Tr(TM_=^{N_1,N_2})=1.
  \]
  Substituting Eq.~\eqref{eq:fixed-query-affine-bound} into $P_{\ME}$ gives
  \begin{align}
    P_{\ME}
     & =
    1-p
    +p\Tr(\Pi_=M_=^{N_1,N_2})
    \nonumber \\
     & \quad
    -(1-p)\Tr(\Pi_=M_{\necase}^{N_1,N_2})
    \nonumber \\
     & \leq
    1-p
    \nonumber \\
     & \quad+
    \left(p-\frac{1-p}{\Theta_d(N_1,N_2)}\right)
    \Tr(\Pi_=M_=^{N_1,N_2}).
    \label{eq:ME-converse-linear}
  \end{align}
  When $p\leq1/(1+\Theta_d(N_1,N_2))$, this remaining coefficient is
  nonpositive, so $P_{\ME}\leq1-p$.
  Otherwise $\Tr(\Pi_=M_=^{N_1,N_2})\leq1$ gives
  $P_{\ME}\leq1-(1-p)/\Theta_d(N_1,N_2)$.

  We have now established the universal upper bound. The next step is
  constructive: we show that a parallel tester can select a maximizing sector
  and saturate the bound.

  \emph{Achievability.}
  In circuit terms, the tester selects the most informative total Schur sector,
  prepares a dimension-weighted input over the admissible coupling-path sectors, and
  tests whether these paths recombine coherently at the output.
  Choose an outer diagram $\nu$ attaining the maximum in
  the definition of $\Theta_d(N_1,N_2)$, so
  \begin{equation}
    \sum_{\lambda\in\Y_{N_1,d}}
    \sum_{\mu\in\Y_{N_2,d}}
    c_{\lambda\mu}^{\nu}D_\lambda D_\mu
    =
    \Theta_d(N_1,N_2)\,D_\nu.
    \label{eq:maximizing-branch}
  \end{equation}
  Such a $\nu$ exists because the maximization is over a finite set. Retain the notation
  of Lemma~\ref{lem:relative-operator-inequality}.
  On the input multiplicity space $\Vcal_\nu^K$, prepare the state
  \begin{equation}
    \rho_\nu
    :=
    \sum_{\lambda,\mu,\alpha}
    \frac{D_\lambda D_\mu}{\Theta_d(N_1,N_2)D_\nu}
    \frac{
      |\alpha\rangle\langle\alpha|^K
      \otimes
      \Id_{\Vcal_\lambda^K\otimes\Vcal_\mu^K}
    }{\dim(\Vcal_\lambda\otimes\Vcal_\mu)}.
    \label{eq:tau-nu}
  \end{equation}
  Let $S^{\mK}$ be the input operator that vanishes off the $\nu$ block and equals
  $\Id_{\Ucal_\nu^K}/D_\nu\otimes\rho_\nu$ on that block. Then
  \[
    \Tr S^{\mK}
    =
    \frac{\Tr\Id_{\Ucal_\nu^K}}{D_\nu}\Tr\rho_\nu
    =
    1,
  \]
  so $S^{\mK}$ satisfies the trace normalization of a parallel tester.

  Using the same block ordering as in Eqs.~\eqref{eq:same-nu-block} and
  \eqref{eq:diff-nu-block},
  \[
    |\Phi_{\Vcal_\nu^K,\Vcal_\nu^H}\rangle
    =
    \sum_{\lambda,\mu,\alpha}|\Phi_{\lambda\mu\alpha}\rangle .
  \]
  On the chosen $\nu$ block, set
  \begin{equation}
    \begin{aligned}
      \Pi_=^{(\nu)}
       & :=
      \frac{\Id_{\Ucal_\nu^K}}{D_\nu}
      \otimes
      \Id_{\Ucal_\nu^H}
      \\
       & \quad\otimes
      \bigl(\rho_\nu\otimes\Id_{\Vcal_\nu^H}\bigr)
      |\Phi_{\Vcal_\nu^K,\Vcal_\nu^H}\rangle
      \langle\Phi_{\Vcal_\nu^K,\Vcal_\nu^H}|
      \bigl(\rho_\nu\otimes\Id_{\Vcal_\nu^H}\bigr),
    \end{aligned}
    \label{eq:Pi-same-nu}
  \end{equation}
  and set $\Pi_=^{(\nu')}=0$ on all other blocks.
  To see that this is a valid parallel tester, note that
  \[
    \bigl\|
    \bigl(\sqrt{\rho_\nu}\otimes\Id_{\Vcal_\nu^H}\bigr)
    |\Phi_{\Vcal_\nu^K,\Vcal_\nu^H}\rangle
    \bigr\|^2
    =
    \Tr\rho_\nu
    =
    1,
  \]
  hence
  \begin{equation}
    \bigl(\sqrt{\rho_\nu}\otimes\Id_{\Vcal_\nu^H}\bigr)
    |\Phi_{\Vcal_\nu^K,\Vcal_\nu^H}\rangle
    \langle\Phi_{\Vcal_\nu^K,\Vcal_\nu^H}|
    \bigl(\sqrt{\rho_\nu}\otimes\Id_{\Vcal_\nu^H}\bigr)
    \leq
    \Id.
    \label{eq:parallel-tester-positivity}
  \end{equation}
  Conjugating by $\sqrt{\rho_\nu}\otimes\Id_{\Vcal_\nu^H}$ gives
  \[
    \bigl(\rho_\nu\otimes\Id_{\Vcal_\nu^H}\bigr)
    |\Phi_{\Vcal_\nu^K,\Vcal_\nu^H}\rangle
    \langle\Phi_{\Vcal_\nu^K,\Vcal_\nu^H}|
    \bigl(\rho_\nu\otimes\Id_{\Vcal_\nu^H}\bigr)
    \leq
    \rho_\nu\otimes\Id_{\Vcal_\nu^H}.
  \]
  Therefore $0\leq\Pi_=\leq S^{\mK}\otimes\Id^{\mH}$ on every block. Thus
  $\Pi_{\necase}:=S^{\mK}\otimes\Id^{\mH}-\Pi_=$, together with $\Pi_=$, satisfies
  Eqs.~\eqref{con1-para} and \eqref{con2-para}.

  It remains only to evaluate the two traces.  First, using Eq.~\eqref{eq:same-nu-block},
  \begin{align}
    \Tr(\Pi_=M_=^{N_1,N_2})
     & =
    \Tr(\Pi_=^{(\nu)}M_=^{(\nu)})
    \nonumber                            \\
     & =
    \bigl|
    \langle\Phi_{\Vcal_\nu^K,\Vcal_\nu^H}|
    (\rho_\nu\otimes\Id_{\Vcal_\nu^H})
    \nonumber \\
     & \qquad\quad
    |\Phi_{\Vcal_\nu^K,\Vcal_\nu^H}\rangle
    \bigr|^2
    \nonumber                            \\
     & =
    (\Tr\rho_\nu)^2
    =
    1,
    \label{eq:achievability-accept-same} \\
    \Tr(\Pi_=M_{\necase}^{N_1,N_2})
     & =
    \Tr(\Pi_=^{(\nu)}M_{\necase}^{(\nu)})
    \nonumber                            \\
     & =
    D_\nu
    \sum_{\lambda,\mu,\alpha}
    \frac{1}{D_\lambda D_\mu}\,
    \bigl|
    \langle\Phi_{\lambda\mu\alpha}|
    (\rho_\nu\otimes\Id_{\Vcal_\nu^H})
    \nonumber \\
     & \qquad\quad
    |\Phi_{\Vcal_\nu^K,\Vcal_\nu^H}\rangle
    \bigr|^2
    \nonumber                            \\
     & =
    D_\nu
    \sum_{\lambda,\mu,\alpha}
    \frac{D_\lambda D_\mu}{\Theta_d(N_1,N_2)^2 D_\nu^2}
    \nonumber                            \\
     & =
    \frac{1}{\Theta_d(N_1,N_2)}.
    \label{eq:achievability-false-positive}
  \end{align}
  In the second trace we used Eq.~\eqref{eq:diff-nu-block}. The trace over the two
  $\Ucal_\nu$ representation spaces gives the prefactor $D_\nu$, and the matched
  terms give the factors $1/(D_\lambda D_\mu)$. The unmatched
  Littlewood--Richardson multiplicity sectors are orthogonal to the coherent
  vector $|\Phi_{\Vcal_\nu^K,\Vcal_\nu^H}\rangle$ after the diagonal action of
  $\rho_\nu$.
  Since the selected block attains the maximum, it yields the nontrivial branch
  $1-(1-p)/\Theta_d(N_1,N_2)$ of the minimum-error formula.  The constant-decision
  branch $1-p$ is obtained by always answering ``different,'' and it is optimal when
  $p\leq1/(1+\Theta_d(N_1,N_2))$.  Thus the valid parallel construction attains the
  upper bound proved for the larger relaxed set, so no valid general strategy
  can improve the optimum.
\end{proof}

As an illustration of Theorem~\ref{thm:minimum-error}, we evaluate the case
$N_1=N_2=1$. Appendix~\ref{appendix:theta-computation} gives
$\Theta_d(1,1)=2d/(d-1)$, so the optimum reduces to
\[
  P_{\ME}^{\star}
  =
  \max\left\{
  1-p,\,
  1-\frac{(1-p)(d-1)}{2d}
  \right\},
\]
which agrees with the unitary-channel specialization of the single-shot
random-channel comparison result of
Ref.~\cite{markiewiczSingleShotComparisonRandom2025}.
The same relative operator bound also controls the one-sided unambiguous
problem treated in the next section.

\section{Optimal unambiguous strategy in the general tester formalism}
\label{sec:optimal-unambiguous-comparison}

Unambiguous unitary comparison is a unitary comparison task without allowing errors. In
the one-sided unambiguous setting considered here, the third outcome ``?'' is introduced
for a tester, where ``?'' represents the inconclusive outcome.  In this section, write
$M_=:=M_=^{N_1,N_2}$ and $M_{\necase}:=M_{\necase}^{N_1,N_2}$.

Operationally, a three-outcome tester
$\{\Pi_=,\Pi_{\necase},\Pi_?\}$ is unambiguous if the conclusive wrong-answer
probabilities vanish pointwise.
\begin{align*}
  \Tr\!\left[
         \Pi_=\left(C_{U_1}^{\otimes N_1}\otimes C_{U_2}^{\otimes N_2}\right)
         \right]
   & =0
  \quad\text{for all }U_1\neq U_2, \\
  \Tr\!\left[
         \Pi_{\necase}\left(C_U^{\otimes N_1}\otimes C_U^{\otimes N_2}\right)
         \right]
   & =0
  \quad\text{for all }U .
\end{align*}
The first condition also holds on the diagonal $U_1=U_2$ by continuity.  Since these
probability functions are nonnegative and continuous on compact unitary groups, the
pointwise no-error conditions are equivalent to the Haar-averaged trace conditions
\begin{equation}
  \Tr(\Pi_=M_{\necase})=0,
  \qquad
  \Tr(\Pi_{\necase}M_=)=0.
  \label{eq:unambiguous-conditions}
\end{equation}
Indeed, an integral of a nonnegative continuous function can vanish only if the
function vanishes everywhere.  We impose both no-error conditions for all priors
$0\leq p\leq1$. If a zero-prior hypothesis were removed from the no-error convention,
the endpoints would become degenerate single-hypothesis tasks rather than comparison.
The inconclusive probability is
\begin{equation}
  P_?
  :=
  p\Pr(?\mid H_=)+(1-p)\Pr(?\mid H_{\necase}).
  \label{eq:inconclusive-probability}
\end{equation}
The result below shows that the outcome $\Pi_=$ is always useless, so the optimal
unambiguous protocol is one-sided in the operational sense that only the independent
hypothesis is ever concluded.

\begin{theorem}
  \label{thm:unambiguous}
  For one-sided unambiguous comparison over valid general testers, the optimal
  inconclusive probability is
  \begin{equation}
    P_?^\star
    =
    p+\frac{1-p}{\Theta_d(N_1,N_2)}.
    \label{eq:unambiguous-main}
  \end{equation}
  A conclusive ``same'' outcome is impossible. The optimal valid tester is obtained
  from the parallel tester of Theorem~\ref{thm:minimum-error} by declaring its
  ``same'' outcome inconclusive and its complement conclusive ``different.''
\end{theorem}

\begin{proof}
  \emph{Upper bound.}
  The upper-bound proof uses only positivity and unitary normalization, and therefore holds for
  the relaxed set.
  The relative operator inequality first rules out a conclusive ``same'' outcome. Indeed,
  unambiguity of such an outcome would require
  \[
    \Tr(\Pi_=M_{\necase})=0.
  \]
  Using Lemma~\ref{lem:relative-operator-inequality} gives
  \[
    0
    \leq
    \Tr(\Pi_=M_=)
    \leq
    \Theta_d(N_1,N_2)\Tr(\Pi_=M_{\necase})
    =0.
  \]
  Hence the conclusive ``same'' outcome never occurs and can be removed without changing
  performance. Only the conclusive ``different'' outcome is relevant.

  Let $T=\Pi_{\necase}+\Pi_?$ be any unambiguous tester. Unambiguity of the conclusive
  ``different'' outcome gives
  \[
    \Tr(\Pi_{\necase}M_=)=0.
  \]
  Therefore, using deterministic normalization and the relative operator inequality,
  \begin{align}
    1
     & =\Tr(TM_=)
    \nonumber                     \\
     & =\Tr((T-\Pi_{\necase})M_=)
    \nonumber                     \\
     & \leq
    \Theta_d(N_1,N_2)
    \Tr((T-\Pi_{\necase})M_{\necase})
    \nonumber                     \\
     & =
    \Theta_d(N_1,N_2)\Tr(\Pi_?M_{\necase}).
    \label{eq:unambiguous-lower-bound-diff}
  \end{align}
  Thus
  \[
    \Tr(\Pi_?M_{\necase})\geq\frac{1}{\Theta_d(N_1,N_2)}.
  \]
  Unambiguity also implies $\Tr(\Pi_?M_=)=1$. Consequently,
  \begin{align}
    P_?
     & =
    p\Tr(\Pi_?M_=)+(1-p)\Tr(\Pi_?M_{\necase})
    \nonumber \\
     & \geq
    p+\frac{1-p}{\Theta_d(N_1,N_2)}.
    \label{eq:unambiguous-converse}
  \end{align}

  \emph{Achievability.}
  Use the parallel tester from the proof of
  Theorem~\ref{thm:minimum-error}. Relabel its ``same'' outcome as inconclusive and its
  complement as conclusive ``different.'' The tester is never conclusive under $H_=$ and
  is inconclusive under $H_{\necase}$ with probability $1/\Theta_d(N_1,N_2)$. This attains
  Eq.~\eqref{eq:unambiguous-main}.
\end{proof}

For $N_1=N_2=1$, this gives
$P_?^\star=p+(1-p)(d-1)/(2d)$, reproducing the one-use unambiguous unitary-comparison
value of Ref.~\cite{sedlakUnambiguousComparisonUnitary2009}.

\section{Saturation of optimal comparison with many queries to \texorpdfstring{$U_2$}{U2}}
\label{sec:saturation}

In this section, we analyze the regime where sufficiently many queries to the second
channel are available. The result shows that, once $N_2\geq(d-1)N_1$, extra queries to
$U_2$ no longer improve the optimal success probabilities.

The following monotonicity property will be used to extend the saturation equality from
$N_2=(d-1)N_1$ to all larger $N_2$.  The proof is given in
Appendix~\ref{appendix:monotonicity-proof}.

\begin{proposition}
  \label{prop:monotonicity}
  For all $a,b\geq1$,
  \begin{equation}
    \Theta_d(a,b)\leq\Theta_d(a,b+1),
    \qquad
    \Theta_d(a,b)\leq\Theta_d(a+1,b).
    \label{eq:Theta-monotone}
  \end{equation}
\end{proposition}

The threshold $(d-1)N_1$ has a simple operational origin. The known-reference test can
be rewritten using $N_1$ queries to $U_2^\ast$, and one query to $U_2^\ast$ can be
simulated exactly using $d-1$ parallel queries to $U_2$.

\begin{theorem}
  \label{thm:saturation-main}
  If $N_2\geq(d-1)N_1$, then
  \begin{equation}
    \Theta_d(N_1,N_2)
    =
    \gamma_{N_1,d}
    :=
    \binom{N_1+d^2-1}{N_1}.
    \label{eq:gamma-main}
  \end{equation}
  Consequently, writing $\gamma=\gamma_{N_1,d}$,
  \begin{align}
    P_{\ME}^{\star}
     & =
    \begin{cases}
      1-p,
       & 0\leq p\leq \dfrac{1}{1+\gamma}, \\[1.1ex]
      1-\dfrac{1-p}{\gamma},
       & \dfrac{1}{1+\gamma}<p\leq1,
    \end{cases}
    \label{eq:ME-saturated} \\
    P_?^\star
     & =
    p+\frac{1-p}{\gamma}.
    \label{eq:unambiguous-saturated}
  \end{align}
\end{theorem}

The key representation-theoretic statement used to prove
Theorem~\ref{thm:saturation-main} is the following lemma. The proof is given in
Appendix~\ref{appendix:saturation-proof}.

\begin{lemma}
  \label{lem:saturation}
  For all $a,b\geq1$,
  \begin{align}
    \Theta_d(a,b)
     & \leq
    \gamma_{a,d},
    \qquad
    \gamma_{a,d}:=
    \sum_{\lambda\in\Y_{a,d}}D_\lambda^2,
    \label{eq:Theta-upper-gamma} \\
    \gamma_{a,d}
     & =
    \binom{a+d^2-1}{a}.
    \label{eq:gamma-binomial}
  \end{align}
  Moreover,
  \begin{equation}
    \Theta_d(a,(d-1)a)\geq\gamma_{a,d}.
    \label{eq:Theta-saturation-lower}
  \end{equation}
  Consequently, if $N_2\geq(d-1)N_1$, then
  \begin{equation}
    \Theta_d(N_1,N_2)=\gamma_{N_1,d}.
    \label{eq:Theta-saturated}
  \end{equation}
\end{lemma}

\begin{proof}
  The assertion of Theorem~\ref{thm:saturation-main} follows immediately by inserting
  Lemma~\ref{lem:saturation} into Theorems~\ref{thm:minimum-error} and
  \ref{thm:unambiguous}.
\end{proof}

\begin{figure}[tbp]
  \centering
  \includegraphics[width=\columnwidth]{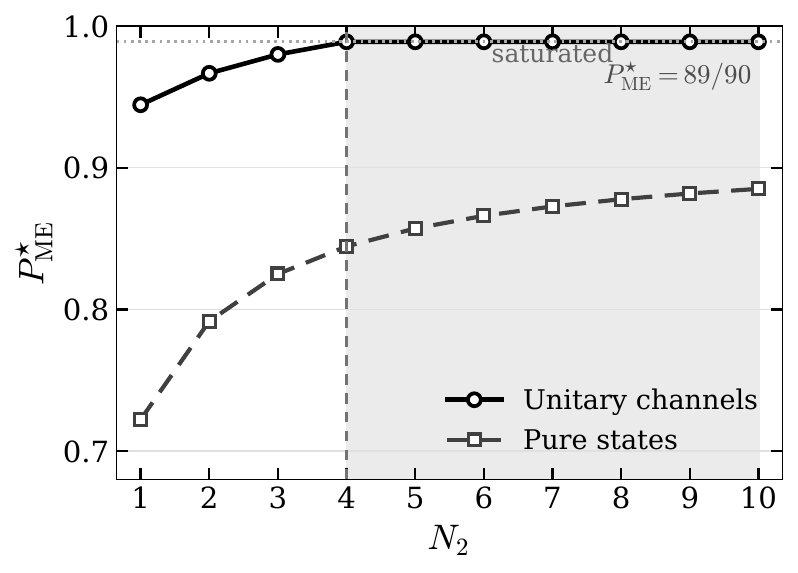}
  \caption{Minimum-error optimal success probability $P_{\ME}^{\star}$ as a
    function of $N_2$ for $d=3$, $N_1=2$, and prior probability $p=1/2$.
    The solid curve shows unitary-channel comparison, and the dashed curve shows
    pure-state comparison for the same $(d,N_1,p)$.
    The point $N_2=1$ lies outside the ordering convention $N_2\geq N_1$ used in
    the main text and is included using the symmetry under exchange of the two
    channels.
    The gray shaded region indicates the saturated regime
    $N_2\geq(d-1)N_1=4$, and the dotted horizontal line marks the saturated value
    $P_{\ME}^{\star}=89/90$ for unitary comparison.}
  \label{fig:non-saturated-d3-n1-2}
\end{figure}

This saturation is in sharp contrast with comparison of unknown pure states.  Suppose
that $N_1$ copies of $|\psi_1\rangle$ and $N_2$ copies of $|\psi_2\rangle$ are given, and
that either the two states are the same Haar-random pure state with probability $p$, or
they are independently Haar-random with probability $1-p$.  The optimal minimum-error
state-comparison success probability is
\begin{equation}
  P_{\mathrm{state},\ME}^{\star}
  =
  \begin{cases}
    1-p,
     & 0\leq p\leq \dfrac{1}{1+\beta_{N_1,N_2,d}}, \\[1.1ex]
    1-\dfrac{1-p}{\beta_{N_1,N_2,d}},
     & \dfrac{1}{1+\beta_{N_1,N_2,d}}<p\leq1,
  \end{cases}
  \label{eq:state-comparison-main}
\end{equation}
where
\begin{equation}
  \beta_{N_1,N_2,d}
  :=
  \frac{d^{\mathrm{sym}}_{N_1,d}d^{\mathrm{sym}}_{N_2,d}}
  {d^{\mathrm{sym}}_{N_1+N_2,d}},
  \qquad
  d^{\mathrm{sym}}_{N,d}:=\binom{N+d-1}{d-1}.
  \label{eq:state-beta-main}
\end{equation}
Appendix~\ref{appendix:state-comparison} gives the derivation.  Unlike
$\Theta_d(N_1,N_2)$, the quantity $\beta_{N_1,N_2,d}$ keeps increasing with $N_2$ for
fixed $N_1$ and approaches $d^{\mathrm{sym}}_{N_1,d}$ only in the limit
$N_2\to\infty$.  Thus finitely many copies of the second state do not reproduce the
known-reference value.  This statement concerns the comparison parameter
$\beta_{N_1,N_2,d}$ itself.  The success probability in
Eq.~\eqref{eq:state-comparison-main} remains the prior-guessing value $1-p$ in the
small-prior regime. In the nontrivial decision regime, where the second branch is
optimal, the increasing parameter gives continued improvement with $N_2$.
Figure~\ref{fig:non-saturated-d3-n1-2} illustrates this contrast for a representative
choice of parameters.

\paragraph*{\textit{Known-reference identity test.}}

The saturated value also has a direct operational interpretation as the optimal value
of a known-reference identity test. Consider the following modified problem. An
unknown channel is promised to be either the identity channel or a Haar-random unitary
channel $U$, with prior probabilities $p$ and $1-p$. The channel is queried $N$ times,
and valid general testers are allowed. As above, the upper bound is
proved on the larger unitary-normalized positive relaxation.

\begin{proposition}
  \label{prop:known-reference-identity-test}
  For the $N$-query identity-versus-Haar-unitary test described above, let
  \[
    \gamma_{N,d}:=\binom{N+d^2-1}{N}.
  \]
  The optimal minimum-error success probability is
  \begin{equation}
    P_{\mathrm{ref},\ME}^{\star}
    =
    \begin{cases}
      1-p,
       & 0\leq p\leq \dfrac{1}{1+\gamma_{N,d}}, \\[1.1ex]
      1-\dfrac{1-p}{\gamma_{N,d}},
       & \dfrac{1}{1+\gamma_{N,d}}<p\leq1.
    \end{cases}
    \label{eq:known-reference-ME}
  \end{equation}
  In the unambiguous version with outcomes
  $\{\Pi_I,\Pi_U,\Pi_?\}$ and no-error conclusive outcomes, following the convention of
  Section~\ref{sec:optimal-unambiguous-comparison}, the optimal inconclusive
  probability is
  \begin{equation}
    P_{\mathrm{ref},?}^{\star}
    =
    p+\frac{1-p}{\gamma_{N,d}}.
    \label{eq:known-reference-unambiguous}
  \end{equation}
\end{proposition}

The proof is given in Appendix~\ref{appendix:known-reference-proof}. The
upper-bound proof has the same tester-theoretic structure as that of
Theorem~\ref{thm:minimum-error}.
Let
\[
  M_I^{(N)}:=C_I^{\otimes N},
  \qquad
  M_U^{(N)}:=\int_{\SU(d)}dU\,C_U^{\otimes N}
\]
be the averaged Choi operators of the identity and Haar-random hypotheses. Schur--Weyl
duality gives the relative operator inequality
\begin{equation}
  M_I^{(N)}\leq\gamma_{N,d}M_U^{(N)}.
  \label{eq:known-reference-relative-bound}
\end{equation}
Thus, if $\Pi_I$ is the tester outcome for deciding ``identity,'' then
\[
  \Tr(\Pi_I M_I^{(N)})
  \leq
  \gamma_{N,d}\Tr(\Pi_I M_U^{(N)}).
\]
In words, a tester that accepts the identity with probability $\alpha$ must also accept
a Haar-random unitary with probability at least $\alpha/\gamma_{N,d}$. The optimum is
therefore obtained either by always answering ``random unitary'' when the prior $p$ is
small, or by using a tester with
\[
  \Pr(I\mid I)=1,
  \qquad
  \Pr(I\mid U\ \mathrm{Haar})=\frac{1}{\gamma_{N,d}}.
\]
Such a tester can be chosen to be parallel.

We now return to comparison. Although unitary comparison aims to compare \textit{both}
unknown unitary channels, the known-reference test provides a useful benchmark. If
$U_2$ is perfectly known, comparison of $U_1$ with $U_2$ is equivalent to the identity
test for
\[
  V:=U_1U_2^\dagger.
\]
Indeed, knowing $U_2$ allows us to apply $U_2^\dagger$ exactly and deterministically.
This operation is not available when $U_2$ is unknown and only finitely many black-box
queries to $U_2$ are given. Hence the known-$U_2$ problem has a larger strategy set than
the original comparison problem, and Proposition~\ref{prop:known-reference-identity-test}
gives the upper bound governed by $\gamma_{N_1,d}$.

Appendix~\ref{appendix:saturation-proof} proves saturation algebraically by pairing
complementary Schur sectors. The complex-conjugation construction below gives the
operational counterpart of the same phenomenon: $(d-1)N_1$ queries to $U_2$ are
sufficient to reproduce the reference transformation needed by the known-unitary test.
The upper bound is therefore reachable in the original comparison problem once
$N_2\geq(d-1)N_1$. The key point is the same normalization operator $S$ that appears in
the parallel tester for the known-reference problem.  In the Schur--Weyl decomposition
of the reference systems for the conjugate representation,
\[
  \Hcal_N^R
  \cong
  \bigoplus_{\lambda\in\Y_{N,d}}
  \overline{\Ucal_\lambda}^{\,R}\otimes\Vcal_\lambda^R,
\]
set
\begin{equation}
  S_N^R
  :=
  \sum_{\lambda\in\Y_{N,d}}
  q_\lambda
  \frac{\Id_{\overline{\Ucal_\lambda}^{\,R}}}{D_\lambda}
  \otimes
  \frac{\Id_{\Vcal_\lambda^R}}{\dim \Vcal_\lambda}.
  \label{eq:known-reference-S}
\end{equation}
Here
\[
  q_\lambda:=\frac{D_\lambda^2}{\gamma_{N,d}},
  \qquad
  D_\lambda:=\dim\Ucal_\lambda,
\]
so that $\Tr S_N^R=1$.  This is precisely the Schur-block version of the optimal
parallel-tester normalization in the known-reference test. Choosing the multiplicity
part of the Schur decomposition to be maximally entangled, the input state can be
written as
\begin{equation}
  |\eta_N\rangle
  =
  \left(\sqrt{S_N^R}\otimes\Id^K\right)|\Phi_N^+\rangle,
  \label{eq:eta-from-S}
\end{equation}
where
\[
  |\Phi_N^+\rangle
  :=
  \bigotimes_{k=1}^{N}
  \sum_{r=1}^{d}|r\rangle^{R_k}|r\rangle^{K_k}
\]
is the unnormalized maximally entangled vector between the reference and channel-input
systems. Since $S_N^R$ is a direct sum of scalar multiples of the identity on the Schur
blocks, it commutes with the conjugate representation.
\begin{equation}
  \left[S_N^R,U^{\ast\otimes N}\right]=0
  \qquad
  \text{for every }U\in\SU(d).
  \label{eq:S-commutes-conjugate}
\end{equation}
If $N_1$ queries to $U_2^\ast$ are available, this commutativity lets us rewrite the
known-reference tester for $V=U_1U_2^\dagger$ without applying $U_2^\dagger$ directly.
Here $U^\ast$ denotes entrywise complex conjugation in the computational basis used to
define the Choi vector $|\Phi\rangle$. The circuit identity behind this reduction is
shown in Fig.~\ref{fig:conjugate-reference-reduction}.
\begin{figure*}[t]
  \centering
  \begin{tikzpicture}[
      x=0.52cm,
      y=0.52cm,
      line cap=round,
      gate/.style={
          draw,
          fill=white,
          rounded corners=1.5pt,
          minimum width=0.76cm,
          minimum height=0.40cm,
          inner sep=0.5pt
        },
      uone/.style={gate, fill=blue!22, draw=blue!70!black},
      utwo/.style={gate, fill=orange!28, draw=orange!80!black},
      inputstate/.style={draw=black!45, fill=black!5},
      tester/.style={draw=black!50, fill=green!7},
      wirelabel/.style={anchor=west,fill=white,inner sep=0.5pt},
      every node/.style={font=\scriptsize}
    ]
    %% top circuit: U_2^\dagger then U_1 on the K registers
    \begin{scope}
      \filldraw[inputstate]
      (-0.95,-0.40) .. controls (-3.25,0.30) and (-3.25,6.70) .. (-0.95,7.40)
      -- (-0.95,-0.40) -- cycle;
      \node at (-1.95,3.50) {$|\eta_{N_1}\rangle$};
      \filldraw[tester] (5.70,-0.40) rectangle (8.85,7.40);
      \node at (7.275,3.50) {$\{\Pi_I,\Pi_U\}$};
      \foreach \y/\lab in {7/K_1,6/K_2,4/K_{N_1},3/R_1,2/R_2,0/R_{N_1}} {
          \draw (-0.95,\y) -- (5.70,\y);
          \node[wirelabel] at (-0.98,\y) {$\lab$};
        }
      \foreach \y in {7,6,4} {
          \node[utwo] at (2.00,\y) {$U_2^\dagger$};
          \node[uone] at (3.70,\y) {$U_1$};
        }
      \node at (2.85,5) {$\vdots$};
      \node at (2.85,1) {$\vdots$};
    \end{scope}

    \draw[<->,thick] (2.40,-0.90) -- (2.40,-2.20)
    node[midway,right,font=\scriptsize,inner sep=2pt] {transpose trick};

    \begin{scope}[shift={(0,-9.9)}]
      \filldraw[inputstate]
      (-0.95,-0.40) .. controls (-3.25,0.30) and (-3.25,6.70) .. (-0.95,7.40)
      -- (-0.95,-0.40) -- cycle;
      \node at (-1.95,3.50) {$|\eta_{N_1}\rangle$};
      \filldraw[tester] (5.70,-0.40) rectangle (8.85,7.40);
      \node at (7.275,3.50) {$\{\Pi_I,\Pi_U\}$};
      \foreach \y/\lab in {7/K_1,6/K_2,4/K_{N_1},3/R_1,2/R_2,0/R_{N_1}} {
          \draw (-0.95,\y) -- (5.70,\y);
          \node[wirelabel] at (-0.98,\y) {$\lab$};
        }
      \foreach \y in {7,6,4} {
          \node[uone] at (2.85,\y) {$U_1$};
        }
      \foreach \y in {3,2,0} {
          \node[utwo] at (2.85,\y) {$U_2^\ast$};
        }
      \node at (2.85,5) {$\vdots$};
      \node at (2.85,1) {$\vdots$};
    \end{scope}
  \end{tikzpicture}
  \caption{The known-reference comparison circuit rewritten using queries to
    $U_2^\ast$.  The top circuit applies $U_2^\dagger$ and then $U_1$ to the $K$
    registers.  The bottom circuit gives the same input to the tester by moving
    $U_2^\dagger$ to the reference registers as $U_2^\ast$.}
  \label{fig:conjugate-reference-reduction}
\end{figure*}
\begin{align}
   &
  \left(\Id^{\otimes N_1}\right)^R
  \otimes
  \left((U_1U_2^\dagger)^{\otimes N_1}\right)^K
  |\eta_{N_1}\rangle
  \nonumber \\
   & =
  \left(\sqrt{S_{N_1}^R}\otimes\Id^K\right)
  \left[
    \left(\Id^{\otimes N_1}\right)^R
    \otimes
    \left((U_1U_2^\dagger)^{\otimes N_1}\right)^K
    \right]
  |\Phi_{N_1}^+\rangle
  \nonumber \\
   & =
  \left[
    \sqrt{S_{N_1}^R}
    \left(U_2^{\ast\otimes N_1}\right)^R
    \otimes
    \left(U_1^{\otimes N_1}\right)^K
    \right]
  |\Phi_{N_1}^+\rangle
  \nonumber \\
   & =
  \left[
    \left(U_2^{\ast\otimes N_1}\right)^R
    \sqrt{S_{N_1}^R}
    \otimes
    \left(U_1^{\otimes N_1}\right)^K
    \right]
  |\Phi_{N_1}^+\rangle
  \nonumber \\
   & =
  \left(U_2^{\ast\otimes N_1}\right)^R
  \otimes
  \left(U_1^{\otimes N_1}\right)^K
  |\eta_{N_1}\rangle .
  \label{eq:eta-comparison-reference-reduction}
\end{align}
Thus, the final expression is exactly the same input as the known-reference tester for
$V=U_1U_2^\dagger$. The second equality uses the transpose trick
$\Id\otimes AB|\Phi_N^+\rangle=B^T\otimes A|\Phi_N^+\rangle$, and the third equality is
exactly the commutativity in Eq.~\eqref{eq:S-commutes-conjugate}. The exact unitary
complex-conjugation protocol of
Ref.~\cite{miyazakiComplexConjugationSupermap2019}, also in line with the optimal
universal complex-conjugation circuits of
Ref.~\cite{eblerOptimalUniversalQuantum2023}, simulates one query to $U^\ast$ from $d-1$
parallel queries to an unknown $d$-dimensional unitary $U$. Thus $(d-1)N_1$ queries to
$U_2$ suffice to simulate the required $N_1$ queries to $U_2^\ast$. This parallel
construction reaches the known-$U_2$ upper bound, matching the
representation-theoretic proof above.

\section{Conclusion}
\label{sec:Conclusion}

In this paper, we analyzed unitary channel comparison, which is a task determining
whether two unknown unitary channels are the same or different by directly detecting the
difference between the two channels without tomography by querying each of the channels
only finite times. We considered the setting that the unknown unitary channels are
uniformly and randomly given under the promise that the two unitary channels are
identical with probability $p$ and independently distributed with probability $1-p$.

For both minimum-error and one-sided unambiguous comparison, the optimal probabilities
are determined by the single finite parameter $\Theta_d(N_1,N_2)$. We distinguished the
valid general tester class from a larger positive relaxation normalized only on
repeated-unitary inputs. The upper bound, based on the relative operator inequality
$M_=^{a,b}\leq\Theta_d(a,b)M_{\necase}^{a,b}$, holds even for this relaxation, while the
bound is attained by an explicit valid parallel tester.

The characteristic property of unitary comparison is that the optimal averaged success
probabilities saturate for all $N_2\geq(d-1)N_1$ when $N_1$ is fixed and cannot
be improved by adding more queries to $U_2$ in that regime. This feature contrasts with
pure-state comparison at the level of the comparison parameter. For states, the
corresponding parameter exhibits no saturation at finite $N_2$ and approaches the
known-reference value only asymptotically. In the nontrivial decision regime, this
continued growth of the parameter steadily improves the optimal averaged success
probability. This highlights a difference between the state and channel versions of the
task, paralleling what is observed in quantum discrimination
\cite{acinStatisticalDistinguishabilityUnitary2001,
  duanEntanglementNotNecessary2007,
  duanPerfectDistinguishabilityQuantum2009}.

\section*{AI disclosure}
For the scientific derivations, the authors used OpenAI ChatGPT with GPT-5.6 Sol
in Pro mode in two separate interactive workflows. In the first workflow, the
authors supplied the definitions, notation, and Schur--Weyl setup used in the
manuscript and asked the tool to derive the Schur--Weyl block form of the
Haar-averaged Choi operator $M_{\necase}^{a,b}$ used in
Lemma~\ref{lem:relative-operator-inequality}. In the second workflow, the authors
supplied the definition of $\Theta_d(a,b)$ and asked the tool to construct a proof
of Proposition~\ref{prop:monotonicity}, which is used in the saturation argument.
The authors independently rederived every retained expression, checked each proof
step against the definitions in the manuscript, and rewrote the final arguments.

For the numerical implementation, the authors used OpenAI Codex with GPT-5.6 Sol
through iterative, author-directed prompts to implement the finite procedure in
Appendix~\ref{appendix:theta-computation} in Python. The authors reviewed the
complete source code and validated its output against hand-checkable cases,
including the worked examples and selected exact values in
Table~\ref{tab:small-theta-values}.

For manuscript preparation, OpenAI Codex with GPT-5.6 Sol assisted through
iterative, author-directed prompts with drafting and organization and with the
TikZ implementation of author-designed schematic figures. The authors determined
the scientific content and design of the figures, inspected the rendered output,
and reviewed and revised all AI-assisted text. All conclusions and final decisions
were made by the authors, who take full responsibility for the content of the
paper.

\section*{Acknowledgments}
This work was supported by the MEXT Quantum Leap Flagship Program (MEXT QLEAP)
JPMXS0118069605 and JPMXS0120351339, Japan Society for the Promotion of Science (JSPS)
KAKENHI Grants No. 21H03394 and No. 23K21643, JST CREST Grant No. JPMJCR25I5, JST ASPIRE
JPMJAP25A3, JST NEXUS JPMJNX26C9, FoPM, WINGS Program, the University of Tokyo, and IBM
Quantum, and JST Moonshot R\&D Program Grant Numbers JPMJMS226C and JPMJMS256K.

\appendix

\section{Explicit computation of \texorpdfstring{$\Theta_d(a,b)$}{Theta\_d(a,b)}}
\label{appendix:theta-computation}

This appendix explains how to evaluate $\Theta_d(a,b)$ exactly for finite $(d,a,b)$.
The computation is a finite search over Schur sectors. The required inputs are
standard: the list of Young diagrams, their representation dimensions, and the
Littlewood--Richardson multiplicities. We first summarize the algorithm and then
review the two representation-theoretic subroutines used in its implementation.

\begin{quote}\small
  \textbf{Procedure for computing $\Theta_d(a,b)$.}
  \begin{enumerate}
    \item Enumerate
          $L=\Y_{a,d}$, $M=\Y_{b,d}$, and $N=\Y_{a+b,d}$.
    \item Compute $D_\lambda$, $D_\mu$, and $D_\nu$ for all
          $\lambda\in L$, $\mu\in M$, and $\nu\in N$ using
          Eq.~\eqref{eq:weyl-dimension-formula}.
    \item Compute each $c_{\lambda\mu}^{\nu}$ by the LR tableau rule described below.
    \item For each $\nu\in N$, set
          \[
            A_\nu\leftarrow
            \sum_{\lambda\in L}\sum_{\mu\in M}
            c_{\lambda\mu}^{\nu}D_\lambda D_\mu.
          \]
    \item Return
          \[
            \Theta_d(a,b)=\max_{\nu\in N}\frac{A_\nu}{D_\nu}.
          \]
  \end{enumerate}
\end{quote}

The two subroutines used in Steps 2 and 3 are as follows.

\paragraph*{Weyl dimension formula.}

For a Young diagram $\lambda=(\lambda_1,\ldots,\lambda_d)$ with at most $d$ rows, let
$D_\lambda:=\dim\Ucal_\lambda$.  We append trailing zero rows if $\lambda$ has fewer
than $d$ rows.  Weyl's dimension formula
\cite{weylClassicalGroupsTheir1939,fultonRepresentationTheory2004}
gives
\begin{equation}
  D_\lambda=\prod_{1\leq i<j\leq d}
  \frac{\lambda_i-\lambda_j+j-i}{j-i}.
  \label{eq:weyl-dimension-formula}
\end{equation}

\paragraph*{Littlewood--Richardson tableau rule.}

The Littlewood--Richardson coefficients defined in Eq.~\eqref{eq:LR-definition} can be
evaluated as follows.  If
$\lambda\not\subseteq\nu$ or $|\nu|-|\lambda|\neq|\mu|$, then
$c_{\lambda\mu}^{\nu}=0$.  Otherwise, form the skew shape $\nu/\lambda$ by removing
the boxes of $\lambda$ from the upper-left corner of $\nu$.  Figure~\ref{fig:skew-shape}
shows this operation for $\lambda=(2)$ and $\nu=(3,1)$.  In the middle panel, the two
gray boxes are the copy of $\lambda$ inside $\nu$.  In the right panel, these boxes are
kept as empty outlines, while the two blue boxes form the skew diagram $\nu/\lambda$.

\begin{figure*}[t]
  \centering
  \begin{tikzpicture}[
      >=Latex,
      arr/.style={->, line width=0.65pt, draw=black!50}
    ]
    \node[anchor=base west, font=\small] at (0.00,1.95) {(a)};
    \node[anchor=base west, font=\scriptsize] at (0.48,1.95) {$\lambda=(2)$};
    \node at (1.85,0.70)
      {\scalebox{1.45}{\ydiagram{\ydold{0}{0}\ydold{1}{0}}}};
    \draw[arr] (3.45,0.70) -- (4.35,0.70);
    \node[anchor=base west, font=\small] at (4.55,1.95) {(b)};
    \node[anchor=base west, font=\scriptsize] at (5.03,1.95)
      {embed in $\nu=(3,1)$};
    \node at (6.85,0.40)
      {\scalebox{1.45}{\ydiagram{\ydold{0}{0}\ydold{1}{0}\ydnew{2}{0}{}\ydnew{0}{1}{}}}};
    \draw[arr] (8.85,0.70) -- (9.75,0.70);
    \node[anchor=base west, font=\small] at (9.95,1.95) {(c)};
    \node[anchor=base west, font=\scriptsize] at (10.43,1.95)
      {remove $\lambda$ to form $\nu/\lambda$};
    \node at (12.65,0.40)
      {\scalebox{1.45}{\ydiagram{\ydblank{0}{0}\ydblank{1}{0}\ydnew{2}{0}{}\ydnew{0}{1}{}}}};
  \end{tikzpicture}
  \caption{\label{fig:skew-shape}
    Construction of the skew diagram $\nu/\lambda$.
    (a)~The diagram $\lambda$.
    (b)~A copy of $\lambda$ is placed in the upper-left corner of $\nu$; the
    remaining boxes of $\nu$ are the skew shape.
    (c)~Those remaining boxes are the only ones filled when computing
    $c_{\lambda\mu}^{\nu}$. Empty outlines keep the ambient shape of $\nu$
    visible.}
\end{figure*}

A Littlewood--Richardson tableau of shape $\nu/\lambda$ and content
$\mu=(\mu_1,\mu_2,\ldots)$ is a filling of this skew shape with $\mu_i$ copies of the
integer $i$ satisfying three conditions.
\begin{enumerate}
  \item rows weakly increase from left to right.
  \item columns strictly increase from top to bottom.
  \item the reading word, obtained by reading each row from right to left and then from
        the top row to the bottom row, is a lattice word.
\end{enumerate}
The lattice condition means that, in every initial segment of the reading word, the
number of $1$'s is at least the number of $2$'s, the number of $2$'s is at least the
number of $3$'s, and so on.  The coefficient $c_{\lambda\mu}^{\nu}$ is the number of
fillings that survive all three tests.  This convention is one standard tableau
formulation of the Littlewood--Richardson rule \cite{fultonYoungTableaux1997}.

Figure~\ref{fig:lr-reading-word} gives a representative example of this LR count.

\begin{figure*}[t]
  \centering
  \begin{tikzpicture}[
      >=Latex,
      arr/.style={->, line width=0.65pt, draw=black!50}
    ]
    \node[anchor=base west, font=\small] at (0.00,2.55) {(a)};
    \node[anchor=base west, font=\scriptsize] at (0.48,2.55)
      {Skew shape $(3,2,1)/(2,1)$};
    \node at (2.25,0.95)
      {\scalebox{1.35}{\ydiagram{\ydblank{0}{0}\ydblank{1}{0}\ydblank{0}{1}\ydnew{2}{0}{}\ydnew{1}{1}{}\ydnew{0}{2}{}}}};
    \draw[arr] (4.35,1.10) -- (5.25,1.10);
    \node[anchor=base west, font=\small] at (5.45,2.55) {(b)};
    \node[anchor=base west, font=\scriptsize] at (5.93,2.55)
      {Valid LR tableaux};
    \node at (7.65,0.95)
      {\scalebox{1.35}{\ydiagram{\ydblank{0}{0}\ydblank{1}{0}\ydblank{0}{1}\ydnew{2}{0}{1}\ydnew{1}{1}{1}\ydnew{0}{2}{2}}}};
    \node[font=\scriptsize] at (7.65,-0.15) {$112$};
    \node at (10.95,0.95)
      {\scalebox{1.35}{\ydiagram{\ydblank{0}{0}\ydblank{1}{0}\ydblank{0}{1}\ydnew{2}{0}{1}\ydnew{1}{1}{2}\ydnew{0}{2}{1}}}};
    \node[font=\scriptsize] at (10.95,-0.15) {$121$};
    \node[font=\scriptsize] at (14.00,1.10)
      {$c_{(2,1),(2,1)}^{(3,2,1)}=2$};
  \end{tikzpicture}
  \caption{\label{fig:lr-reading-word}
    Complete LR count for $\lambda=(2,1)$, $\mu=(2,1)$, and $\nu=(3,2,1)$.
    (a)~The skew shape $\nu/\lambda$ has three boxes.
    (b)~The content $\mu=(2,1)$ requires two $1$'s and one $2$.
    Reading each row from right to left, top to bottom, gives the lattice words
    $112$ and $121$. The only other word of this content is $211$, whose first
    initial segment violates the lattice condition. Hence exactly two LR tableaux
    survive.}
\end{figure*}

Table~\ref{tab:small-theta-values} lists small exact values generated by this
procedure.  The displayed $\nu$ is one maximizing outer Young diagram. When several
diagrams tie, any one of them can be used in the parallel construction of
Theorem~\ref{thm:minimum-error}.

\begin{table}[t]
  \caption{\label{tab:small-theta-values}
    Small values of $\Theta_d(a,b)$ and one maximizing Young diagram $\nu$.}
  \begin{ruledtabular}
    \begin{tabular}{cccc}
      $d$ & $(a,b)$ & maximizing $\nu$ & $\Theta_d(a,b)$ \\
      \hline
      $2$ & $(1,1)$ & $(1,1)$          & $4$             \\
      $2$ & $(2,2)$ & $(2,2)$          & $10$            \\
      $3$ & $(1,1)$ & $(1,1)$          & $3$             \\
      $3$ & $(1,2)$ & $(1,1,1)$        & $9$             \\
      $3$ & $(2,2)$ & $(2,1,1)$        & $15$            \\
      $3$ & $(2,4)$ & $(2,2,2)$        & $45$            \\
      $4$ & $(1,1)$ & $(1,1)$          & $8/3$           \\
      $4$ & $(2,2)$ & $(1,1,1,1)$      & $36$            \\
      $4$ & $(2,4)$ & $(2,2,1,1)$      & $61$            \\
    \end{tabular}
  \end{ruledtabular}
\end{table}

We give two small examples in which all diagrams and dimensions can be checked by hand.

\paragraph*{Example 1. General $d$ and $N_1=N_2=1$.}

In this case the procedure is evaluated at $(a,b)=(1,1)$.  The only input diagrams are
$\lambda=\mu=(1)$, and
\[
  \Ucal_{(1)}\otimes\Ucal_{(1)}
  \cong
  \Ucal_{(2)}\oplus\Ucal_{(1,1)}.
\]
The two Littlewood--Richardson coefficients used here are
$c_{(1),(1)}^{(2)}=c_{(1),(1)}^{(1,1)}=1$.  In this case, the
Littlewood--Richardson rule reduces to adding one box to $(1)$. The two possible
resulting diagrams are $(2)$ and $(1,1)$, each with multiplicity one.
Here
\[
  D_{(1)}=d,\qquad
  D_{(2)}=\frac{d(d+1)}{2},\qquad
  D_{(1,1)}=\frac{d(d-1)}{2}.
\]
Hence
\begin{equation}
  \label{eq:theta-one-one}
  \Theta_d(1,1)
  =
  \max\left\{
  \frac{d^2}{d(d+1)/2},
  \frac{d^2}{d(d-1)/2}
  \right\}
  =
  \frac{2d}{d-1}.
\end{equation}
Thus, for all $d\geq2$, $\Theta_d(1,1)=2d/(d-1)$.
The two candidate outer sectors share the same numerator $d^2$, coming from the
single coupling $(1)\otimes(1)$. The antisymmetric sector $(1,1)$ therefore
wins solely because it has the smaller representation dimension, which
enlarges the coherent-to-incoherent ratio.

\paragraph*{Example 2. $d=3$, $N_1=1$, $N_2=2$.}

For $a=1$ and $b=2$, we have
$\Y_{1,3}=\{(1)\}$ and $\Y_{2,3}=\{(2),(1,1)\}$.  The relevant products are
\[
  \begin{aligned}
    \Ucal_{(1)}\otimes\Ucal_{(2)}
     & \cong
    \Ucal_{(3)}\oplus\Ucal_{(2,1)}, \\
    \Ucal_{(1)}\otimes\Ucal_{(1,1)}
     & \cong
    \Ucal_{(2,1)}\oplus\Ucal_{(1,1,1)}.
  \end{aligned}
\]
Equivalently, the nonzero Littlewood--Richardson coefficients in these two
products are all equal to one.
\[
  c_{(1),(2)}^{(3)}
  =
  c_{(1),(2)}^{(2,1)}
  =
  c_{(1),(1,1)}^{(2,1)}
  =
  c_{(1),(1,1)}^{(1,1,1)}
  =
  1.
\]
The same one-box LR rule shows that adding one box to $(2)$ or $(1,1)$ produces
exactly the displayed diagrams, each with multiplicity one.
Using
\[
  \begin{gathered}
    D_{(1)}=3,\quad
    D_{(2)}=6,\quad
    D_{(1,1)}=3,\\
    D_{(3)}=10,\quad
    D_{(2,1)}=8,\quad
    D_{(1,1,1)}=1,
  \end{gathered}
\]
we compute the numerators $A_\nu$ directly from the definition
$A_\nu=\sum_{\lambda,\mu}c_{\lambda\mu}^{\nu}D_\lambda D_\mu$.  In this example
$\lambda=(1)$ is fixed and $\mu$ is either $(2)$ or $(1,1)$, so
\[
  \begin{aligned}
    A_{(3)}
     & =
    c_{(1),(2)}^{(3)}D_{(1)}D_{(2)}
    +c_{(1),(1,1)}^{(3)}D_{(1)}D_{(1,1)}
    \\
     & =
    1\cdot3\cdot6+0\cdot3\cdot3
    =
    18,  \\
    A_{(2,1)}
     & =
    c_{(1),(2)}^{(2,1)}D_{(1)}D_{(2)}
    +c_{(1),(1,1)}^{(2,1)}D_{(1)}D_{(1,1)}
    \\
     & =
    1\cdot3\cdot6+1\cdot3\cdot3
    =
    27,  \\
    A_{(1,1,1)}
     & =
    c_{(1),(2)}^{(1,1,1)}D_{(1)}D_{(2)}
    +c_{(1),(1,1)}^{(1,1,1)}D_{(1)}D_{(1,1)}
    \\
     & =
    0\cdot3\cdot6+1\cdot3\cdot3
    =
    9.
  \end{aligned}
\]
Dividing by the corresponding $D_\nu$ gives
\[
  \frac{A_{(3)}}{D_{(3)}}=\frac{18}{10}=\frac95,
  \qquad
  \frac{A_{(2,1)}}{D_{(2,1)}}=
  \frac{27}{8},
  \qquad
  \frac{A_{(1,1,1)}}{D_{(1,1,1)}}=\frac{9}{1}=9.
\]
Therefore
\[
  \Theta_3(1,2)=9.
\]
The completely antisymmetric sector $(1,1,1)$ wins because it is
one-dimensional, so the weight $A_{(1,1,1)}=9$ occupies a single
one-dimensional outer sector. The mixed sector $(2,1)$ carries a larger
raw weight $A_{(2,1)}=27$, but that weight is spread over an
eight-dimensional representation, so the ratio $27/8$ is smaller.
Thus the relevant parameter is $\Theta_3(1,2)=9$, which exceeds the smaller-budget
value $\Theta_3(1,1)=3$, as required by monotonicity.  This value also coincides with
the saturated value $\gamma_{1,3}=\binom{1+3^2-1}{1}=9$, since
$N_2=2\geq(d-1)N_1=2$.

\section{Monotonicity of \texorpdfstring{$\Theta_d(a,b)$}{Theta\_d(a,b)}}
\label{appendix:monotonicity-proof}

This appendix proves Proposition~\ref{prop:monotonicity}.  Throughout we abbreviate the
numerator of Eq.~\eqref{eq:Theta-definition} by
\begin{align}
  A_\nu^{(a,b)}
   & :=
  \sum_{\lambda\in\Y_{a,d}}
  \sum_{\mu\in\Y_{b,d}}
  c_{\lambda\mu}^{\nu}D_\lambda D_\mu,
  \nonumber \\
  \Theta_d(a,b)
   & =
  \max_{\nu\in\Y_{a+b,d}}
  \frac{A_\nu^{(a,b)}}{D_\nu},
  \label{eq:Anu-ab-monotone}
\end{align}
and we let $V:=\Ucal_{(1)}\cong\C^d$ denote the defining representation of
$\mathrm{U}(d)$, with dual $V^\ast$.  We use polynomial $\mathrm{U}(d)$
representations, so that $c_{\lambda\mu}^{\nu}=\dim\Hom_{\mathrm{U}(d)}(\Ucal_\nu,
  \Ucal_\lambda\otimes\Ucal_\mu)$.

Since $c_{\lambda\mu}^{\nu}=c_{\mu\lambda}^{\nu}$, the numerator is symmetric,
i.e.,
$A_\nu^{(a,b)}=A_\nu^{(b,a)}$, and hence
\begin{equation}
  \Theta_d(a,b)=\Theta_d(b,a).
  \label{eq:Theta-symmetric}
\end{equation}
Therefore it suffices to prove monotonicity in the second argument,
$\Theta_d(a,b)\leq\Theta_d(a,b+1)$. The statement
$\Theta_d(a,b)\leq\Theta_d(a+1,b)$ then follows by Eq.~\eqref{eq:Theta-symmetric}.

\begin{proof}
  \emph{Step 1.}
  We first claim that for every $\nu\in\Y_{a+b,d}$, $\lambda\in\Y_{a,d}$, and
  $\rho\in\Y_{b+1,d}$,
  \begin{equation}
    \sum_{\kappa\in\Y_{a+b+1,d}}
    c_{\nu,(1)}^{\kappa}\,c_{\lambda,\rho}^{\kappa}
    \geq
    \sum_{\mu\in\Y_{b,d}}
    c_{\lambda,\mu}^{\nu}\,c_{\mu,(1)}^{\rho}.
    \label{eq:LR-inequality}
  \end{equation}
  To see this, decompose $\Ucal_\nu\otimes V\cong\bigoplus_\kappa
    \Ucal_\kappa\otimes\C^{c_{\nu,(1)}^{\kappa}}$ and use
  $c_{\lambda,\rho}^{\kappa}=\dim\Hom_{\mathrm{U}(d)}(\Ucal_\kappa,
    \Ucal_\lambda\otimes\Ucal_\rho)$ to express the left-hand side as a single Hom
  space and then apply duality.
  \begin{equation}
    \begin{aligned}
      \sum_{\kappa}c_{\nu,(1)}^{\kappa}c_{\lambda,\rho}^{\kappa}
       & =
      \dim\Hom_{\mathrm{U}(d)}
      \bigl(\Ucal_\nu\otimes V,\,\Ucal_\lambda\otimes\Ucal_\rho\bigr)
      \\
       & =
      \dim\Hom_{\mathrm{U}(d)}
      \bigl(\Ucal_\nu,\,\Ucal_\lambda\otimes\Ucal_\rho\otimes V^\ast\bigr).
    \end{aligned}
    \label{eq:LR-hom-rewrite}
  \end{equation}
  By Frobenius reciprocity (adjunction between $-\otimes V$ and $-\otimes V^\ast$),
  for each $\mu$
  \begin{align}
    \dim\Hom_{\mathrm{U}(d)}
    \bigl(\Ucal_\mu,\Ucal_\rho\otimes V^\ast\bigr)
     & =
    \dim\Hom_{\mathrm{U}(d)}
    \bigl(\Ucal_\mu\otimes V,\Ucal_\rho\bigr)
    \nonumber \\
     & =
    c_{\mu,(1)}^{\rho},
    \label{eq:frobenius-step}
  \end{align}
  so the constituents of $\Ucal_\rho\otimes V^\ast$ that are polynomial irreducibles
  $\Ucal_\mu$ with $\mu\in\Y_{b,d}$ contribute a subrepresentation
  \begin{equation}
    \Ucal_\rho\otimes V^\ast
    \supseteq
    \bigoplus_{\mu\in\Y_{b,d}}
    \Ucal_\mu\otimes\C^{\,c_{\mu,(1)}^{\rho}}.
    \label{eq:rho-Vstar-decomp}
  \end{equation}
  Tensoring with $\Ucal_\lambda$ and substituting into
  Eq.~\eqref{eq:LR-hom-rewrite} gives
  \begin{equation}
    \begin{aligned}
       & \dim\Hom_{\mathrm{U}(d)}
      \bigl(\Ucal_\nu,\Ucal_\lambda\otimes\Ucal_\rho\otimes V^\ast\bigr)
      \\
       & \quad\geq
      \sum_{\mu\in\Y_{b,d}}
      c_{\mu,(1)}^{\rho}\,
      \dim\Hom_{\mathrm{U}(d)}(\Ucal_\nu,\Ucal_\lambda\otimes\Ucal_\mu)
      \\
       & \quad=
      \sum_{\mu\in\Y_{b,d}}
      c_{\mu,(1)}^{\rho}\,c_{\lambda,\mu}^{\nu},
    \end{aligned}
    \label{eq:LR-inequality-derivation}
  \end{equation}
  which is Eq.~\eqref{eq:LR-inequality}.
  \emph{Step 2.}
  Multiplying Eq.~\eqref{eq:LR-inequality} by $D_\lambda D_\rho$ and summing over
  $\lambda\in\Y_{a,d}$ and $\rho\in\Y_{b+1,d}$ gives
  \begin{equation}
    \begin{aligned}
       & \sum_{\lambda,\rho}D_\lambda D_\rho
      \sum_{\kappa\in\Y_{a+b+1,d}}
      c_{\nu,(1)}^{\kappa}\,c_{\lambda,\rho}^{\kappa}
      \\
       & \quad\geq
      \sum_{\lambda,\rho}D_\lambda D_\rho
      \sum_{\mu\in\Y_{b,d}}
      c_{\lambda,\mu}^{\nu}\,c_{\mu,(1)}^{\rho}.
    \end{aligned}
    \label{eq:weighted-LR-inequality}
  \end{equation}
  The left-hand side of Eq.~\eqref{eq:weighted-LR-inequality} is
  \begin{equation}
    \begin{aligned}
       & \sum_{\lambda,\rho}D_\lambda D_\rho
      \sum_{\kappa}c_{\nu,(1)}^{\kappa}c_{\lambda,\rho}^{\kappa}
      \\
       & \quad=
      \sum_{\kappa}c_{\nu,(1)}^{\kappa}
      \sum_{\lambda,\rho}c_{\lambda,\rho}^{\kappa}D_\lambda D_\rho
      \\
       & \quad=
      \sum_{\kappa}c_{\nu,(1)}^{\kappa}A_\kappa^{(a,b+1)}.
    \end{aligned}
    \label{eq:weighted-LR-left}
  \end{equation}
  For the right-hand side of Eq.~\eqref{eq:weighted-LR-inequality},
  the Pieri rule
  $\Ucal_\mu\otimes V\cong\bigoplus_{\rho\in\Y_{b+1,d}}
    \Ucal_\rho\otimes\C^{c_{\mu,(1)}^{\rho}}$ gives, on comparing dimensions,
  \begin{equation}
    \sum_{\rho\in\Y_{b+1,d}}c_{\mu,(1)}^{\rho}D_\rho
    =
    \dim(\Ucal_\mu\otimes V)
    =
    dD_\mu .
    \label{eq:pieri-dim}
  \end{equation}
  Hence
  \begin{equation}
    \begin{aligned}
       & \sum_{\lambda,\rho,\mu}D_\lambda D_\rho
      c_{\lambda,\mu}^{\nu}c_{\mu,(1)}^{\rho}
      \\
       & \quad=
      \sum_{\lambda,\mu}D_\lambda c_{\lambda,\mu}^{\nu}
      \Bigl(\sum_\rho c_{\mu,(1)}^{\rho}D_\rho\Bigr)
      \\
       & \quad=
      d\sum_{\lambda,\mu}c_{\lambda,\mu}^{\nu}D_\lambda D_\mu
      =
      dA_\nu^{(a,b)}.
    \end{aligned}
    \label{eq:weighted-LR-right}
  \end{equation}
  Substituting Eqs.~\eqref{eq:weighted-LR-left} and
  \eqref{eq:weighted-LR-right} into Eq.~\eqref{eq:weighted-LR-inequality} gives
  \begin{equation}
    \sum_{\kappa\in\Y_{a+b+1,d}}c_{\nu,(1)}^{\kappa}A_\kappa^{(a,b+1)}
    \geq
    dA_\nu^{(a,b)}.
    \label{eq:numerator-inequality}
  \end{equation}
  \emph{Step 3.}
  Comparing dimensions in $\Ucal_\nu\otimes V\cong\bigoplus_\kappa
    \Ucal_\kappa\otimes\C^{c_{\nu,(1)}^{\kappa}}$ gives
  \begin{equation}
    \sum_{\kappa\in\Y_{a+b+1,d}}c_{\nu,(1)}^{\kappa}D_\kappa=dD_\nu .
    \label{eq:nu-dim}
  \end{equation}
  For fixed $\nu$, define
  \[
    p_\kappa^{(\nu)}
    :=
    \frac{c_{\nu,(1)}^{\kappa}D_\kappa}{dD_\nu}.
  \]
  By Eq.~\eqref{eq:nu-dim} these are nonnegative and sum to one over
  $\kappa\in\Y_{a+b+1,d}$.  Using Eq.~\eqref{eq:numerator-inequality},
  \[
    \sum_{\kappa}p_\kappa^{(\nu)}\frac{A_\kappa^{(a,b+1)}}{D_\kappa}
    =
    \frac{1}{dD_\nu}\sum_{\kappa}c_{\nu,(1)}^{\kappa}A_\kappa^{(a,b+1)}
    \geq
    \frac{A_\nu^{(a,b)}}{D_\nu}.
  \]
  The left-hand side is a convex combination of the ratios
  $A_\kappa^{(a,b+1)}/D_\kappa$, so it is at most their maximum.
  \[
    \Theta_d(a,b+1)
    =
    \max_{\kappa\in\Y_{a+b+1,d}}\frac{A_\kappa^{(a,b+1)}}{D_\kappa}
    \geq
    \frac{A_\nu^{(a,b)}}{D_\nu}.
  \]
  Choosing $\nu=\nu_\ast$ to attain
  $\Theta_d(a,b)=A_{\nu_\ast}^{(a,b)}/D_{\nu_\ast}$ yields
  $\Theta_d(a,b+1)\geq\Theta_d(a,b)$.  By the symmetry
  Eq.~\eqref{eq:Theta-symmetric}, $\Theta_d(a+1,b)\geq\Theta_d(a,b)$ as well.
\end{proof}

\section{Proof of the saturation formula}
\label{appendix:saturation-proof}

The saturation proof has two complementary parts. The upper bound shows that, for any
outer sector, the total contribution from the $a$-query side cannot exceed the sum of
the squared Schur dimensions. To show that this bound is tight, we choose a rectangular
outer diagram. At $b=(d-1)a$, every $a$-box diagram has a unique complementary diagram
inside this rectangle, with the same representation dimension. The rectangle therefore
collects the full upper-bound weight in a single one-dimensional outer sector.
Operationally, enough queries to the second channel let every Schur sector on the first
channel be paired with a complementary sector of the same weight, so the known-reference
bound is attained.

\begin{proof}
  We prove Eqs.~\eqref{eq:Theta-upper-gamma} and \eqref{eq:Theta-saturation-lower}, and
  then derive Eq.~\eqref{eq:Theta-saturated}, by matching the upper bound with this
  rectangular contribution.

  \emph{Upper bound.}
  Fix $\nu\in\Y_{a+b,d}$ and write
  \[
    A_\nu^{(a,b)}
    :=
    \sum_{\lambda\in\Y_{a,d}}
    \sum_{\mu\in\Y_{b,d}}
    c_{\lambda\mu}^{\nu}D_\lambda D_\mu.
  \]
  It suffices to prove $A_\nu^{(a,b)}\leq D_\nu\gamma_{a,d}$ for every $\nu$.  We first
  make explicit how the Littlewood--Richardson coefficient enters this estimate.
  In this paragraph we use polynomial $\mathrm{U}(d)$ representations.  The coefficient
  $c_{\lambda\mu}^{\nu}$ is the multiplicity with which $\Ucal_\nu$ appears in
  $\Ucal_\lambda\otimes\Ucal_\mu$, equivalently
  \[
    c_{\lambda\mu}^{\nu}
    =
    \dim\Hom_{\mathrm{U}(d)}(\Ucal_\nu,\Ucal_\lambda\otimes\Ucal_\mu).
  \]
  The estimate below requires summing over $\mu$ while keeping $\lambda$ and $\nu$
  fixed.  For this purpose, transfer the factor $\Ucal_\lambda$ by duality.  For
  finite-dimensional representations,
  \[
    \Hom_{\mathrm{U}(d)}(\Ucal_\nu,\Ucal_\lambda\otimes\Ucal_\mu)
    \cong
    \Hom_{\mathrm{U}(d)}(\Ucal_\mu,\Ucal_\lambda^\ast\otimes\Ucal_\nu).
  \]
  Thus the same coefficient can be read in the opposite direction. For fixed
  $\lambda$ and $\nu$, $c_{\lambda\mu}^{\nu}$ is the multiplicity of $\Ucal_\mu$ in the
  fixed representation $\Ucal_\lambda^\ast\otimes\Ucal_\nu$.
  This latter representation need not be polynomial. Its irreducible $\mathrm{U}(d)$
  summands may have highest weights with negative entries.  Therefore
  \[
    \bigoplus_{\mu\in\Y_{b,d}}
    \Ucal_\mu\otimes\C^{c_{\lambda\mu}^{\nu}}
  \]
  should be understood as the polynomial part of the decomposition, not as the whole
  decomposition of $\Ucal_\lambda^\ast\otimes\Ucal_\nu$.
  Since finite-dimensional $\mathrm{U}(d)$ representations are completely reducible,
  the quantity
  \[
    \sum_{\mu\in\Y_{b,d}}c_{\lambda\mu}^{\nu}D_\mu
  \]
  is the total dimension of the direct sum of those irreducible summands
  $\Ucal_\mu$ with $\mu\in\Y_{b,d}$, counted with multiplicity, inside
  $\Ucal_\lambda^\ast\otimes\Ucal_\nu$.  This subrepresentation is contained in the full
  irreducible decomposition, so its dimension cannot exceed the full dimension of
  $\Ucal_\lambda^\ast\otimes\Ucal_\nu$, namely $D_\lambda D_\nu$.
  Therefore
  \begin{align}
    A_\nu^{(a,b)}
     & =
    \sum_{\lambda\in\Y_{a,d}}
    D_\lambda
    \sum_{\mu\in\Y_{b,d}}c_{\lambda\mu}^{\nu}D_\mu
    \nonumber \\
     & \leq
    \sum_{\lambda\in\Y_{a,d}}D_\lambda(D_\lambda D_\nu)
    =
    D_\nu\sum_{\lambda\in\Y_{a,d}}D_\lambda^2
    =
    D_\nu\gamma_{a,d}.
    \label{eq:Anu-gamma-bound}
  \end{align}
  Dividing by $D_\nu$ and maximizing over $\nu$ proves
  Eq.~\eqref{eq:Theta-upper-gamma}.
  The binomial identity Eq.~\eqref{eq:gamma-binomial} follows from the Schur decomposition
  of the symmetric subspace of $\C^d\otimes\C^d$.
  \[
    \operatorname{Sym}^a(\C^d\otimes\C^d)
    \cong
    \bigoplus_{\lambda\in\Y_{a,d}}\Ucal_\lambda\otimes\Ucal_\lambda,
  \]
  whose right-hand side has dimension $\sum_{\lambda\in\Y_{a,d}}D_\lambda^2$.  Since
  $\C^d\otimes\C^d$ has dimension $d^2$, the left-hand side has dimension
  $\binom{a+d^2-1}{a}$.

  \emph{Lower bound at the saturated point.}
  We show that the upper bound is sharp when $b=(d-1)a$.  The goal is to
  find one term in the maximum defining $\Theta_d(a,(d-1)a)$ whose numerator already
  contains the full sum $\sum_{\lambda\in\Y_{a,d}}D_\lambda^2$.  For this purpose, take
  \[
    \nu_{\mathrm{rect}}:=(\underbrace{a,\ldots,a}_{d\ \mathrm{entries}}),
  \]
  the $d\times a$ rectangle.  This diagram has $da$ boxes, as required when
  $b=(d-1)a$.
  The argument uses four facts about this rectangle.  First, for each
  $\lambda\in\Y_{a,d}$, the complement of $\lambda$ inside the rectangle is a diagram
  with $(d-1)a$ boxes.  Second, this complement has the same representation dimension as
  $\lambda$.  Third, the pair consisting of $\lambda$ and its complement fills the
  rectangle with Littlewood--Richardson multiplicity one.  Fourth, the rectangular
  diagram itself has dimension one.

  For $\lambda=(\lambda_1,\ldots,\lambda_d)\in\Y_{a,d}$, with trailing zeros added if
  necessary, define the complementary diagram
  \[
    \lambda^c:=(a-\lambda_d,\ldots,a-\lambda_1).
  \]
  This is the Young diagram obtained by taking the boxes of the rectangle not occupied
  by $\lambda$ and rotating that complementary skew shape by $180^\circ$.  Its number of
  boxes is, using $|\lambda|=a$,
  \[
    |\lambda^c|
    =
    \sum_{i=1}^d(a-\lambda_i)
    =
    da-a
    =
    (d-1)a,
  \]
  so $\lambda^c\in\Y_{(d-1)a,d}$.

  The second fact, $D_{\lambda^c}=D_\lambda$, follows directly from Weyl's dimension formula,
  Eq.~\eqref{eq:weyl-dimension-formula}.  Put
  $\lambda_i^c=a-\lambda_{d+1-i}$.  For $1\leq i<j\leq d$,
  \[
    \lambda_i^c-\lambda_j^c+j-i
    =
    \lambda_{d+1-j}-\lambda_{d+1-i}+j-i.
  \]
  If $p=d+1-j$ and $q=d+1-i$, then $p<q$ and $q-p=j-i$, so this factor becomes
  $\lambda_p-\lambda_q+q-p$.  As $(i,j)$ ranges over all pairs with $i<j$, the pair
  $(p,q)$ also ranges once over all pairs with $p<q$.  Therefore the Weyl products for
  $D_{\lambda^c}$ and $D_\lambda$ are identical.

  The third fact is the rectangle-complement case of the Littlewood--Richardson rule
  (see Appendix~\ref{appendix:theta-computation} for the tableau formulation).
  \[
    c_{\lambda,\lambda^c}^{\nu_{\mathrm{rect}}}=1.
  \]
  Equivalently, as $\mathrm{U}(d)$ representations one has
  \[
    \Ucal_{\lambda^c}\cong \det^a\otimes\Ucal_\lambda^\ast,
    \qquad
    \Ucal_{\nu_{\mathrm{rect}}}\cong\det^a.
  \]
  Hence
  \[
    \begin{aligned}
      c_{\lambda,\lambda^c}^{\nu_{\mathrm{rect}}}
       & =
      \dim\Hom_{\mathrm{U}(d)}
      (\det^a,\Ucal_\lambda\otimes\det^a\otimes\Ucal_\lambda^\ast) \\
       & =
      \dim\operatorname{End}_{\mathrm{U}(d)}(\Ucal_\lambda)
      =
      1
    \end{aligned}
  \]
  by Schur's lemma.
  After removing $\lambda$ from the rectangle $\nu_{\mathrm{rect}}$, the remaining skew
  shape is filled by the complementary diagram $\lambda^c$ in exactly one
  Littlewood--Richardson-compatible way.

  Finally, Weyl's dimension formula gives $D_{\nu_{\mathrm{rect}}}=1$ directly. Since
  $\nu_{\mathrm{rect}}=(a,\ldots,a)$,
  \[
    D_{\nu_{\mathrm{rect}}}
    =
    \prod_{1\leq i<j\leq d}
    \frac{a-a+j-i}{j-i}
    =
    1.
  \]
  We apply the defining formula for $\Theta_d(a,(d-1)a)$ in
  Eq.~\eqref{eq:Theta-definition}.  Since $\nu_{\mathrm{rect}}\in\Y_{da,d}$ and
  $D_{\nu_{\mathrm{rect}}}=1$, evaluating the maximum at this particular outer diagram
  gives the first lower bound.  Nonnegativity of the summands then allows us to retain
  only the terms with $\mu=\lambda^c$.
  \begin{align}
    \Theta_d(a,(d-1)a)
     & =
    \max_{\nu\in\Y_{da,d}}
    \frac{
      \displaystyle
      \sum_{\lambda\in\Y_{a,d}}
      \sum_{\mu\in\Y_{(d-1)a,d}}
      c_{\lambda\mu}^{\nu}D_\lambda D_\mu
    }{D_\nu}
    \nonumber \\
     & \geq
    \frac{
      \displaystyle
      \sum_{\lambda\in\Y_{a,d}}
      \sum_{\mu\in\Y_{(d-1)a,d}}
      c_{\lambda\mu}^{\nu_{\mathrm{rect}}}D_\lambda D_\mu
    }{D_{\nu_{\mathrm{rect}}}}
    \nonumber \\
     & =
    \sum_{\lambda\in\Y_{a,d}}
    \sum_{\mu\in\Y_{(d-1)a,d}}
    c_{\lambda\mu}^{\nu_{\mathrm{rect}}}D_\lambda D_\mu
    \nonumber \\
     & \geq
    \sum_{\lambda\in\Y_{a,d}}
    c_{\lambda,\lambda^c}^{\nu_{\mathrm{rect}}}
    D_\lambda D_{\lambda^c}
    \nonumber \\
     & =
    \sum_{\lambda\in\Y_{a,d}}D_\lambda D_{\lambda^c}
    \nonumber \\
     & =
    \sum_{\lambda\in\Y_{a,d}}D_\lambda^2
    =
    \gamma_{a,d}.
    \label{eq:rectangle-lower-bound}
  \end{align}

  \emph{Extension by monotonicity.}
  Assume $N_2\geq(d-1)N_1$.
  Monotonicity of $\Theta_d$ in its second argument
  (Proposition~\ref{prop:monotonicity}) gives
  \[
    \Theta_d(N_1,N_2)
    \geq
    \Theta_d(N_1,(d-1)N_1)
    \geq
    \gamma_{N_1,d},
  \]
  where the last step is Eq.~\eqref{eq:Theta-saturation-lower}.  Conversely,
  Eq.~\eqref{eq:Theta-upper-gamma} gives $\Theta_d(N_1,N_2)\leq\gamma_{N_1,d}$.
  The two inequalities combine to give Eq.~\eqref{eq:Theta-saturated}.
\end{proof}

\section{Optimal minimum-error comparison of pure states}
\label{appendix:state-comparison}

This appendix derives Eq.~\eqref{eq:state-comparison-main}.  Consider the comparison of
two unknown pure states with $N_1$ copies of the first state and $N_2$ copies of the
second state.  Under the identical-state hypothesis, a single Haar-random state
$|\psi\rangle$ is prepared on all copies.  Under the independent-state hypothesis,
$|\psi_1\rangle$ and $|\psi_2\rangle$ are sampled independently from the Haar measure.
The corresponding averaged states are
\begin{align}
  \rho_=
   & :=
  \int d\psi\,
  \bigl(|\psi\rangle\langle\psi|\bigr)^{\otimes(N_1+N_2)},
  \label{eq:state-rho-same} \\
  \rho_{\necase}
   & :=
  \int d\psi_1d\psi_2\,
  \bigl(|\psi_1\rangle\langle\psi_1|\bigr)^{\otimes N_1}
  \otimes
  \bigl(|\psi_2\rangle\langle\psi_2|\bigr)^{\otimes N_2}.
  \label{eq:state-rho-different}
\end{align}
Thus the comparison task is the binary discrimination of $\rho_=$ and $\rho_{\necase}$
with prior probabilities $p$ and $1-p$ \cite{watrousTheoryQuantumInformation2018}.

Let $P^{\mathrm{sym}}_{N,d}$ be the projector onto the symmetric subspace of
$(\C^d)^{\otimes N}$ and let
$d^{\mathrm{sym}}_{N,d}=\Tr P^{\mathrm{sym}}_{N,d}=\binom{N+d-1}{d-1}$.  The standard
Haar integral identity gives
\begin{align}
  \rho_=
   & =
  \frac{P^{\mathrm{sym}}_{N_1+N_2,d}}
  {d^{\mathrm{sym}}_{N_1+N_2,d}},
  \label{eq:state-rho-same-sym} \\
  \rho_{\necase}
   & =
  \frac{P^{\mathrm{sym}}_{N_1,d}}{d^{\mathrm{sym}}_{N_1,d}}
  \otimes
  \frac{P^{\mathrm{sym}}_{N_2,d}}{d^{\mathrm{sym}}_{N_2,d}}.
  \label{eq:state-rho-different-sym}
\end{align}
Because the fully symmetric subspace is contained in the tensor product of the two
blockwise symmetric subspaces,
\[
  P^{\mathrm{sym}}_{N_1+N_2,d}
  \leq
  P^{\mathrm{sym}}_{N_1,d}\otimes P^{\mathrm{sym}}_{N_2,d}.
\]
Equations~\eqref{eq:state-rho-same-sym} and \eqref{eq:state-rho-different-sym} imply
the relative bound
\begin{equation}
  \rho_=
  \leq
  \beta_{N_1,N_2,d}\rho_{\necase},
  \qquad
  \beta_{N_1,N_2,d}
  =
  \frac{d^{\mathrm{sym}}_{N_1,d}d^{\mathrm{sym}}_{N_2,d}}
  {d^{\mathrm{sym}}_{N_1+N_2,d}}.
  \label{eq:state-relative-bound}
\end{equation}

Let $\{\Pi_{=},\Pi_{\necase}\}$ be a POVM.  The average success probability is
\[
  P_{\mathrm{state},\ME}
  =
  \Tr\!\left[p\rho_=\Pi_{=}+(1-p)\rho_{\necase}\Pi_{\necase}\right].
\]
Writing $\Pi_{\necase}=I-\Pi_{=}$, this can be rewritten as
\begin{equation}
  P_{\mathrm{state},\ME}
  =
  1-p+
  \Tr\!\left[\left(p\rho_=-(1-p)\rho_{\necase}\right)\Pi_{=}\right].
  \label{eq:state-success-rewrite}
\end{equation}

If $0\leq p\leq1/(1+\beta_{N_1,N_2,d})$, then Eq.~\eqref{eq:state-relative-bound}
implies
\[
  p\rho_=
  \leq
  p\beta_{N_1,N_2,d}\rho_{\necase}
  \leq
  (1-p)\rho_{\necase}.
\]
The second term in Eq.~\eqref{eq:state-success-rewrite} is therefore nonpositive for
every POVM, so $P_{\mathrm{state},\ME}\leq1-p$.  This upper bound is attained by the
trivial POVM
\[
  \Pi_{=}=0,
  \qquad
  \Pi_{\necase}=I.
\]

Now suppose that $p>1/(1+\beta_{N_1,N_2,d})$.  For an arbitrary POVM, set
\[
  q:=\Tr(\rho_=\Pi_{=}),
  \qquad
  r:=\Tr(\rho_{\necase}\Pi_{=}).
\]
The relative bound Eq.~\eqref{eq:state-relative-bound} gives $q\leq
  \beta_{N_1,N_2,d}r$, or $r\geq q/\beta_{N_1,N_2,d}$.  Hence
\begin{align}
  P_{\mathrm{state},\ME}
   & =
  pq+(1-p)(1-r)
  \nonumber \\
   & \leq
  1-p+
  \left(p-\frac{1-p}{\beta_{N_1,N_2,d}}\right)q
  \nonumber \\
   & \leq
  1-\frac{1-p}{\beta_{N_1,N_2,d}},
  \label{eq:state-large-p-upper}
\end{align}
where the last inequality uses $q\leq1$ and
$p-(1-p)/\beta_{N_1,N_2,d}>0$.  The upper bound is attained by
\[
  \Pi_{=}=P^{\mathrm{sym}}_{N_1+N_2,d},
  \qquad
  \Pi_{\necase}=\Id-P^{\mathrm{sym}}_{N_1+N_2,d}.
\]
Indeed, this POVM gives $q=1$ and
\[
  r
  =
  \Tr\!\left(\rho_{\necase}P^{\mathrm{sym}}_{N_1+N_2,d}\right)
  =
  \frac{1}{\beta_{N_1,N_2,d}},
\]
because $P^{\mathrm{sym}}_{N_1+N_2,d}$ is contained in the support of
$P^{\mathrm{sym}}_{N_1,d}\otimes P^{\mathrm{sym}}_{N_2,d}$.  Combining the two regimes
proves Eq.~\eqref{eq:state-comparison-main}.

\section{Proof of the known-reference identity test}
\label{appendix:known-reference-proof}

\begin{proof}
  \emph{Upper bound.}
  Let
  \[
    M_I^{(N)}:=C_I^{\otimes N},
    \qquad
    M_U^{(N)}:=\int_{\SU(d)}dU\,C_U^{\otimes N}.
  \]
  We use the Schur--Weyl decompositions
  \[
    \Hcal_N^K
    \cong
    \bigoplus_{\lambda\in\Y_{N,d}}
    \overline{\Ucal_\lambda}^{\,K}\otimes\Vcal_\lambda^K,
    \qquad
    \Hcal_N^H
    \cong
    \bigoplus_{\lambda\in\Y_{N,d}}
    \Ucal_\lambda^H\otimes\Vcal_\lambda^H.
  \]
  As in Eq.~\eqref{eq:single-twirl},
  \begin{equation}
    M_U^{(N)}
    =
    \sum_{\lambda\in\Y_{N,d}}
    \frac{\Id_{\Ucal_\lambda^K}\otimes\Id_{\Ucal_\lambda^H}}{D_\lambda}
    \otimes
    |\Phi_{\Vcal_\lambda^K,\Vcal_\lambda^H}\rangle
    \langle\Phi_{\Vcal_\lambda^K,\Vcal_\lambda^H}|.
    \label{eq:known-reference-Haar-block}
  \end{equation}
  On the other hand,
  \begin{align}
    M_I^{(N)}
     & =
    |\Phi\rangle^{\otimes N}\langle\Phi|^{\otimes N},
    \nonumber       \\
    |\Phi\rangle^{\otimes N}
     & =
    \sum_{\lambda\in\Y_{N,d}}
    |\Phi_{\Ucal_\lambda^K,\Ucal_\lambda^H}\rangle
    \nonumber       \\
     & \quad\otimes
    |\Phi_{\Vcal_\lambda^K,\Vcal_\lambda^H}\rangle.
    \label{eq:known-reference-identity-block}
  \end{align}
  We now prove the operator inequality
  $M_I^{(N)}\leq\gamma_{N,d}M_U^{(N)}$ by comparing quadratic forms.  It is enough to
  show
  $\langle\xi|M_I^{(N)}|\xi\rangle
    \leq
    \gamma_{N,d}\langle\xi|M_U^{(N)}|\xi\rangle$
  for every vector $\xi$.  The full Schur--Weyl decomposition of
  $\Hcal_N^K\otimes\Hcal_N^H$ has $(\lambda,\mu)$ sectors.  However, both
  $M_I^{(N)}$ and $M_U^{(N)}$ are supported on the diagonal subspace
  \[
    \mathcal D_N
    :=
    \bigoplus_{\lambda\in\Y_{N,d}}
    \overline{\Ucal_\lambda}^{\,K}
    \otimes\Vcal_\lambda^K
    \otimes
    \Ucal_\lambda^H
    \otimes\Vcal_\lambda^H .
  \]
  Indeed, $|\Phi\rangle^{\otimes N}\in\mathcal D_N$ in
  Eq.~\eqref{eq:known-reference-identity-block}, and
  Eq.~\eqref{eq:known-reference-Haar-block} also has only diagonal $\lambda$ blocks.
  Hence the off-diagonal components with $\lambda\neq\mu$ do not contribute to either
  quadratic form.  It is therefore enough to consider $\xi\in\mathcal D_N$ and write it as
  $\xi=\bigoplus_{\lambda\in\Y_{N,d}}\xi_\lambda$, where $\xi_\lambda$ is its component
  in
  \[
    \overline{\Ucal_\lambda}^{\,K}
    \otimes\Vcal_\lambda^K
    \otimes
    \Ucal_\lambda^H
    \otimes\Vcal_\lambda^H .
  \]
  For each sector, define
  \[
    y_\lambda
    :=
    \left(
    \Id_{\Ucal_\lambda^K\otimes\Ucal_\lambda^H}
    \otimes
    \langle\Phi_{\Vcal_\lambda^K,\Vcal_\lambda^H}|
    \right)\xi_\lambda .
  \]
  Then
  \[
    y_\lambda\in
    \overline{\Ucal_\lambda}^{\,K}\otimes\Ucal_\lambda^H .
  \]
  Thus $y_\lambda$ is obtained from the $\lambda$ block of $\xi$ by contracting only the
  multiplicity spaces $\Vcal_\lambda^K\otimes\Vcal_\lambda^H$ with the maximally
  entangled vector appearing in Eqs.~\eqref{eq:known-reference-Haar-block} and
  \eqref{eq:known-reference-identity-block}.  Applying the block form of
  $M_U^{(N)}$ gives
  \[
    \langle\xi|M_U^{(N)}|\xi\rangle
    =
    \sum_{\lambda\in\Y_{N,d}}\frac{\|y_\lambda\|^2}{D_\lambda}.
  \]
  Set
  \[
    \gamma_{N,d}:=\sum_{\lambda\in\Y_{N,d}}D_\lambda^2 .
  \]
  The identity block has the same multiplicity contraction, but also contracts
  $y_\lambda$ with the representation-space maximally entangled vector.
  \begin{align}
    \langle\xi|M_I^{(N)}|\xi\rangle
     & =
    \bigl|\langle\Phi|^{\otimes N}|\xi\rangle\bigr|^2
    \nonumber \\
     & =
    \left|
    \sum_{\lambda\in\Y_{N,d}}
    \langle\Phi_{\Ucal_\lambda^K,\Ucal_\lambda^H}|y_\lambda\rangle
    \right|^2
    \nonumber \\
     & =
    \left|
    \sum_{\lambda\in\Y_{N,d}}
    D_\lambda
    \frac{
      \langle\Phi_{\Ucal_\lambda^K,\Ucal_\lambda^H}|y_\lambda\rangle
    }{D_\lambda}
    \right|^2
    \nonumber \\
     & \leq
    \left(\sum_{\lambda\in\Y_{N,d}}D_\lambda^2\right)
    \left(
    \sum_{\lambda\in\Y_{N,d}}
    \frac{
      |\langle\Phi_{\Ucal_\lambda^K,\Ucal_\lambda^H}|y_\lambda\rangle|^2
    }{D_\lambda^2}
    \right)
    \nonumber \\
     & \leq
    \left(\sum_{\lambda\in\Y_{N,d}}D_\lambda^2\right)
    \left(
    \sum_{\lambda\in\Y_{N,d}}
    \frac{\|y_\lambda\|^2}{D_\lambda}
    \right)
    \nonumber \\
     & =
    \gamma_{N,d}\langle\xi|M_U^{(N)}|\xi\rangle .
    \label{eq:known-reference-CS}
  \end{align}
  In the first inequality we used Cauchy--Schwarz.  In the second, we used
  $\||\Phi_{\Ucal_\lambda^K,\Ucal_\lambda^H}\rangle\|^2=D_\lambda$, so
  $|\langle\Phi_{\Ucal_\lambda^K,\Ucal_\lambda^H}|y_\lambda\rangle|^2
    \leq D_\lambda\|y_\lambda\|^2$.
  Therefore
  \begin{equation}
    M_I^{(N)}
    \leq
    \gamma_{N,d}M_U^{(N)}.
    \label{eq:known-reference-relative-proof}
  \end{equation}

  Let $\Pi_I$ be the tester outcome corresponding to the decision ``identity,'' and set
  \[
    \alpha:=\Tr(\Pi_I M_I^{(N)}),
    \qquad
    \beta:=\Tr(\Pi_I M_U^{(N)}).
  \]
  By Eq.~\eqref{eq:known-reference-relative-proof}, $\alpha\leq\gamma_{N,d}\beta$.
  Deterministic normalization of the tester gives $\alpha,\beta\leq1$. Hence the
  minimum-error success probability satisfies
  \begin{align}
    P_{\ME}
     & =
    p\alpha+(1-p)(1-\beta)
    \nonumber \\
     & =
    1-p+p\alpha-(1-p)\beta.
    \label{eq:known-reference-ME-converse-proof}
  \end{align}
  If $p\leq1/(1+\gamma_{N,d})$, then
  $p\alpha-(1-p)\beta\leq(p\gamma_{N,d}-(1-p))\beta\leq0$, so
  $P_{\ME}\leq1-p$. If $p>1/(1+\gamma_{N,d})$, then
  $\beta\geq\alpha/\gamma_{N,d}$ implies
  \[
    P_{\ME}
    \leq
    1-p+
    \left(p-\frac{1-p}{\gamma_{N,d}}\right)\alpha
    \leq
    1-\frac{1-p}{\gamma_{N,d}}.
  \]
  This proves the upper-bound part of Eq.~\eqref{eq:known-reference-ME}.

  \emph{Achievability.}
  We show that the bounds are tight.  Let $S_N^K$ be the input-side copy of the operator
  $S_N^R$ in Eq.~\eqref{eq:known-reference-S}.  It is positive semidefinite and
  \begin{equation}
    \Tr S_N^K
    =
    \sum_{\lambda\in\Y_{N,d}}q_\lambda
    \frac{\Tr\Id_{\overline{\Ucal_\lambda}^{\,K}}}{D_\lambda}
    \frac{\Tr\Id_{\Vcal_\lambda^K}}{\dim\Vcal_\lambda}
    =
    \sum_{\lambda\in\Y_{N,d}}q_\lambda
    =
    1.
    \label{eq:known-reference-S-trace}
  \end{equation}
  For the nontrivial branch, use the parallel tester whose identity-acceptance tester
  element is
  \begin{equation}
    \Pi_I
    :=
    \left(S_N^K\otimes\Id^H\right)
    |\Phi\rangle^{\otimes N}\langle\Phi|^{\otimes N}
    \left(S_N^K\otimes\Id^H\right),
    \label{eq:known-reference-achieving-Pi}
  \end{equation}
  and set $\Pi_U=S_N^K\otimes\Id^H-\Pi_I$.  Then
  \begin{equation}
    \Pi_I+\Pi_U=S_N^K\otimes\Id^H,
    \qquad
    \Tr S_N^K=1,
    \label{eq:known-reference-parallel-normalization}
  \end{equation}
  which is exactly the parallel-tester normalization in
  Eqs.~\eqref{con1-para} and \eqref{con2-para}.  Moreover, if
  $|\widetilde\eta_N\rangle:=(\sqrt{S_N^K}\otimes\Id^H)|\Phi\rangle^{\otimes N}$, then
  $\langle\widetilde\eta_N|\widetilde\eta_N\rangle=\Tr S_N^K=1$ and
  \[
    \Pi_U
    =
    (\sqrt{S_N^K}\otimes\Id^H)
    (\Id-|\widetilde\eta_N\rangle\langle\widetilde\eta_N|)
    (\sqrt{S_N^K}\otimes\Id^H)
    \geq0.
  \]
  Hence this is a valid parallel tester. In circuit language, this is the same
  construction as the parallel input $|\eta_N\rangle$ in Eq.~\eqref{eq:eta-from-S},
  with the reference system identified with the input-side copy.
  Directly evaluating the two relevant traces gives
  \begin{align}
    \Tr(\Pi_I M_I^{(N)})
     & =
    \left(
    \langle\Phi|^{\otimes N}
    \bigl(S_N^K\otimes\Id^H\bigr)
    |\Phi\rangle^{\otimes N}
    \right)^2
    \nonumber \\
     & =
    (\Tr S_N^K)^2=1,
    \label{eq:known-reference-direct-identity} \\
    \Tr(\Pi_I M_U^{(N)})
     & =
    \sum_{\lambda\in\Y_{N,d}}\frac{q_\lambda^2}{D_\lambda^2}
    =
    \frac{1}{\gamma_{N,d}^2}
    \sum_{\lambda\in\Y_{N,d}}D_\lambda^2
    \nonumber \\
     & =
    \frac{1}{\gamma_{N,d}},
    \label{eq:known-reference-direct-haar}
  \end{align}
  In the second line we used the block form of $M_U^{(N)}$ in
  Eq.~\eqref{eq:known-reference-Haar-block} and
  $q_\lambda=D_\lambda^2/\gamma_{N,d}$.  Hence the tester attains
  $P_{\mathrm{ref},\ME}=p+(1-p)(1-1/\gamma_{N,d})=1-(1-p)/\gamma_{N,d}$, i.e., the
  second branch of Eq.~\eqref{eq:known-reference-ME}. The first branch is attained by
  always answering ``random unitary.''

  \emph{One-sided unambiguous case.}
  For the one-sided unambiguous task, the conclusive ``random unitary'' outcome must
  have zero probability under the identity hypothesis. Since the relative inequality
  implies that a conclusive ``identity'' outcome cannot occur without also occurring
  under the Haar-random hypothesis, only the conclusive ``random unitary'' outcome is
  useful. The same argument as in Theorem~\ref{thm:unambiguous} gives the lower bound
  $P_{\mathrm{ref},?}\geq p+(1-p)/\gamma_{N,d}$. The parallel tester above attains it by declaring
  ``random unitary'' on the complement of the identity-acceptance projector and declaring
  the identity-acceptance outcome inconclusive.
\end{proof}

\bibliography{unitarycomparison}

\begin{thebibliography}{57}%
\makeatletter
\providecommand \@ifxundefined [1]{%
 \@ifx{#1\undefined}
}%
\providecommand \@ifnum [1]{%
 \ifnum #1\expandafter \@firstoftwo
 \else \expandafter \@secondoftwo
 \fi
}%
\providecommand \@ifx [1]{%
 \ifx #1\expandafter \@firstoftwo
 \else \expandafter \@secondoftwo
 \fi
}%
\providecommand \natexlab [1]{#1}%
\providecommand \enquote  [1]{``#1''}%
\providecommand \bibnamefont  [1]{#1}%
\providecommand \bibfnamefont [1]{#1}%
\providecommand \citenamefont [1]{#1}%
\providecommand \href@noop [0]{\@secondoftwo}%
\providecommand \href [0]{\begingroup \@sanitize@url \@href}%
\providecommand \@href[1]{\@@startlink{#1}\@@href}%
\providecommand \@@href[1]{\endgroup#1\@@endlink}%
\providecommand \@sanitize@url [0]{\catcode `\\12\catcode `\$12\catcode `\&12\catcode `\#12\catcode `\^12\catcode `\_12\catcode `\%12\relax}%
\providecommand \@@startlink[1]{}%
\providecommand \@@endlink[0]{}%
\providecommand \url  [0]{\begingroup\@sanitize@url \@url }%
\providecommand \@url [1]{\endgroup\@href {#1}{\urlprefix }}%
\providecommand \urlprefix  [0]{URL }%
\providecommand \Eprint [0]{\href }%
\providecommand \doibase [0]{https://doi.org/}%
\providecommand \selectlanguage [0]{\@gobble}%
\providecommand \bibinfo  [0]{\@secondoftwo}%
\providecommand \bibfield  [0]{\@secondoftwo}%
\providecommand \translation [1]{[#1]}%
\providecommand \BibitemOpen [0]{}%
\providecommand \bibitemStop [0]{}%
\providecommand \bibitemNoStop [0]{.\EOS\space}%
\providecommand \EOS [0]{\spacefactor3000\relax}%
\providecommand \BibitemShut  [1]{\csname bibitem#1\endcsname}%
\let\auto@bib@innerbib\@empty
%</preamble>
\bibitem [{\citenamefont {Holevo}(1973)}]{holevoStatisticalDecisionTheory1973}%
  \BibitemOpen
  \bibfield  {author} {\bibinfo {author} {\bibfnamefont {A.~S.}\ \bibnamefont {Holevo}},\ }\bibfield  {title} {\bibinfo {title} {Statistical decision theory for quantum systems},\ }\href {https://doi.org/10.1016/0047-259X(73)90028-6} {\bibfield  {journal} {\bibinfo  {journal} {Journal of Multivariate Analysis}\ }\textbf {\bibinfo {volume} {3}},\ \bibinfo {pages} {337} (\bibinfo {year} {1973})}\BibitemShut {NoStop}%
\bibitem [{\citenamefont {Helstrom}(1969)}]{helstromQuantumDetectionEstimation1969}%
  \BibitemOpen
  \bibfield  {author} {\bibinfo {author} {\bibfnamefont {C.~W.}\ \bibnamefont {Helstrom}},\ }\bibfield  {title} {\bibinfo {title} {Quantum detection and estimation theory},\ }\href {https://doi.org/10.1007/BF01007479} {\bibfield  {journal} {\bibinfo  {journal} {Journal of Statistical Physics}\ }\textbf {\bibinfo {volume} {1}},\ \bibinfo {pages} {231} (\bibinfo {year} {1969})}\BibitemShut {NoStop}%
\bibitem [{\citenamefont {Yuen}\ \emph {et~al.}(1975)\citenamefont {Yuen}, \citenamefont {Kennedy},\ and\ \citenamefont {Lax}}]{yuenOptimumTestingMultiple1975}%
  \BibitemOpen
  \bibfield  {author} {\bibinfo {author} {\bibfnamefont {H.}~\bibnamefont {Yuen}}, \bibinfo {author} {\bibfnamefont {R.}~\bibnamefont {Kennedy}},\ and\ \bibinfo {author} {\bibfnamefont {M.}~\bibnamefont {Lax}},\ }\bibfield  {title} {\bibinfo {title} {Optimum testing of multiple hypotheses in quantum detection theory},\ }\href {https://doi.org/10.1109/TIT.1975.1055351} {\bibfield  {journal} {\bibinfo  {journal} {IEEE Transactions on Information Theory}\ }\textbf {\bibinfo {volume} {21}},\ \bibinfo {pages} {125} (\bibinfo {year} {1975})}\BibitemShut {NoStop}%
\bibitem [{\citenamefont {Ac{\'i}n}(2001)}]{acinStatisticalDistinguishabilityUnitary2001}%
  \BibitemOpen
  \bibfield  {author} {\bibinfo {author} {\bibfnamefont {A.}~\bibnamefont {Ac{\'i}n}},\ }\bibfield  {title} {\bibinfo {title} {Statistical {{Distinguishability}} between {{Unitary Operations}}},\ }\href {https://doi.org/10.1103/PhysRevLett.87.177901} {\bibfield  {journal} {\bibinfo  {journal} {Phys. Rev. Lett.}\ }\textbf {\bibinfo {volume} {87}},\ \bibinfo {pages} {177901} (\bibinfo {year} {2001})}\BibitemShut {NoStop}%
\bibitem [{\citenamefont {Sacchi}(2005)}]{sacchiOptimalDiscriminationQuantum2005}%
  \BibitemOpen
  \bibfield  {author} {\bibinfo {author} {\bibfnamefont {M.~F.}\ \bibnamefont {Sacchi}},\ }\bibfield  {title} {\bibinfo {title} {Optimal discrimination of quantum operations},\ }\href {https://doi.org/10.1103/PhysRevA.71.062340} {\bibfield  {journal} {\bibinfo  {journal} {Phys. Rev. A}\ }\textbf {\bibinfo {volume} {71}},\ \bibinfo {pages} {062340} (\bibinfo {year} {2005})}\BibitemShut {NoStop}%
\bibitem [{\citenamefont {Childs}\ \emph {et~al.}(2000)\citenamefont {Childs}, \citenamefont {Preskill},\ and\ \citenamefont {Renes}}]{childsQuantumInformationPrecision2000}%
  \BibitemOpen
  \bibfield  {author} {\bibinfo {author} {\bibfnamefont {A.~M.}\ \bibnamefont {Childs}}, \bibinfo {author} {\bibfnamefont {J.}~\bibnamefont {Preskill}},\ and\ \bibinfo {author} {\bibfnamefont {J.}~\bibnamefont {Renes}},\ }\bibfield  {title} {\bibinfo {title} {Quantum information and precision measurement},\ }\href {https://doi.org/10.1080/09500340008244034} {\bibfield  {journal} {\bibinfo  {journal} {Journal of Modern Optics}\ }\textbf {\bibinfo {volume} {47}},\ \bibinfo {pages} {155} (\bibinfo {year} {2000})},\ \Eprint {https://arxiv.org/abs/quant-ph/9904021} {arXiv:quant-ph/9904021} \BibitemShut {NoStop}%
\bibitem [{\citenamefont {Pirandola}\ \emph {et~al.}(2019)\citenamefont {Pirandola}, \citenamefont {Laurenza}, \citenamefont {Lupo},\ and\ \citenamefont {Pereira}}]{pirandolaFundamentalLimitsQuantum2019}%
  \BibitemOpen
  \bibfield  {author} {\bibinfo {author} {\bibfnamefont {S.}~\bibnamefont {Pirandola}}, \bibinfo {author} {\bibfnamefont {R.}~\bibnamefont {Laurenza}}, \bibinfo {author} {\bibfnamefont {C.}~\bibnamefont {Lupo}},\ and\ \bibinfo {author} {\bibfnamefont {J.~L.}\ \bibnamefont {Pereira}},\ }\bibfield  {title} {\bibinfo {title} {Fundamental limits to quantum channel discrimination},\ }\href {https://doi.org/10.1038/s41534-019-0162-y} {\bibfield  {journal} {\bibinfo  {journal} {npj Quantum Inf}\ }\textbf {\bibinfo {volume} {5}},\ \bibinfo {pages} {50} (\bibinfo {year} {2019})}\BibitemShut {NoStop}%
\bibitem [{\citenamefont {Buhrman}\ \emph {et~al.}(2001)\citenamefont {Buhrman}, \citenamefont {Cleve}, \citenamefont {Watrous},\ and\ \citenamefont {de~Wolf}}]{buhrmanQuantumFingerprinting2001}%
  \BibitemOpen
  \bibfield  {author} {\bibinfo {author} {\bibfnamefont {H.}~\bibnamefont {Buhrman}}, \bibinfo {author} {\bibfnamefont {R.}~\bibnamefont {Cleve}}, \bibinfo {author} {\bibfnamefont {J.}~\bibnamefont {Watrous}},\ and\ \bibinfo {author} {\bibfnamefont {R.}~\bibnamefont {de~Wolf}},\ }\bibfield  {title} {\bibinfo {title} {Quantum {{Fingerprinting}}},\ }\href {https://doi.org/10.1103/PhysRevLett.87.167902} {\bibfield  {journal} {\bibinfo  {journal} {Phys. Rev. Lett.}\ }\textbf {\bibinfo {volume} {87}},\ \bibinfo {pages} {167902} (\bibinfo {year} {2001})},\ \Eprint {https://arxiv.org/abs/quant-ph/0102001} {arXiv:quant-ph/0102001} \BibitemShut {NoStop}%
\bibitem [{\citenamefont {Gottesman}\ and\ \citenamefont {Chuang}(2001)}]{gottesmanQuantumDigitalSignatures2001}%
  \BibitemOpen
  \bibfield  {author} {\bibinfo {author} {\bibfnamefont {D.}~\bibnamefont {Gottesman}}\ and\ \bibinfo {author} {\bibfnamefont {I.}~\bibnamefont {Chuang}},\ }\href@noop {} {\bibinfo {title} {Quantum {{Digital Signatures}}}} (\bibinfo {year} {2001}),\ \Eprint {https://arxiv.org/abs/quant-ph/0105032} {arXiv:quant-ph/0105032} \BibitemShut {NoStop}%
\bibitem [{\citenamefont {Gour}(2019)}]{gourComparisonQuantumChannels2019}%
  \BibitemOpen
  \bibfield  {author} {\bibinfo {author} {\bibfnamefont {G.}~\bibnamefont {Gour}},\ }\bibfield  {title} {\bibinfo {title} {Comparison of {{Quantum Channels}} by {{Superchannels}}},\ }\href {https://doi.org/10.1109/TIT.2019.2907989} {\bibfield  {journal} {\bibinfo  {journal} {IEEE Trans. Inform. Theory}\ }\textbf {\bibinfo {volume} {65}},\ \bibinfo {pages} {5880} (\bibinfo {year} {2019})},\ \Eprint {https://arxiv.org/abs/1808.02607} {arXiv:1808.02607} \BibitemShut {NoStop}%
\bibitem [{\citenamefont {Barnett}\ \emph {et~al.}(2003)\citenamefont {Barnett}, \citenamefont {Chefles},\ and\ \citenamefont {Jex}}]{barnettComparisonTwoUnknown2003}%
  \BibitemOpen
  \bibfield  {author} {\bibinfo {author} {\bibfnamefont {S.~M.}\ \bibnamefont {Barnett}}, \bibinfo {author} {\bibfnamefont {A.}~\bibnamefont {Chefles}},\ and\ \bibinfo {author} {\bibfnamefont {I.}~\bibnamefont {Jex}},\ }\bibfield  {title} {\bibinfo {title} {Comparison of two unknown pure quantum states},\ }\href {https://doi.org/10.1016/S0375-9601(02)01602-X} {\bibfield  {journal} {\bibinfo  {journal} {Physics Letters A}\ }\textbf {\bibinfo {volume} {307}},\ \bibinfo {pages} {189} (\bibinfo {year} {2003})},\ \Eprint {https://arxiv.org/abs/quant-ph/0202087} {arXiv:quant-ph/0202087} \BibitemShut {NoStop}%
\bibitem [{\citenamefont {Jex}\ \emph {et~al.}(2004)\citenamefont {Jex}, \citenamefont {Andersson},\ and\ \citenamefont {Chefles}}]{jexComparingStatesMany2004}%
  \BibitemOpen
  \bibfield  {author} {\bibinfo {author} {\bibfnamefont {I.}~\bibnamefont {Jex}}, \bibinfo {author} {\bibfnamefont {E.}~\bibnamefont {Andersson}},\ and\ \bibinfo {author} {\bibfnamefont {A.}~\bibnamefont {Chefles}},\ }\bibfield  {title} {\bibinfo {title} {Comparing the states of many quantum systems},\ }\href {https://doi.org/10.1080/09500340408238064} {\bibfield  {journal} {\bibinfo  {journal} {Journal of Modern Optics}\ }\textbf {\bibinfo {volume} {51}},\ \bibinfo {pages} {505} (\bibinfo {year} {2004})},\ \Eprint {https://arxiv.org/abs/quant-ph/0305120} {arXiv:quant-ph/0305120} \BibitemShut {NoStop}%
\bibitem [{\citenamefont {Chefles}\ \emph {et~al.}(2004)\citenamefont {Chefles}, \citenamefont {Andersson},\ and\ \citenamefont {Jex}}]{cheflesUnambiguousComparisonStates2004}%
  \BibitemOpen
  \bibfield  {author} {\bibinfo {author} {\bibfnamefont {A.}~\bibnamefont {Chefles}}, \bibinfo {author} {\bibfnamefont {E.}~\bibnamefont {Andersson}},\ and\ \bibinfo {author} {\bibfnamefont {I.}~\bibnamefont {Jex}},\ }\bibfield  {title} {\bibinfo {title} {Unambiguous comparison of the states of multiple quantum systems},\ }\href {https://doi.org/10.1088/0305-4470/37/29/009} {\bibfield  {journal} {\bibinfo  {journal} {J. Phys. A: Math. Gen.}\ }\textbf {\bibinfo {volume} {37}},\ \bibinfo {pages} {7315} (\bibinfo {year} {2004})}\BibitemShut {NoStop}%
\bibitem [{\citenamefont {Kleinmann}\ \emph {et~al.}(2005)\citenamefont {Kleinmann}, \citenamefont {Kampermann},\ and\ \citenamefont {Bru{\ss}}}]{kleinmannGeneralizationQuantumstateComparison2005}%
  \BibitemOpen
  \bibfield  {author} {\bibinfo {author} {\bibfnamefont {M.}~\bibnamefont {Kleinmann}}, \bibinfo {author} {\bibfnamefont {H.}~\bibnamefont {Kampermann}},\ and\ \bibinfo {author} {\bibfnamefont {D.}~\bibnamefont {Bru{\ss}}},\ }\bibfield  {title} {\bibinfo {title} {Generalization of quantum-state comparison},\ }\href {https://doi.org/10.1103/PhysRevA.72.032308} {\bibfield  {journal} {\bibinfo  {journal} {Phys. Rev. A}\ }\textbf {\bibinfo {volume} {72}},\ \bibinfo {pages} {032308} (\bibinfo {year} {2005})}\BibitemShut {NoStop}%
\bibitem [{\citenamefont {Sedl{\'a}k}\ \emph {et~al.}(2008)\citenamefont {Sedl{\'a}k}, \citenamefont {Ziman}, \citenamefont {Bu{\v z}ek},\ and\ \citenamefont {Hillery}}]{sedlakUnambiguousComparisonEnsembles2008}%
  \BibitemOpen
  \bibfield  {author} {\bibinfo {author} {\bibfnamefont {M.}~\bibnamefont {Sedl{\'a}k}}, \bibinfo {author} {\bibfnamefont {M.}~\bibnamefont {Ziman}}, \bibinfo {author} {\bibfnamefont {V.}~\bibnamefont {Bu{\v z}ek}},\ and\ \bibinfo {author} {\bibfnamefont {M.}~\bibnamefont {Hillery}},\ }\bibfield  {title} {\bibinfo {title} {Unambiguous comparison of ensembles of quantum states},\ }\href {https://doi.org/10.1103/PhysRevA.77.042304} {\bibfield  {journal} {\bibinfo  {journal} {Phys. Rev. A}\ }\textbf {\bibinfo {volume} {77}},\ \bibinfo {pages} {042304} (\bibinfo {year} {2008})}\BibitemShut {NoStop}%
\bibitem [{\citenamefont {Pang}\ and\ \citenamefont {Wu}(2011)}]{pangComparisonMixedQuantum2011}%
  \BibitemOpen
  \bibfield  {author} {\bibinfo {author} {\bibfnamefont {S.}~\bibnamefont {Pang}}\ and\ \bibinfo {author} {\bibfnamefont {S.}~\bibnamefont {Wu}},\ }\bibfield  {title} {\bibinfo {title} {Comparison of mixed quantum states},\ }\href {https://doi.org/10.1103/PhysRevA.84.012336} {\bibfield  {journal} {\bibinfo  {journal} {Phys. Rev. A}\ }\textbf {\bibinfo {volume} {84}},\ \bibinfo {pages} {012336} (\bibinfo {year} {2011})}\BibitemShut {NoStop}%
\bibitem [{\citenamefont {Hayashi}\ \emph {et~al.}(2018)\citenamefont {Hayashi}, \citenamefont {Hashimoto},\ and\ \citenamefont {Horibe}}]{hayashiQuantumstateComparisonDiscrimination2018}%
  \BibitemOpen
  \bibfield  {author} {\bibinfo {author} {\bibfnamefont {A.}~\bibnamefont {Hayashi}}, \bibinfo {author} {\bibfnamefont {T.}~\bibnamefont {Hashimoto}},\ and\ \bibinfo {author} {\bibfnamefont {M.}~\bibnamefont {Horibe}},\ }\bibfield  {title} {\bibinfo {title} {Quantum-state comparison and discrimination},\ }\href {https://doi.org/10.1103/PhysRevA.97.052323} {\bibfield  {journal} {\bibinfo  {journal} {Phys. Rev. A}\ }\textbf {\bibinfo {volume} {97}},\ \bibinfo {pages} {052323} (\bibinfo {year} {2018})},\ \Eprint {https://arxiv.org/abs/1803.09030} {arXiv:1803.09030} \BibitemShut {NoStop}%
\bibitem [{\citenamefont {Fanizza}\ \emph {et~al.}(2020)\citenamefont {Fanizza}, \citenamefont {Rosati}, \citenamefont {Skotiniotis}, \citenamefont {Calsamiglia},\ and\ \citenamefont {Giovannetti}}]{fanizzaSwapTestOptimal2020}%
  \BibitemOpen
  \bibfield  {author} {\bibinfo {author} {\bibfnamefont {M.}~\bibnamefont {Fanizza}}, \bibinfo {author} {\bibfnamefont {M.}~\bibnamefont {Rosati}}, \bibinfo {author} {\bibfnamefont {M.}~\bibnamefont {Skotiniotis}}, \bibinfo {author} {\bibfnamefont {J.}~\bibnamefont {Calsamiglia}},\ and\ \bibinfo {author} {\bibfnamefont {V.}~\bibnamefont {Giovannetti}},\ }\bibfield  {title} {\bibinfo {title} {Beyond the {{Swap Test}}: {{Optimal Estimation}} of {{Quantum State Overlap}}},\ }\href {https://doi.org/10.1103/PhysRevLett.124.060503} {\bibfield  {journal} {\bibinfo  {journal} {Phys. Rev. Lett.}\ }\textbf {\bibinfo {volume} {124}},\ \bibinfo {pages} {060503} (\bibinfo {year} {2020})}\BibitemShut {NoStop}%
\bibitem [{\citenamefont {B\u{a}descu}\ \emph {et~al.}(2019)\citenamefont {B\u{a}descu}, \citenamefont {O'Donnell},\ and\ \citenamefont {Wright}}]{badescuQuantumStateCertification2019}%
  \BibitemOpen
  \bibfield  {author} {\bibinfo {author} {\bibfnamefont {C.}~\bibnamefont {B\u{a}descu}}, \bibinfo {author} {\bibfnamefont {R.}~\bibnamefont {O'Donnell}},\ and\ \bibinfo {author} {\bibfnamefont {J.}~\bibnamefont {Wright}},\ }\bibfield  {title} {\bibinfo {title} {Quantum state certification},\ }in\ \href {https://doi.org/10.1145/3313276.3316344} {\emph {\bibinfo {booktitle} {Proceedings of the 51st Annual ACM SIGACT Symposium on Theory of Computing}}}\ (\bibinfo {year} {2019})\ pp.\ \bibinfo {pages} {503--514}\BibitemShut {NoStop}%
\bibitem [{\citenamefont {Ziman}\ \emph {et~al.}(2009)\citenamefont {Ziman}, \citenamefont {Heinosaari},\ and\ \citenamefont {Sedlak}}]{zimanUnambiguousComparisonQuantum2009}%
  \BibitemOpen
  \bibfield  {author} {\bibinfo {author} {\bibfnamefont {M.}~\bibnamefont {Ziman}}, \bibinfo {author} {\bibfnamefont {T.}~\bibnamefont {Heinosaari}},\ and\ \bibinfo {author} {\bibfnamefont {M.}~\bibnamefont {Sedlak}},\ }\bibfield  {title} {\bibinfo {title} {Unambiguous comparison of quantum measurements},\ }\href {https://doi.org/10.1103/PhysRevA.80.052102} {\bibfield  {journal} {\bibinfo  {journal} {Phys. Rev. A}\ }\textbf {\bibinfo {volume} {80}},\ \bibinfo {pages} {052102} (\bibinfo {year} {2009})},\ \Eprint {https://arxiv.org/abs/0905.4445} {arXiv:0905.4445} \BibitemShut {NoStop}%
\bibitem [{\citenamefont {Shimbo}\ \emph {et~al.}(2018)\citenamefont {Shimbo}, \citenamefont {Soeda},\ and\ \citenamefont {Murao}}]{shimboEquivalenceDeterminationUnitary2018}%
  \BibitemOpen
  \bibfield  {author} {\bibinfo {author} {\bibfnamefont {A.}~\bibnamefont {Shimbo}}, \bibinfo {author} {\bibfnamefont {A.}~\bibnamefont {Soeda}},\ and\ \bibinfo {author} {\bibfnamefont {M.}~\bibnamefont {Murao}},\ }\href@noop {} {\bibinfo {title} {Equivalence determination of unitary operations}} (\bibinfo {year} {2018}),\ \Eprint {https://arxiv.org/abs/1803.11414} {arXiv:1803.11414 [quant-ph]} \BibitemShut {NoStop}%
\bibitem [{\citenamefont {Soeda}\ \emph {et~al.}(2021)\citenamefont {Soeda}, \citenamefont {Shimbo},\ and\ \citenamefont {Murao}}]{soedaOptimalQuantumDiscrimination2021}%
  \BibitemOpen
  \bibfield  {author} {\bibinfo {author} {\bibfnamefont {A.}~\bibnamefont {Soeda}}, \bibinfo {author} {\bibfnamefont {A.}~\bibnamefont {Shimbo}},\ and\ \bibinfo {author} {\bibfnamefont {M.}~\bibnamefont {Murao}},\ }\bibfield  {title} {\bibinfo {title} {Optimal quantum discrimination of single-qubit unitary gates between two candidates},\ }\href {https://doi.org/10.1103/PhysRevA.104.022422} {\bibfield  {journal} {\bibinfo  {journal} {Phys. Rev. A}\ }\textbf {\bibinfo {volume} {104}},\ \bibinfo {pages} {022422} (\bibinfo {year} {2021})}\BibitemShut {NoStop}%
\bibitem [{\citenamefont {Andersson}\ \emph {et~al.}(2003)\citenamefont {Andersson}, \citenamefont {Jex},\ and\ \citenamefont {Barnett}}]{anderssonComparisonUnitaryTransforms2003}%
  \BibitemOpen
  \bibfield  {author} {\bibinfo {author} {\bibfnamefont {E.}~\bibnamefont {Andersson}}, \bibinfo {author} {\bibfnamefont {I.}~\bibnamefont {Jex}},\ and\ \bibinfo {author} {\bibfnamefont {S.~M.}\ \bibnamefont {Barnett}},\ }\bibfield  {title} {\bibinfo {title} {Comparison of unitary transforms},\ }\href {https://doi.org/10.1088/0305-4470/36/9/310} {\bibfield  {journal} {\bibinfo  {journal} {J. Phys. A: Math. Gen.}\ }\textbf {\bibinfo {volume} {36}},\ \bibinfo {pages} {2325} (\bibinfo {year} {2003})}\BibitemShut {NoStop}%
\bibitem [{\citenamefont {Chefles}(1998)}]{cheflesUnambiguousDiscriminationLinearlyIndependent1998}%
  \BibitemOpen
  \bibfield  {author} {\bibinfo {author} {\bibfnamefont {A.}~\bibnamefont {Chefles}},\ }\bibfield  {title} {\bibinfo {title} {Unambiguous {{Discrimination Between Linearly-Independent Quantum States}}},\ }\href {https://doi.org/10.1016/S0375-9601(98)00064-4} {\bibfield  {journal} {\bibinfo  {journal} {Physics Letters A}\ }\textbf {\bibinfo {volume} {239}},\ \bibinfo {pages} {339} (\bibinfo {year} {1998})},\ \Eprint {https://arxiv.org/abs/quant-ph/9807022} {arXiv:quant-ph/9807022} \BibitemShut {NoStop}%
\bibitem [{\citenamefont {Feng}\ \emph {et~al.}(2004)\citenamefont {Feng}, \citenamefont {Duan},\ and\ \citenamefont {Ying}}]{fengUnambiguousDiscriminationMixed2004}%
  \BibitemOpen
  \bibfield  {author} {\bibinfo {author} {\bibfnamefont {Y.}~\bibnamefont {Feng}}, \bibinfo {author} {\bibfnamefont {R.}~\bibnamefont {Duan}},\ and\ \bibinfo {author} {\bibfnamefont {M.}~\bibnamefont {Ying}},\ }\bibfield  {title} {\bibinfo {title} {Unambiguous discrimination between mixed quantum states},\ }\href {https://doi.org/10.1103/PhysRevA.70.012308} {\bibfield  {journal} {\bibinfo  {journal} {Phys. Rev. A}\ }\textbf {\bibinfo {volume} {70}},\ \bibinfo {pages} {012308} (\bibinfo {year} {2004})}\BibitemShut {NoStop}%
\bibitem [{\citenamefont {Wang}\ and\ \citenamefont {Ying}(2006)}]{wangUnambiguousDiscriminationQuantum2006}%
  \BibitemOpen
  \bibfield  {author} {\bibinfo {author} {\bibfnamefont {G.}~\bibnamefont {Wang}}\ and\ \bibinfo {author} {\bibfnamefont {M.}~\bibnamefont {Ying}},\ }\bibfield  {title} {\bibinfo {title} {Unambiguous discrimination among quantum operations},\ }\href {https://doi.org/10.1103/PhysRevA.73.042301} {\bibfield  {journal} {\bibinfo  {journal} {Phys. Rev. A}\ }\textbf {\bibinfo {volume} {73}},\ \bibinfo {pages} {042301} (\bibinfo {year} {2006})},\ \Eprint {https://arxiv.org/abs/quant-ph/0512142} {arXiv:quant-ph/0512142} \BibitemShut {NoStop}%
\bibitem [{\citenamefont {Sedl{\'a}k}\ and\ \citenamefont {Ziman}(2009)}]{sedlakUnambiguousComparisonUnitary2009}%
  \BibitemOpen
  \bibfield  {author} {\bibinfo {author} {\bibfnamefont {M.}~\bibnamefont {Sedl{\'a}k}}\ and\ \bibinfo {author} {\bibfnamefont {M.}~\bibnamefont {Ziman}},\ }\bibfield  {title} {\bibinfo {title} {Unambiguous comparison of unitary channels},\ }\href {https://doi.org/10.1103/PhysRevA.79.012303} {\bibfield  {journal} {\bibinfo  {journal} {Phys. Rev. A}\ }\textbf {\bibinfo {volume} {79}},\ \bibinfo {pages} {012303} (\bibinfo {year} {2009})}\BibitemShut {NoStop}%
\bibitem [{\citenamefont {Hillery}\ \emph {et~al.}(2010)\citenamefont {Hillery}, \citenamefont {Andersson}, \citenamefont {Barnett},\ and\ \citenamefont {Oi}}]{hilleryDecisionProblemsQuantum2010}%
  \BibitemOpen
  \bibfield  {author} {\bibinfo {author} {\bibfnamefont {M.}~\bibnamefont {Hillery}}, \bibinfo {author} {\bibfnamefont {E.}~\bibnamefont {Andersson}}, \bibinfo {author} {\bibfnamefont {S.~M.}\ \bibnamefont {Barnett}},\ and\ \bibinfo {author} {\bibfnamefont {D.}~\bibnamefont {Oi}},\ }\bibfield  {title} {\bibinfo {title} {Decision problems with quantum black boxes},\ }\href {https://doi.org/10.1080/09500340903203129} {\bibfield  {journal} {\bibinfo  {journal} {Journal of Modern Optics}\ }\textbf {\bibinfo {volume} {57}},\ \bibinfo {pages} {244} (\bibinfo {year} {2010})},\ \Eprint {https://arxiv.org/abs/1109.4823} {arXiv:1109.4823 [quant-ph]} \BibitemShut {NoStop}%
\bibitem [{\citenamefont {Markiewicz}\ \emph {et~al.}(2025)\citenamefont {Markiewicz}, \citenamefont {Pawela},\ and\ \citenamefont {Pucha{\l}a}}]{markiewiczSingleShotComparisonRandom2025}%
  \BibitemOpen
  \bibfield  {author} {\bibinfo {author} {\bibfnamefont {M.}~\bibnamefont {Markiewicz}}, \bibinfo {author} {\bibfnamefont {{\L}.}~\bibnamefont {Pawela}},\ and\ \bibinfo {author} {\bibfnamefont {Z.}~\bibnamefont {Pucha{\l}a}},\ }\href {https://doi.org/10.48550/arXiv.2501.17317} {\bibinfo {title} {Single-shot comparison of random quantum channels and measurements}} (\bibinfo {year} {2025}),\ \Eprint {https://arxiv.org/abs/2501.17317} {arXiv:2501.17317 [quant-ph]} \BibitemShut {NoStop}%
\bibitem [{\citenamefont {Chiribella}\ \emph {et~al.}(2008{\natexlab{a}})\citenamefont {Chiribella}, \citenamefont {D'Ariano},\ and\ \citenamefont {Perinotti}}]{chiribellaOptimalCloningUnitary2008}%
  \BibitemOpen
  \bibfield  {author} {\bibinfo {author} {\bibfnamefont {G.}~\bibnamefont {Chiribella}}, \bibinfo {author} {\bibfnamefont {G.~M.}\ \bibnamefont {D'Ariano}},\ and\ \bibinfo {author} {\bibfnamefont {P.}~\bibnamefont {Perinotti}},\ }\bibfield  {title} {\bibinfo {title} {Optimal cloning of unitary transformations},\ }\href {https://doi.org/10.1103/PhysRevLett.101.180504} {\bibfield  {journal} {\bibinfo  {journal} {Phys. Rev. Lett.}\ }\textbf {\bibinfo {volume} {101}},\ \bibinfo {pages} {180504} (\bibinfo {year} {2008}{\natexlab{a}})},\ \Eprint {https://arxiv.org/abs/0804.0129} {arXiv:0804.0129} \BibitemShut {NoStop}%
\bibitem [{\citenamefont {Harrow}\ \emph {et~al.}(2010)\citenamefont {Harrow}, \citenamefont {Hassidim}, \citenamefont {Leung},\ and\ \citenamefont {Watrous}}]{harrowAdaptiveNonadaptiveStrategies2010}%
  \BibitemOpen
  \bibfield  {author} {\bibinfo {author} {\bibfnamefont {A.~W.}\ \bibnamefont {Harrow}}, \bibinfo {author} {\bibfnamefont {A.}~\bibnamefont {Hassidim}}, \bibinfo {author} {\bibfnamefont {D.~W.}\ \bibnamefont {Leung}},\ and\ \bibinfo {author} {\bibfnamefont {J.}~\bibnamefont {Watrous}},\ }\bibfield  {title} {\bibinfo {title} {Adaptive versus nonadaptive strategies for quantum channel discrimination},\ }\href {https://doi.org/10.1103/PhysRevA.81.032339} {\bibfield  {journal} {\bibinfo  {journal} {Phys. Rev. A}\ }\textbf {\bibinfo {volume} {81}},\ \bibinfo {pages} {032339} (\bibinfo {year} {2010})}\BibitemShut {NoStop}%
\bibitem [{\citenamefont {Chiribella}\ \emph {et~al.}(2008{\natexlab{b}})\citenamefont {Chiribella}, \citenamefont {D'Ariano},\ and\ \citenamefont {Perinotti}}]{chiribellaMemoryEffectsQuantum2008}%
  \BibitemOpen
  \bibfield  {author} {\bibinfo {author} {\bibfnamefont {G.}~\bibnamefont {Chiribella}}, \bibinfo {author} {\bibfnamefont {G.~M.}\ \bibnamefont {D'Ariano}},\ and\ \bibinfo {author} {\bibfnamefont {P.}~\bibnamefont {Perinotti}},\ }\bibfield  {title} {\bibinfo {title} {Memory effects in quantum channel discrimination},\ }\href {https://doi.org/10.1103/PhysRevLett.101.180501} {\bibfield  {journal} {\bibinfo  {journal} {Phys. Rev. Lett.}\ }\textbf {\bibinfo {volume} {101}},\ \bibinfo {pages} {180501} (\bibinfo {year} {2008}{\natexlab{b}})},\ \Eprint {https://arxiv.org/abs/0803.3237} {arXiv:0803.3237} \BibitemShut {NoStop}%
\bibitem [{\citenamefont {Bavaresco}\ \emph {et~al.}(2022)\citenamefont {Bavaresco}, \citenamefont {Murao},\ and\ \citenamefont {Quintino}}]{bavarescoUnitaryChannelDiscrimination2022}%
  \BibitemOpen
  \bibfield  {author} {\bibinfo {author} {\bibfnamefont {J.}~\bibnamefont {Bavaresco}}, \bibinfo {author} {\bibfnamefont {M.}~\bibnamefont {Murao}},\ and\ \bibinfo {author} {\bibfnamefont {M.~T.}\ \bibnamefont {Quintino}},\ }\bibfield  {title} {\bibinfo {title} {Unitary channel discrimination beyond group structures: {{Advantages}} of sequential and indefinite-causal-order strategies},\ }\href {https://doi.org/10.1063/5.0075919} {\bibfield  {journal} {\bibinfo  {journal} {J. Math. Phys.}\ }\textbf {\bibinfo {volume} {63}},\ \bibinfo {pages} {042203} (\bibinfo {year} {2022})},\ \Eprint {https://arxiv.org/abs/2105.13369} {arXiv:2105.13369} \BibitemShut {NoStop}%
\bibitem [{\citenamefont {Chiribella}\ \emph {et~al.}(2008{\natexlab{c}})\citenamefont {Chiribella}, \citenamefont {D'Ariano},\ and\ \citenamefont {Perinotti}}]{chiribellaTransformingQuantumOperations2008}%
  \BibitemOpen
  \bibfield  {author} {\bibinfo {author} {\bibfnamefont {G.}~\bibnamefont {Chiribella}}, \bibinfo {author} {\bibfnamefont {G.~M.}\ \bibnamefont {D'Ariano}},\ and\ \bibinfo {author} {\bibfnamefont {P.}~\bibnamefont {Perinotti}},\ }\bibfield  {title} {\bibinfo {title} {Transforming quantum operations: {{Quantum}} supermaps},\ }\href {https://doi.org/10.1209/0295-5075/83/30004} {\bibfield  {journal} {\bibinfo  {journal} {EPL (Europhysics Letters)}\ }\textbf {\bibinfo {volume} {83}},\ \bibinfo {pages} {30004} (\bibinfo {year} {2008}{\natexlab{c}})}\BibitemShut {NoStop}%
\bibitem [{\citenamefont {Chiribella}\ \emph {et~al.}(2008{\natexlab{d}})\citenamefont {Chiribella}, \citenamefont {D'Ariano},\ and\ \citenamefont {Perinotti}}]{chiribellaQuantumCircuitsArchitecture2008}%
  \BibitemOpen
  \bibfield  {author} {\bibinfo {author} {\bibfnamefont {G.}~\bibnamefont {Chiribella}}, \bibinfo {author} {\bibfnamefont {G.~M.}\ \bibnamefont {D'Ariano}},\ and\ \bibinfo {author} {\bibfnamefont {P.}~\bibnamefont {Perinotti}},\ }\bibfield  {title} {\bibinfo {title} {Quantum {{Circuit Architecture}}},\ }\href {https://doi.org/10.1103/PhysRevLett.101.060401} {\bibfield  {journal} {\bibinfo  {journal} {Phys. Rev. Lett.}\ }\textbf {\bibinfo {volume} {101}},\ \bibinfo {pages} {060401} (\bibinfo {year} {2008}{\natexlab{d}})},\ \Eprint {https://arxiv.org/abs/0712.1325} {arXiv:0712.1325} \BibitemShut {NoStop}%
\bibitem [{\citenamefont {Chiribella}\ \emph {et~al.}(2009)\citenamefont {Chiribella}, \citenamefont {D'Ariano},\ and\ \citenamefont {Perinotti}}]{chiribellaTheoreticalFrameworkQuantum2009}%
  \BibitemOpen
  \bibfield  {author} {\bibinfo {author} {\bibfnamefont {G.}~\bibnamefont {Chiribella}}, \bibinfo {author} {\bibfnamefont {G.~M.}\ \bibnamefont {D'Ariano}},\ and\ \bibinfo {author} {\bibfnamefont {P.}~\bibnamefont {Perinotti}},\ }\bibfield  {title} {\bibinfo {title} {Theoretical framework for quantum networks},\ }\href {https://doi.org/10.1103/PhysRevA.80.022339} {\bibfield  {journal} {\bibinfo  {journal} {Phys. Rev. A}\ }\textbf {\bibinfo {volume} {80}},\ \bibinfo {pages} {022339} (\bibinfo {year} {2009})},\ \Eprint {https://arxiv.org/abs/0904.4483} {arXiv:0904.4483} \BibitemShut {NoStop}%
\bibitem [{\citenamefont {Ziman}(2008)}]{zimanProcessPOVMMathematical2008}%
  \BibitemOpen
  \bibfield  {author} {\bibinfo {author} {\bibfnamefont {M.}~\bibnamefont {Ziman}},\ }\bibfield  {title} {\bibinfo {title} {Process {{Positive-Operator-Valued Measure}}: {{A}} mathematical framework for the description of process tomography experiments},\ }\href {https://doi.org/10.1103/PhysRevA.77.062112} {\bibfield  {journal} {\bibinfo  {journal} {Phys. Rev. A}\ }\textbf {\bibinfo {volume} {77}},\ \bibinfo {pages} {062112} (\bibinfo {year} {2008})},\ \Eprint {https://arxiv.org/abs/0802.3862} {arXiv:0802.3862} \BibitemShut {NoStop}%
\bibitem [{\citenamefont {Gutoski}\ and\ \citenamefont {Watrous}(2007)}]{gutoskiTowardGeneralTheory2007}%
  \BibitemOpen
  \bibfield  {author} {\bibinfo {author} {\bibfnamefont {G.}~\bibnamefont {Gutoski}}\ and\ \bibinfo {author} {\bibfnamefont {J.}~\bibnamefont {Watrous}},\ }\bibfield  {title} {\bibinfo {title} {Toward a general theory of quantum games},\ }in\ \href {https://doi.org/10.1145/1250790.1250873} {\emph {\bibinfo {booktitle} {Proceedings of the Thirty-Ninth Annual ACM Symposium on Theory of Computing}}}\ (\bibinfo {year} {2007})\ pp.\ \bibinfo {pages} {565--574},\ \Eprint {https://arxiv.org/abs/quant-ph/0611234} {arXiv:quant-ph/0611234} \BibitemShut {NoStop}%
\bibitem [{\citenamefont {Oreshkov}\ \emph {et~al.}(2012)\citenamefont {Oreshkov}, \citenamefont {Costa},\ and\ \citenamefont {Brukner}}]{oreshkovQuantumCorrelationsNo2012}%
  \BibitemOpen
  \bibfield  {author} {\bibinfo {author} {\bibfnamefont {O.}~\bibnamefont {Oreshkov}}, \bibinfo {author} {\bibfnamefont {F.}~\bibnamefont {Costa}},\ and\ \bibinfo {author} {\bibfnamefont {{\v C}.}~\bibnamefont {Brukner}},\ }\bibfield  {title} {\bibinfo {title} {Quantum correlations with no causal order},\ }\href {https://doi.org/10.1038/ncomms2076} {\bibfield  {journal} {\bibinfo  {journal} {Nat Commun}\ }\textbf {\bibinfo {volume} {3}},\ \bibinfo {pages} {1092} (\bibinfo {year} {2012})}\BibitemShut {NoStop}%
\bibitem [{\citenamefont {Chiribella}\ \emph {et~al.}(2013)\citenamefont {Chiribella}, \citenamefont {D'Ariano}, \citenamefont {Perinotti},\ and\ \citenamefont {Valiron}}]{chiribellaQuantumComputationsDefinite2013}%
  \BibitemOpen
  \bibfield  {author} {\bibinfo {author} {\bibfnamefont {G.}~\bibnamefont {Chiribella}}, \bibinfo {author} {\bibfnamefont {G.~M.}\ \bibnamefont {D'Ariano}}, \bibinfo {author} {\bibfnamefont {P.}~\bibnamefont {Perinotti}},\ and\ \bibinfo {author} {\bibfnamefont {B.}~\bibnamefont {Valiron}},\ }\bibfield  {title} {\bibinfo {title} {Quantum computations without definite causal structure},\ }\href {https://doi.org/10.1103/PhysRevA.88.022318} {\bibfield  {journal} {\bibinfo  {journal} {Phys. Rev. A}\ }\textbf {\bibinfo {volume} {88}},\ \bibinfo {pages} {022318} (\bibinfo {year} {2013})}\BibitemShut {NoStop}%
\bibitem [{\citenamefont {Baumeler}\ and\ \citenamefont {Wolf}(2016)}]{baumelerSpaceLogicallyConsistent2016}%
  \BibitemOpen
  \bibfield  {author} {\bibinfo {author} {\bibfnamefont {{\"A}.}~\bibnamefont {Baumeler}}\ and\ \bibinfo {author} {\bibfnamefont {S.}~\bibnamefont {Wolf}},\ }\bibfield  {title} {\bibinfo {title} {The space of logically consistent classical processes without causal order},\ }\href {https://doi.org/10.1088/1367-2630/18/1/013036} {\bibfield  {journal} {\bibinfo  {journal} {New J. Phys.}\ }\textbf {\bibinfo {volume} {18}},\ \bibinfo {pages} {013036} (\bibinfo {year} {2016})},\ \Eprint {https://arxiv.org/abs/1507.01714} {arXiv:1507.01714 [quant-ph]} \BibitemShut {NoStop}%
\bibitem [{\citenamefont {Wechs}\ \emph {et~al.}(2021)\citenamefont {Wechs}, \citenamefont {Dourdent}, \citenamefont {Abbott},\ and\ \citenamefont {Branciard}}]{wechsQuantumCircuitsClassical2021a}%
  \BibitemOpen
  \bibfield  {author} {\bibinfo {author} {\bibfnamefont {J.}~\bibnamefont {Wechs}}, \bibinfo {author} {\bibfnamefont {H.}~\bibnamefont {Dourdent}}, \bibinfo {author} {\bibfnamefont {A.~A.}\ \bibnamefont {Abbott}},\ and\ \bibinfo {author} {\bibfnamefont {C.}~\bibnamefont {Branciard}},\ }\bibfield  {title} {\bibinfo {title} {Quantum {{Circuits}} with {{Classical Versus Quantum Control}} of {{Causal Order}}},\ }\href {https://doi.org/10.1103/PRXQuantum.2.030335} {\bibfield  {journal} {\bibinfo  {journal} {PRX Quantum}\ }\textbf {\bibinfo {volume} {2}},\ \bibinfo {pages} {030335} (\bibinfo {year} {2021})}\BibitemShut {NoStop}%
\bibitem [{\citenamefont {Bavaresco}\ \emph {et~al.}(2021)\citenamefont {Bavaresco}, \citenamefont {Murao},\ and\ \citenamefont {Quintino}}]{bavarescoStrictHierarchy2021}%
  \BibitemOpen
  \bibfield  {author} {\bibinfo {author} {\bibfnamefont {J.}~\bibnamefont {Bavaresco}}, \bibinfo {author} {\bibfnamefont {M.}~\bibnamefont {Murao}},\ and\ \bibinfo {author} {\bibfnamefont {M.~T.}\ \bibnamefont {Quintino}},\ }\bibfield  {title} {\bibinfo {title} {Strict hierarchy between parallel, sequential, and indefinite-causal-order strategies for channel discrimination},\ }\href {https://doi.org/10.1103/PhysRevLett.127.200504} {\bibfield  {journal} {\bibinfo  {journal} {Phys. Rev. Lett.}\ }\textbf {\bibinfo {volume} {127}},\ \bibinfo {pages} {200504} (\bibinfo {year} {2021})},\ \Eprint {https://arxiv.org/abs/2011.08300} {arXiv:2011.08300 [quant-ph]} \BibitemShut {NoStop}%
\bibitem [{\citenamefont {Andr{\'e}}\ \emph {et~al.}(2026)\citenamefont {Andr{\'e}}, \citenamefont {Bavaresco},\ and\ \citenamefont {Mehboudi}}]{andreStrategyOptimizationBayesian2026}%
  \BibitemOpen
  \bibfield  {author} {\bibinfo {author} {\bibfnamefont {E.~L.}\ \bibnamefont {Andr{\'e}}}, \bibinfo {author} {\bibfnamefont {J.}~\bibnamefont {Bavaresco}},\ and\ \bibinfo {author} {\bibfnamefont {M.}~\bibnamefont {Mehboudi}},\ }\href {https://doi.org/10.48550/arXiv.2602.09655} {\bibinfo {title} {Strategy optimization for {{Bayesian}} quantum parameter estimation with finite copies: {{Adaptive}} greedy, parallel, sequential, and general strategies}} (\bibinfo {year} {2026}),\ \Eprint {https://arxiv.org/abs/2602.09655} {arXiv:2602.09655 [quant-ph]} \BibitemShut {NoStop}%
\bibitem [{\citenamefont {Choi}(1975)}]{choiCompletelyPositiveLinear1975}%
  \BibitemOpen
  \bibfield  {author} {\bibinfo {author} {\bibfnamefont {M.-D.}\ \bibnamefont {Choi}},\ }\bibfield  {title} {\bibinfo {title} {Completely positive linear maps on complex matrices},\ }\href {https://doi.org/10.1016/0024-3795(75)90075-0} {\bibfield  {journal} {\bibinfo  {journal} {Linear Algebra and its Applications}\ }\textbf {\bibinfo {volume} {10}},\ \bibinfo {pages} {285} (\bibinfo {year} {1975})}\BibitemShut {NoStop}%
\bibitem [{\citenamefont {Jamio{\l}kowski}(1972)}]{jamiolkowskiLinearTransformationsWhich1972}%
  \BibitemOpen
  \bibfield  {author} {\bibinfo {author} {\bibfnamefont {A.}~\bibnamefont {Jamio{\l}kowski}},\ }\bibfield  {title} {\bibinfo {title} {Linear transformations which preserve trace and positive semidefiniteness of operators},\ }\href {https://doi.org/10.1016/0034-4877(72)90011-0} {\bibfield  {journal} {\bibinfo  {journal} {Reports on Mathematical Physics}\ }\textbf {\bibinfo {volume} {3}},\ \bibinfo {pages} {275} (\bibinfo {year} {1972})}\BibitemShut {NoStop}%
\bibitem [{\citenamefont {Ara{\'u}jo}\ \emph {et~al.}(2015)\citenamefont {Ara{\'u}jo}, \citenamefont {Branciard}, \citenamefont {Costa}, \citenamefont {Feix}, \citenamefont {Giarmatzi},\ and\ \citenamefont {Brukner}}]{araujoWitnessingCausalNonseparability2015}%
  \BibitemOpen
  \bibfield  {author} {\bibinfo {author} {\bibfnamefont {M.}~\bibnamefont {Ara{\'u}jo}}, \bibinfo {author} {\bibfnamefont {C.}~\bibnamefont {Branciard}}, \bibinfo {author} {\bibfnamefont {F.}~\bibnamefont {Costa}}, \bibinfo {author} {\bibfnamefont {A.}~\bibnamefont {Feix}}, \bibinfo {author} {\bibfnamefont {C.}~\bibnamefont {Giarmatzi}},\ and\ \bibinfo {author} {\bibfnamefont {{\v C}.}~\bibnamefont {Brukner}},\ }\bibfield  {title} {\bibinfo {title} {Witnessing causal nonseparability},\ }\href {https://doi.org/10.1088/1367-2630/17/10/102001} {\bibfield  {journal} {\bibinfo  {journal} {New J. Phys.}\ }\textbf {\bibinfo {volume} {17}},\ \bibinfo {pages} {102001} (\bibinfo {year} {2015})},\ \Eprint {https://arxiv.org/abs/1506.03776} {arXiv:1506.03776 [quant-ph]} \BibitemShut {NoStop}%
\bibitem [{\citenamefont {Oreshkov}\ and\ \citenamefont {Giarmatzi}(2016)}]{oreshkovCausalCausallySeparable2016}%
  \BibitemOpen
  \bibfield  {author} {\bibinfo {author} {\bibfnamefont {O.}~\bibnamefont {Oreshkov}}\ and\ \bibinfo {author} {\bibfnamefont {C.}~\bibnamefont {Giarmatzi}},\ }\bibfield  {title} {\bibinfo {title} {Causal and causally separable processes},\ }\href {https://doi.org/10.1088/1367-2630/18/9/093020} {\bibfield  {journal} {\bibinfo  {journal} {New J. Phys.}\ }\textbf {\bibinfo {volume} {18}},\ \bibinfo {pages} {093020} (\bibinfo {year} {2016})},\ \Eprint {https://arxiv.org/abs/1506.05449} {arXiv:1506.05449 [quant-ph]} \BibitemShut {NoStop}%
\bibitem [{\citenamefont {Schur}(1901)}]{schurUeberKlasseMatrizen1901}%
  \BibitemOpen
  \bibfield  {author} {\bibinfo {author} {\bibfnamefont {I.}~\bibnamefont {Schur}},\ }\emph {\bibinfo {title} {{\"U}ber eine {{Klasse}} von {{Matrizen}}, die sich einer gegebenen {{Matrix}} zuordnen lassen}},\ \href@noop {} {Ph.D. thesis},\ \bibinfo  {school} {Universit{\"a}t Berlin} (\bibinfo {year} {1901})\BibitemShut {NoStop}%
\bibitem [{\citenamefont {Weyl}(1939)}]{weylClassicalGroupsTheir1939}%
  \BibitemOpen
  \bibfield  {author} {\bibinfo {author} {\bibfnamefont {H.}~\bibnamefont {Weyl}},\ }\href@noop {} {\emph {\bibinfo {title} {The {{Classical Groups}}: {{Their Invariants}} and {{Representations}}}}}\ (\bibinfo  {publisher} {Princeton University Press},\ \bibinfo {address} {Princeton, NJ},\ \bibinfo {year} {1939})\BibitemShut {NoStop}%
\bibitem [{\citenamefont {Fulton}\ and\ \citenamefont {Harris}(2004)}]{fultonRepresentationTheory2004}%
  \BibitemOpen
  \bibfield  {author} {\bibinfo {author} {\bibfnamefont {W.}~\bibnamefont {Fulton}}\ and\ \bibinfo {author} {\bibfnamefont {J.}~\bibnamefont {Harris}},\ }\href {https://doi.org/10.1007/978-1-4612-0979-9} {\emph {\bibinfo {title} {Representation {{Theory}}}}},\ \bibinfo {series} {Graduate {{Texts}} in {{Mathematics}}}, Vol.\ \bibinfo {volume} {129}\ (\bibinfo  {publisher} {Springer New York},\ \bibinfo {address} {New York, NY},\ \bibinfo {year} {2004})\BibitemShut {NoStop}%
\bibitem [{\citenamefont {Miyazaki}\ \emph {et~al.}(2019)\citenamefont {Miyazaki}, \citenamefont {Soeda},\ and\ \citenamefont {Murao}}]{miyazakiComplexConjugationSupermap2019}%
  \BibitemOpen
  \bibfield  {author} {\bibinfo {author} {\bibfnamefont {J.}~\bibnamefont {Miyazaki}}, \bibinfo {author} {\bibfnamefont {A.}~\bibnamefont {Soeda}},\ and\ \bibinfo {author} {\bibfnamefont {M.}~\bibnamefont {Murao}},\ }\bibfield  {title} {\bibinfo {title} {Complex conjugation supermap of unitary quantum maps and its universal implementation protocol},\ }\href {https://doi.org/10.1103/PhysRevResearch.1.013007} {\bibfield  {journal} {\bibinfo  {journal} {Physical Review Research}\ }\textbf {\bibinfo {volume} {1}},\ \bibinfo {pages} {013007} (\bibinfo {year} {2019})}\BibitemShut {NoStop}%
\bibitem [{\citenamefont {Ebler}\ \emph {et~al.}(2023)\citenamefont {Ebler}, \citenamefont {Horodecki}, \citenamefont {Marciniak}, \citenamefont {M{\l}ynik}, \citenamefont {Quintino},\ and\ \citenamefont {Studzi{\'n}ski}}]{eblerOptimalUniversalQuantum2023}%
  \BibitemOpen
  \bibfield  {author} {\bibinfo {author} {\bibfnamefont {D.}~\bibnamefont {Ebler}}, \bibinfo {author} {\bibfnamefont {M.}~\bibnamefont {Horodecki}}, \bibinfo {author} {\bibfnamefont {M.}~\bibnamefont {Marciniak}}, \bibinfo {author} {\bibfnamefont {T.}~\bibnamefont {M{\l}ynik}}, \bibinfo {author} {\bibfnamefont {M.~T.}\ \bibnamefont {Quintino}},\ and\ \bibinfo {author} {\bibfnamefont {M.}~\bibnamefont {Studzi{\'n}ski}},\ }\bibfield  {title} {\bibinfo {title} {Optimal universal quantum circuits for unitary complex conjugation},\ }\href {https://doi.org/10.1109/TIT.2023.3263771} {\bibfield  {journal} {\bibinfo  {journal} {IEEE Transactions on Information Theory}\ }\textbf {\bibinfo {volume} {69}},\ \bibinfo {pages} {5069} (\bibinfo {year} {2023})},\ \Eprint {https://arxiv.org/abs/2206.00107} {arXiv:2206.00107} \BibitemShut {NoStop}%
\bibitem [{\citenamefont {Duan}\ \emph {et~al.}(2007)\citenamefont {Duan}, \citenamefont {Feng},\ and\ \citenamefont {Ying}}]{duanEntanglementNotNecessary2007}%
  \BibitemOpen
  \bibfield  {author} {\bibinfo {author} {\bibfnamefont {R.}~\bibnamefont {Duan}}, \bibinfo {author} {\bibfnamefont {Y.}~\bibnamefont {Feng}},\ and\ \bibinfo {author} {\bibfnamefont {M.}~\bibnamefont {Ying}},\ }\bibfield  {title} {\bibinfo {title} {Entanglement {{Is Not Necessary}} for {{Perfect Discrimination}} between {{Unitary Operations}}},\ }\href {https://doi.org/10.1103/PhysRevLett.98.100503} {\bibfield  {journal} {\bibinfo  {journal} {Phys. Rev. Lett.}\ }\textbf {\bibinfo {volume} {98}},\ \bibinfo {pages} {100503} (\bibinfo {year} {2007})},\ \Eprint {https://arxiv.org/abs/quant-ph/0601150} {arXiv:quant-ph/0601150} \BibitemShut {NoStop}%
\bibitem [{\citenamefont {Duan}\ \emph {et~al.}(2009)\citenamefont {Duan}, \citenamefont {Feng},\ and\ \citenamefont {Ying}}]{duanPerfectDistinguishabilityQuantum2009}%
  \BibitemOpen
  \bibfield  {author} {\bibinfo {author} {\bibfnamefont {R.}~\bibnamefont {Duan}}, \bibinfo {author} {\bibfnamefont {Y.}~\bibnamefont {Feng}},\ and\ \bibinfo {author} {\bibfnamefont {M.}~\bibnamefont {Ying}},\ }\bibfield  {title} {\bibinfo {title} {Perfect {{Distinguishability}} of {{Quantum Operations}}},\ }\href {https://doi.org/10.1103/PhysRevLett.103.210501} {\bibfield  {journal} {\bibinfo  {journal} {Phys. Rev. Lett.}\ }\textbf {\bibinfo {volume} {103}},\ \bibinfo {pages} {210501} (\bibinfo {year} {2009})}\BibitemShut {NoStop}%
\bibitem [{\citenamefont {Fulton}(1996)}]{fultonYoungTableaux1997}%
  \BibitemOpen
  \bibfield  {author} {\bibinfo {author} {\bibfnamefont {W.}~\bibnamefont {Fulton}},\ }\href {https://doi.org/10.1017/CBO9780511626241} {\emph {\bibinfo {title} {Young Tableaux: With Applications to Representation Theory and Geometry}}},\ \bibinfo {series} {London Mathematical Society Student Texts}, Vol.~\bibinfo {volume} {35}\ (\bibinfo  {publisher} {Cambridge University Press},\ \bibinfo {year} {1996})\BibitemShut {NoStop}%
\bibitem [{\citenamefont {Watrous}(2018)}]{watrousTheoryQuantumInformation2018}%
  \BibitemOpen
  \bibfield  {author} {\bibinfo {author} {\bibfnamefont {J.}~\bibnamefont {Watrous}},\ }\href {https://doi.org/10.1017/9781316848142} {\emph {\bibinfo {title} {The {{Theory}} of {{Quantum Information}}}}},\ \bibinfo {edition} {1st}\ ed.\ (\bibinfo  {publisher} {Cambridge University Press},\ \bibinfo {address} {Cambridge, UK},\ \bibinfo {year} {2018})\BibitemShut {NoStop}%
\end{thebibliography}%

\end{document}